\documentclass[onecolumn,nofootinbib,amsmath,showpacs]{revtex4}
\usepackage{graphicx}
\usepackage{dcolumn}
\usepackage{bm}
\newcommand{\hs}{\hspace*{0.5cm}}

\newcommand{\be}{\begin{equation}}
\newcommand{\ee}{\end{equation}}
\newcommand{\bea}{\begin{eqnarray}}
\newcommand{\eea}{\end{eqnarray}}
\newcommand{\ben}{\begin{enumerate}}
\newcommand{\een}{\end{enumerate}}
\newcommand{\bde}{\begin{widetext}}
\newcommand{\ede}{\end{widetext}}
\newcommand{\nn}{\nonumber}
\newcommand{\crn}{\nonumber \\}

\newcommand{\al}{\alpha}
\newcommand{\la}{\lambda}

\newcommand{\ga}{\gamma}

\newcommand{\pa}{\partial}

\newcommand{\fr}{\frac}
\newcommand{\bc}{\begin{center}}
\newcommand{\ec}{\end{center}}

\newcommand{\La}{\Lambda}

\usepackage[usenames,dvipsnames]{color}
\definecolor{mightnightblue}{RGB}{25,25,112}
\definecolor{brown}{rgb}{0.59, 0.29, 0.0}

\def\21{$\mathrm{SU(2)_L \otimes U(1)_Y}$}

\newcommand{\AdrHEPC}{Phenikaa Institute for Advanced Study and Faculty of Basic Science, Phenikaa University, Hanoi 100000, Vietnam}
\begin{document}
\allowdisplaybreaks
\title{Kinetic mixing effect in noncommutative $B-L$ gauge theory}

\author{Duong Van Loi}\affiliation{\AdrHEPC} 

\author{Le Xuan Thuy}
\affiliation{Graduate University of Science and Technology, Vietnam Academy of Science and Technology, Hanoi 100000, Vietnam}

\author{Phung Van Dong\footnote{Corresponding author.\\ dong.phungvan@phenikaa-uni.edu.vn}}
\affiliation{\AdrHEPC}

\date{\today}

\begin{abstract}
It is well established that the $SU(P)_L$ gauge symmetry for $P\geq 3$ can address the question of fermion generation number due to the anomaly cancellation, but it neither commutes nor closes algebraically with electric and baryon-minus-lepton charges. Hence, two $U(1)$ factors that determine such charges are required, yielding a complete gauge symmetry, $SU(P)_L\otimes U(1)_X\otimes U(1)_N$, apart from the color group. The resulting theory manifestly provides neutrino mass, dark matter, inflation, and baryon asymmetry of the universe. Furthermore, this gauge structure may present kinetic mixing effects associated to the $U(1)$ gauge fields, which affect the electroweak precision test such as the $\rho$ parameter and $Z$ couplings as well as the new physics processes. We will construct the model, examine the interplay between the kinetic mixing and those due to the symmetry breaking, and obtain the physical results in detail.      
\end{abstract}

\maketitle

\section{Introduction}

The standard model of fundamental particles and interactions has been very successful in describing the observed phenomena, but it is incomplete. First of all, the experimental evidences of neutrino oscillations caused by nonzero small neutrino masses and flavor mixing require new physics beyond the standard model \cite{neutrino}. Additionally, the cosmological challenges of particle physics such as inflation, dark matter, and baryon asymmetry also acquire the standard model extension \cite{pdg2018}. Hence, it is worthwhile to look for a theory that addresses all these puzzles. 

The standard model actually contains a hidden/accident symmetry $U(1)_{B-L}$. If one includes, e.g., three right-handed neutrinos it behaves as a gauge symmetry free from all the anomalies. The resulting theory based on $SU(3)_C\otimes SU(2)_L\otimes U(1)_Y\otimes U(1)_{B-L}$ can provide consistent neutrino masses via induced seesaw mechanism \cite{seesaw}. This theory also generates suitable baryon asymmetry converted from the leptogenesis resulting from the seesaw mechanism \cite{leptog}. However, within this framework, it is not naturally to understand dark matter. Indeed, a matter parity can be induced as residual gauge symmetry due to the $U(1)_{B-L}$ breaking. However, the theory does not contain any odd field responsible for dark matter candidate. Let us note that the Majoron associated with $B-L$ breaking is actually eaten by the new neutral gauge boson, which should rapidly decay into quarks and leptons. The $B-L$ Higgs field and right-handed neutrinos are also unstable, since they decay to ordinary particles.  

In this work, we discuss a class of models based upon gauge symmetry, $SU(3)_C\otimes SU(P)_L\otimes U(1)_X\otimes U(1)_{N}$, called 3-$P$-1-1, for $P=3,4$. Here, $SU(2)_L$ is extended to $SU(P)_L$ which offers a natural solution for the question of generation number \cite{331}. It is easily verified that the electric charge $Q$ and baryon-minus-lepton charge $B-L$ neither commute nor close algebraically with $SU(P)_L$ \cite{3311,dong2015}. Hence, the two Abelian factors $U(1)_{X,N}$ are resulted from algebraic closure condition, in which the new charges $X$ and $N$ are related to $Q$ and $B-L$ via the Cartan generators of $SU(P)_L$, respectively. Besides the answer of generation number, the model manifestly accommodates dark matter which is unified with normal matter to form $SU(P)_L$ multiplets. This is a consequence of the noncommutative $B-L$ symmetry and matter partiy as a residual gauge symmetry. Such dark fields have ``wrong'' $B-L$ charge in comparison to the standard model definition, that is old under the matter parity, providing dark matter candidates. They may be a fermion, scalar or gauge boson. The abundance of dark matter observed today can either be thermally produced as a WIMP or results from a standard leptogenesis similarly to the baryon asymmetry \cite{a3311}. Therefore, in the second case both the dark and normal matter asymmetries are produced due to the CP-violating decay of the lightest right-handed neutrino. In a scenario, the $U(1)_{N}$ breaking field successfully inflates the early universe, and its decay reheats the universe producing such right-handed neutrinos, as desirable \cite{a3311}.       

The 3-3-1-1 model has been extensively investigated in the literature \cite{3311,dong2015,a3311}, but the 3-4-1-1 model has not considered yet. In this work, we construct the 3-4-1-1 model with general fermion and scalar contents, obtain the matter parity, and interpret dark matter candidates. Since the theory contains two $U(1)$ factors, the kinetic mixing between the corresponding gauge bosons is not avoidable \cite{kineticmixing}. Therefore, we diagonalize the gauge boson sector when including the kinetic mixing term. The effect of the kinetic mixing is present in the $\rho$ parameter and the coupling of $Z$ with fermions, which can alter the electroweak precision test. It significantly modifies the neutral meson mixings and rare meson decays. The last aim of this work is to probe the new physics of the model at the LHC. This work also revisits the kinetic mixing effect in the 3-3-1-1 model, which was previously studied \cite{dong2016}.

The rest of this work is organized as follows. In Sec. \ref{model}, we introduce the model and show dark matter. In Sec. \ref{gauge}, we diagonalize the gauge sector. In Sec. \ref{pheno}, we examine the $\rho$ parameter, mixing parameters, and the $Z$ couplings. In Sec. \ref{pheno1}, we investigate the FCNCs. The search for the new physics is presented in Sec. \ref{lhc}. In Sec. \ref{3311}, the kinetic mixing effect in a previous study is revisited. Finally, we conclude this work in Sec. \ref{con}. 

\section{\label{model} The model}

In this section we propose the 3-4-1-1 model, while the 3-3-1-1 model \cite{dong2015,dong2016} was well established and skipped.  

\subsection{Gauge symmetry}

As stated, the gauge symmetry is given by
\be SU(3)_C\otimes SU(4)_L\otimes U(1)_X\otimes U(1)_N.\ee
The electric and baryon-minus-lepton charges are embedded as 
\bea && Q=T_3+\beta T_8 + \ga T_{15}+X,\\
&& B-L= b T_8 +c T_{15}+N,\eea where $T_i$ ($i=1,2,3,...,15$), $X$, and $N$ are $SU(4)_L$, $U(1)_X$, and $U(1)_N$ charges, respectively.

Nontrivial commutation relations are obtained by 
\bea && [Q,T_1\pm i T_2]=\pm(T_1\pm i T_2),\crn &&[Q,T_4\pm i T_5]=\mp q(T_4\pm i T_5),\crn 
&& [Q,T_6\pm i T_7]=\mp(1+q)(T_6\pm i T_7),\crn
&& [Q,T_9\pm i T_{10}]=\mp p(T_9\pm i T_{10}),\crn 
&& [Q,T_{11}\pm i T_{12}]=\mp(1+p)(T_{11}\pm i T_{12}),\crn
&& [Q,T_{13}\pm i T_{14}]=\mp(p-q)(T_{13}\pm i T_{14}),\crn
&& [B-L,T_4\pm i T_5]=\mp (1+n) (T_4\pm i T_5),\crn
&& [B-L,T_6\pm i T_7]=\mp(1+n)(T_6\pm i T_7),\crn 
&& [B-L,T_9\pm i T_{10}]=\mp (1+m)(T_9\pm i T_{10}),\crn 
&& [B-L,T_{11}\pm i T_{12}]=\mp(1+m)(T_{11}\pm i T_{12}),\crn
&& [B-L,T_{13}\pm i T_{14}]=\mp(m-n)(T_{13}\pm i T_{14}),\eea where we define the basic electric charges as $q=-(1+\sqrt{3}\beta)/2$ and $p=-(1+\sqrt{6}\ga-q)/3$ and the basic baryon-minus-lepton charges as $n=-(2+\sqrt{3}b)/2$ and $m=-(2+\sqrt{6}c-n)/3$. Hence, ($q,p$) and ($n,m$) will determine the $Q$ and $B-L$ charges of new particles, respectively.   

\subsection{Particle presentation}

The fermions transform under the 3-4-1-1 gauge symmetry as 
\bea && \psi_{aL}\equiv \left(\begin{array}{c}\nu\\ e \\ E^{q,n}\\ F^{p,m}\end{array}\right)_{aL}\sim \left(1,4,\fr{p+q-1}{4},\fr{m+n-2}{4}\right), \\ 
&& Q_{\al L}\equiv \left(\begin{array}{c}d \\ -u \\ J^{-q-1/3,-n-2/3}\\ K^{-p-1/3,-m-2/3} \end{array}\right)_{\al L}\sim \left(3,4^*,-\fr{p+q+1/3}{4},-\fr{m+n+2/3}{4}\right),\\
&& Q_{3 L}\equiv \left(\begin{array}{c}u \\ d \\ J^{q+2/3,n+4/3}\\ K^{p+2/3,m+4/3}\end{array}\right)_{3L}\sim \left(3,4,\fr{p+q+5/3}{4},\fr{m+n+10/3}{4}\right),\\ 
&& \nu_{aR}\sim \left(1,1,0,-1\right),\hs e_{aR}\sim \left(1,1,-1,-1\right),\\ 
&& E_{aR}\sim \left(1,1,q,n\right), \hs F_{aR}\sim \left(1,1,p,m\right),\\  
&& u_{a R}\sim \left(3,1,2/3,1/3\right),\hs d_{aR}\sim \left(3,1,-1/3,1/3\right),\\
&& J_{\al R}\sim \left(3,1,-q-1/3,-n-2/3\right),\hs K_{\al R}\sim \left(3,1,-p-1/3,-m-2/3\right), \\
&&J_{3R}\sim \left(3,1,q+2/3,n+4/3\right),\hs K_{3R}\sim \left(3,1,p+2/3,m+4/3\right),\eea
where $a=1, 2, 3$ and $\alpha =1, 2$ denote generation indices. Additionally, $\nu_{R}, E, F, J$, and $K$ are new fields, included to complete the representations. This fermion content is independent of all the anomalies (cf. Appendix \ref{appa}).

In order for gauge symmetry breaking and mass generation, we introduce the scalar content,
\bea \eta &=&
\left(
\begin{array}{l}
\eta^{0,0}_1\\
\eta^{-1,0}_2\\
\eta^{q,n+1}_3\\
\eta^{p,m+1}_4
\end{array}\right)\sim \left(1,4,\fr{p+q-1}{4},\fr{m+n+2}{4}\right),\\
\rho &=&
\left(
\begin{array}{l}
\rho^{1,0}_1\\
\rho^{0,0}_2\\
\rho^{q+1,n+1}_3\\
\rho^{p+1,m+1}_4
\end{array}\right)\sim \left(1,4,\fr{p+q+3}{4},\fr{m+n+2}{4}\right),\\
\chi &=&
\left(
\begin{array}{l}
\chi^{-q,-n-1}_1\\
\chi^{-q-1,-n-1}_2\\
\chi^{0,0}_3\\
\chi^{p-q,m-n}_4
\end{array}\right)\sim\left(1,4,\fr{p-3q-1}{4},\fr{m-3n-2}{4}\right), \\
 \Xi &=&
\left(
\begin{array}{l}
\Xi^{-p,-m-1}_1\\
\Xi^{-p-1,-m-1}_2\\
\Xi^{-p+q,-m+n}_3\\
\Xi^{0,0}_4
\end{array}\right)\sim\left(1,4,\fr{-3p+q-1}{4},\fr{-3m+n-2}{4}\right),\\
\phi &\sim& (1,1,0,2),\eea
where the superscipts stand for ($Q,B-L$) respectively, while the subscripts indicate $SU(4)_L$ components. The scalars obtain such quantum numbers, provided that they couple left-handed fermions to corresponding right-handed counterparts, except that $\phi$ couples to $\nu_R\nu_R$ (see below).
\subsection{Total Lagrangian}

The total Lagrangian has the form, 
\bea
\mathcal{L}= \mathcal{L}_{\mathrm{kinetic}} + \mathcal{L}_{\mathrm{Yukawa}}-V,
\eea where the first part combines kinetic terms and gauge interactions, given by 
\bea \mathcal{L}_{\mathrm{kinetic}} &=& \sum_F \bar{F}i\ga^\mu D_\mu F+\sum_S (D^\mu S)^\dagger (D_\mu S)\crn
&& -\fr 1 4 G_{r \mu\nu} G_r^{\mu\nu}-\fr 1 4 A_{i \mu\nu} A_i^{\mu\nu}-\fr 1 4 B_{\mu\nu} B^{\mu\nu}-\fr 1 4 C_{\mu\nu} C^{\mu\nu}-\fr{\delta}{2} B_{\mu\nu} C^{\mu\nu}.\eea The covariant derivative is 
\be D_\mu = \pa_\mu + i g_s T_r G_{r\mu}+ i g T_i A_{i \mu}+ i g_X X B_\mu + i g_N N C_\mu,\ee and we denote the coupling constants $(g_s,g,g_X,g_N)$, generators $(T_r,T_i,X,N)$, and gauge bosons $(G_r, A_i, B,C)$ corresponding to the 3-4-1-1 subgroups, respectively. Above, $F$ and $S$ run over the fermion and scalar multiplets, while the parameter $\delta$ is dimensionless, called kinetic mixing.\footnote{This kinetic mixing term is always presented due to the gauge invariance and cannot be removed by rescaling the corresponding fields. Even if its tree-level value vanishes, it can be radiatively induced \cite{kineticmixing}.}      

The second and last parts are the Yukawa interactions and scalar potential, given respectively by 
\bea \mathcal{L}_{\mathrm{Yukawa}}&=&h^\nu_{ab}\bar{\psi}_{aL}\eta\nu_{bR}+ h^e_{ab}\bar{\psi}_{aL}\rho e_{bR} + h^E_{ab}\bar{\psi}_{aL}\chi E_{bR}+ h^{F}_{ab}\bar{\psi}_{aL}\Xi F_{bR}+h'^\nu_{ab}\bar{\nu}^c_{aR}\nu_{bR}\phi\crn
&& + h^J_{33}\bar{Q}_{3L}\chi J_{3R}+ h^{K}_{33}\bar{Q}_{3L}\Xi K_{3R} + h^J_{\al \beta}\bar{Q}_{\al L} \chi^* J_{\beta R}+ h^{K}_{\al \beta}\bar{Q}_{\al L} \Xi^* K_{\beta R} \crn
&&+ h^u_{3a} \bar{Q}_{3L}\eta u_{aR}+h^u_{\al a } \bar{Q}_{\al L}\rho^* u_{aR} +h^d_{3a}\bar{Q}_{3L}\rho d_{aR} + h^d_{\al a} \bar{Q}_{\al L}\eta^* d_{aR} +H.c.,\\ 
V &=& \mu^2_1\eta^\dagger \eta + \mu^2_2 \rho^\dagger \rho + \mu^2_3 \chi^\dagger \chi + \mu^2_4 \Xi^\dagger \Xi + \mu^2_5 \phi^\dagger \phi +\la_1 (\eta^\dagger \eta)^2 + \la_2 (\rho^\dagger \rho)^2 \crn
&&+ \la_3 (\chi^\dagger \chi)^2 + \la_4 (\Xi^\dagger \Xi)^2+ \la_5 (\phi^\dagger \phi)^2 + \la_6 (\eta^\dagger \eta)(\rho^\dagger \rho) +\la_7 (\eta^\dagger \eta)(\chi^\dagger \chi)\crn
&&+\la_8 (\eta^\dagger \eta)(\Xi^\dagger \Xi)+\la_{9} (\eta^\dagger\eta)(\phi^\dagger \phi)+\la_{10} (\rho^\dagger \rho)(\chi^\dagger \chi) +\la_{11} (\rho^\dagger \rho)(\Xi^\dagger \Xi) \crn
&& +\la_{12}(\rho^\dagger \rho)(\phi^\dagger \phi)+\la_{13}(\chi^\dagger \chi)(\Xi^\dagger \Xi)+\la_{14}(\chi^\dagger \chi)(\phi^\dagger \phi) +\la_{15}(\Xi^\dagger \Xi)(\phi^\dagger \phi) \crn
&&+\la_{16} (\eta^\dagger \rho)(\rho^\dagger \eta) +\la_{17} (\eta^\dagger \chi)(\chi^\dagger \eta)+\la_{18} (\eta^\dagger \Xi)(\Xi^\dagger \eta)+\la_{19} (\rho^\dagger \chi)(\chi^\dagger \rho)\crn
&& +\la_{20} (\rho^\dagger \Xi)(\Xi^\dagger \rho)+\la_{21} (\chi^\dagger \Xi)(\Xi^\dagger \chi)+ (\la\eta \rho \chi \Xi+H.c.), 
\eea where the Yukawa ($h$'s) and scalar ($\la$'s) couplings are dimensionless, while the $\mu$'s parameters have the mass dimension.

\subsection{Matter parity}

Since $Q$ is conserved, only the neutral components $\eta_1, \rho_2, \chi_3, \Xi_4$, and $\phi$ develop vacuum expectation values (VEVs),
\bea \langle \eta \rangle &=& \fr{1}{\sqrt{2}}\left(
\begin{array}{c}
u \\
0\\
0\\
0
\end{array}\right),\hs
\langle \rho\rangle =
\fr{1}{\sqrt{2}} \left(
\begin{array}{c}
0\\
v \\
0\\
0
\end{array}\right),\hs
\langle \chi\rangle =
\fr{1}{\sqrt{2}} \left(
\begin{array}{c}
0\\
0\\
w\\
0
\end{array}\right),\hs
\langle \Xi\rangle =
\fr{1}{\sqrt{2}} \left(
\begin{array}{c}
0\\
0\\
0\\
V
\end{array}\right),\hs
 \langle \phi\rangle = \fr{1}{\sqrt{2}} \La.\eea
 
The VEVs $V, w, u, v$ break the 3-4-1-1 symmetry to $SU(3)_C\otimes U(1)_Q\otimes U(1)_{B-L}$, while the VEV $\La$ breaks $B-L$ to the matter parity, $U(1)_{B-L}\rightarrow P$, where \be P=(-1)^{3(B-L)+2s}=(-1)^{3(bT_8+cT_{15}+N)+2s}\ee is multiplied by the spin parity $(-1)^{2s}$ as conserved by the Lorentz symmetry, similar to the 3-3-1-1 model \cite{3311,dong2015}.\footnote{This kind of the matter parity is also recognized in the class of the left-right extensions \cite{3331}.} 

Because $w, V, \La$ provide the masses of new particles, whereas $u,v$ do so for the ordinary particles, we assume $u,v\ll w, V, \La$, to keep a consistency with the standard model.
 
The matter parity $P$ divides particles into two types: \ben \item Normal particles according to $P=1$ (even): $\nu$, $e$, $u$, $d$, $\eta_{1,2}$, $\rho_{1,2}$, $\chi_3$, $\Xi_4$, $\phi$, $\gamma$, $W$, $Z_{1,2,3,4}$, which include the standard model particles. \item Wrong particles according to $P=P^\pm_n\equiv (-1)^{\pm(3n+1)}$ or $P=P^\pm_m\equiv (-1)^{\pm(3m+1)}$: $E$, $F$, $J$, $K$, $\eta_{3,4}$, $\rho_{3,4}$, $\chi_{1,2,4}$, $\Xi_{1,2,3}$, $W_{13}^{\mp q,\mp (n+1)}$, $W_{14}^{\mp p,\mp (m+1)}$, $W_{23}^{\mp (q+1),\mp (n+1)}$, $W_{24}^{\mp (p+1),\mp (m+1)}$, where the $W$'s fields are non-Hermitian gauge bosons which couple to the mentioned weight-raising/lowering operators. The remainders $\chi_4$, $\Xi_3$, and $W_{34}^{\pm (q-p),\pm (n-m)}$ have $P=P^+_n P^-_m$ or conjugated. \een

Generally, the wrong fields transform nontrivially under the matter parity for $n,m\neq (2k-1)/3$ and $n-m\neq 2k/3$ for every $k$ integer. However, an alternative case is that both $P_{n,m}=-1$ are odd, i.e. $n,m=2k/3$. In this case all the wrong fields are odd, except that $\chi_4$, $\Xi_3$, and $W_{34}$ are even which belong to the first type of normal particles.
  
\subsection{Dark matter} 

It is easily to prove that the wrong particles always couple in pairs or self-interacted due to the matter parity conservation, which is analogous to superparticles in supersymmetry \cite{dong2015} (see also Dong and Huong in \cite{3331}). Hence, the lightest wrong particle (LWP) is stabilized, responsible for dark matter. 

Since the candidate must be color and electrically neutral, we have several dark matter models: (i) $q=0$ including $E^0$, $\eta^0_3$, $\chi^0_1$, $W^0_{13}$; (ii) $p=0$ including $F^0$, $\eta^0_4$, $\Xi^0_1$, $W^0_{14}$; (iii) $q=-1$ consisting of $\rho^0_3$, $\chi^0_2$, $W^0_{23}$; (iv) $p=-1$ consisting of $\rho^0_4$, $\chi^0_2$, $W^0_{24}$. In each case, the remaining basic electric charge is left arbitrary.        

The specific dark matter models that combine above cases are  
 \begin{enumerate}
\item $q=p=-1$: The candidate is a scalar combination of $\rho_{3,4}$, $\chi_2$, and $\Xi_2$, or a gauge boson combination of $W_{23}$ and $W_{24}$.
\item $q=-1, p=0$: The candidate is a fermion combination of $F_{1,2,3}$, a scalar combination of $\eta_4$, $\rho_3$, $\chi_2$, and $\Xi_1$, or a gauge boson combination of $W_{14}$ and $W_{23}$.
\item $q=0, p=-1$: The candidate is a fermion combination of $E_{1,2,3}$, a scalar combination of $\eta_3$, $\rho_4$, $\chi_1$, and $\Xi_2$, or a gauge boson combination of  $W_{13}$ and $W_{24}$.
\item $q=p=0$: The candidate is a fermion combination of $E_{1,2,3}$ and $F_{1,2,3}$, a scalar combination of $\eta_{3,4}$, $\chi_1$, and $\Xi_1$, or a gauge boson combination of  $W_{13}$ and $W_{14}$.
\end{enumerate}

The last model is for $p=q\neq 0,-1$. The candidate includes a scalar combination of $\chi^0_4$ and $\Xi^0_3$, or a vector $W^0_{34}$.

\subsection{Fermion mass} 
When the scalars develop VEVs, the fermions gain masses and we write Dirac masses as $-\bar{f}_L m_f f_R+H.c.$ and Majorana masses as $-\fr 1 2 \bar{f}^c_{L,R} m^{L,R}_f f_{L,R}+H.c.$ 

The mass matrices of new fermions $E_a$, $F_a$, $J_a$, and $K_a$ are given by \bea && [m_E]_{ab}=-h^E_{ab}\fr{w}{\sqrt{2}},\hs  [m_{F}]_{ab}=-h^{F}_{ab}\fr{V}{\sqrt{2}},\\ 
&& [m_{J}]_{33}=-h^{J}_{33}\fr{w}{\sqrt{2}},\hs  [m_{J}]_{\alpha\beta}=-h^J_{\alpha\beta}\fr{w}{\sqrt{2}},\\ && 
[m_{K}]_{33}=-h^{K}_{33}\fr{V}{\sqrt{2}},\hs
[m_{K}]_{\alpha\beta}=-h^{K}_{\alpha\beta}\fr{V}{\sqrt{2}},\eea which all have masses at $w$, $V$ scale. 

The mass matrices of charged-leptons and quarks $e_a$, $u_a$ and $d_a$ are obtained as \bea && [m_e]_{ab}=-h^e_{ab}\fr{v}{\sqrt{2}},\crn
&& [m_u]_{3 a}=-h^u_{3 a}\fr{u}{\sqrt{2}},\hs [m_u]_{\al a}=h^u_{\al a}\fr{v}{\sqrt{2}},\crn 
&& [m_d]_{3 a}=-h^d_{3 a}\fr{v}{\sqrt{2}},\hs [m_d]_{\al a}=-h^d_{\al a}\fr{u}{\sqrt{2}},\eea which provide appropriate masses at $u,v$ scale. 

For the neutrinos, $\nu_{aL,R}$, the Dirac and Majorana masses are $[m_\nu]_{ab}=-h^\nu_{ab}\fr{u}{\sqrt{2}}$ and $[m^R_\nu]_{ab}=-\sqrt{2}h'^\nu_{ab}\La$, respectively. Since $u\ll \La$, the observed neutrinos $(\sim \nu_{aL})$ achieve masses via the type I seesaw mechanism, \be m^L_\nu \simeq - m_\nu (m^R_\nu)^{-1} (m_\nu)^T\sim u^2/\La,\ee which is small, as expected. The sterile neutrinos $(\sim \nu_{aR})$ obtain large masses, such as $m^R_\nu$.

\section{\label{gauge}Kinetic mixing}

\subsection{Canonical basis}

Let us write down the kinetic terms of the two $U(1)$ gauge fields as
\be \mathcal{L}_{\mathrm{kinetic}}\supset -\fr 1 4 B^2_{\mu\nu}-\fr 1 4 C^2_{\mu\nu}-\fr{\delta}{2}  B_{\mu\nu}C^{\mu\nu}= -\fr 1 4 (B_{\mu\nu}+\delta C_{\mu\nu})^2-\fr 1 4 (1-\delta^2) C^2_{\mu\nu},\ee where $B_{\mu\nu}=\pa_\mu B_\nu-\pa_\nu B_\mu$ and $C_{\mu\nu}=\pa_\mu C_\nu-\pa_\nu C_\mu$ are the corresponding field strength tensors.   

Because of the kinetic mixing term ($\delta$), the two gauge bosons $B_\mu$ and $C_\mu$ are generally not orthonormalized. We change to the canonical basis by a nonunitary transformation $(B_\mu,C_\mu)\rightarrow (B'_\mu,C'_\mu)$, where
\be B'=B+\delta C,\hs C'=\sqrt{1-\delta^2}C.\ee 

We substitute $B,C$ in terms of $B',C'$ into the covariant derivative. It becomes       
\be 
D_\mu \supset ig_X X B_\mu+ig_N N C_\mu =  i g_X X B'_\mu+\fr{i}{\sqrt{1-\delta^2}}(g_N N-g_X X\delta)  C'_\mu, 
\ee which is given in terms of the orthonormalized (canonical) fields $(B'_\mu,C'_\mu)$. 

\subsection{Gauge boson mass}
The 3-4-1-1 symmetry breaking leads to mixings among $A_{3}$, $A_8$, $A_{15}$, $B'$, and $C'$. Their mass Lagrangian arises from $\sum_S (D_\mu\langle S\rangle)^\dagger (D^\mu\langle S\rangle)$, such that  
\bea 
 \mathcal{L}_{\mathrm{mass}}^{\mathrm{neutral}}  =
  \fr1 2\left(A_3 \  A_8  \ A_{15} \  B' \ C'  \right) M^2 \left(A_3 \  A_8  \ A_{15} \  B' \ C'  \right)^T, \eea 
 where the mass matrix $M^2=\{m^2_{ij}\}$ is symmetric, possessing the elements,  
 \bea
m^2_{11}&= & \fr {g^2} 4 (u^2 + v^2),\hs m^2_{12}=  \fr {g^2} {4\sqrt3} (u^2-v^2),\hs m^2_{13} = \fr {g^2} {4\sqrt6} (u^2-v^2),\crn
m^2_{14} &=& -\fr{g^2t_X}{4\sqrt6}[\beta_1u^2+(2\sqrt6-\beta_1)v^2], \crn 
m^2_{15}&=& \fr{g^2}{4\sqrt{6(1-\delta^2)}}\{[\delta\beta_1t_X-(\sqrt2b+ c )t_N] u^2 +[\delta (2\sqrt6-\beta_1)t_X+(\sqrt2b+c )t_N] v^2 \} ,\crn
m^2_{22}&=&\fr {g^2} {12}(u^2+v^2+4w^2),\hs m^2_{23}=\fr {g^2} {12\sqrt2}(u^2+v^2-2w^2),\crn
m^2_{24}&=& -\fr{g^2t_X}{12\sqrt2}[\beta_1u^2-(2\sqrt6-\beta_1)v^2+2(2\sqrt2\beta-\gamma) w^2], \crn
m^2_{25}&=& \fr{g^2} { 12\sqrt{2(1-\delta^2)}}\{[\delta \beta_1t_X-(\sqrt2b+c )t_N] u^2-[\delta (2\sqrt6-\beta_1)t_X+(\sqrt2b+c )t_N] v^2\crn
&& +2[\delta(2\sqrt2\beta-\gamma) t_X-(2\sqrt2 b-c) t_N]w^2\},\crn
m^2_{33}&=&\fr {g^2} {24}(u^2+v^2+w^2+9V^2),\hs m^2_{34}= -\fr{g^2t_X}{24}[\beta_1 u^2-(2\sqrt6-\beta_1) v^2 -(2\sqrt2\beta-\gamma)w^2+9\gamma V^2],\crn
m^2_{35}&=& \fr{g^2} {24\sqrt{1-\delta^2}}\{[\delta\beta_1t_X-(\sqrt2b+c )t_N] u^2-[\delta (2\sqrt6-\beta_1)t_X+(\sqrt2b+c )t_N] v^2\crn
&&-[\delta(2\sqrt2\beta-\gamma) t_X-(2\sqrt2 b-c) t_N]w^2+9 (\delta\gamma t_X-ct_N)V^2\},\crn
m^2_{44}&=& \fr{g^2t^2_X} {24}[\beta_1^2u^2+(2\sqrt6-\beta_1)^2 v^2+(2\sqrt2\beta-\gamma)^2w^2+9 \gamma^2V^2],\crn
m^2_{45}&=&-\fr{g^2t_X} {24\sqrt{1-\delta^2}}\{[\delta\beta_1t_X-(\sqrt2 b+c)t_N] \beta_1u^2+[\delta (2\sqrt6-\beta_1)t_X+(\sqrt2 b+c)t_N] (2\sqrt6-\beta_1)v^2\crn
&&+[\delta (2\sqrt2\beta-\gamma )t_X-(2\sqrt2 b-c)t_N](2\sqrt2\beta- \gamma)w^2+9\gamma (\delta\gamma t_X-ct_N) V^2\},\crn
m^2_{55}&=& \fr{g^2} {24(1-\delta^2)}\{[\delta \beta_1t_X-(\sqrt2 b+c )t_N]^2 u^2+[\delta (2\sqrt6-\beta_1)t_X+(\sqrt2 b+c )t_N]^2v^2\crn
&&+[\delta (2\sqrt2 \beta-\gamma)t_X-(2\sqrt2 b-c )t_N]^2w^2+9(\delta\gamma t_X-ct_N)^2 V^2+96t_N^2\Lambda^2\},\nn
\eea where we have defined $t_X=g_X/g$, $t_N=g_N/g$, and $\beta_1=\sqrt6+\sqrt2\beta+\gamma$.

The mass matrix always provides a zero eigenvalue with corresponding eigenstate (photon field),
\be
 A = s_W A_3 + c_W \left(\beta t_W A_8 +\gamma t_W A_{15} +\frac {t_W}{t_X} B'\right),\label{amu}\ee where $s_W=e/g=t_X/\sqrt{1+(1+\beta^2+\gamma^2)t_X^2}$ is the sine of the Weinberg's angle \cite{donglong}. Since the field in parentheses of (\ref{amu}) is properly the hypercharge field coupled to $Y=Q-T_3$, we define the standard model $Z$ as
 \bea 
Z=c_W A_3 -s_W \left(\beta t_W A_8 +\gamma t_W A_{15} +\frac {t_W}{t_X} B'\right).\eea 
The new neutral gauge bosons, called $Z'_2, Z'_3$, orthogonal to the hypercharge field take the forms,  
\bea
Z'_2 &=& \frac{1}{\sqrt{1-\beta^2 t_W^2}} \left[(1-\beta^2 t_W^2) A_8-\beta\gamma t_W^2 A_{15}-\frac{\beta t_W^2}{t_X}  B'\right], \\
Z'_3 &=& \frac{1}{\sqrt{1+\gamma^2t_X^2}}\left(A_{15}- 
 \gamma t_X B'\right).
\eea
At this stage, $C'$ is always orthogonal to $A, Z, Z'_2, Z'_3$.

Let us change to the new basis $A, Z, Z'_2, Z'_3$, and $C'$, such that $(A_3\, A_8\, A_{15}\, B'\, C')^T=U_1(A\, Z\, Z'_2\, Z'_3\, C')^T$, where
\bea
U_1= \left(%
\begin{array}{ccccc}
s_W&c_W&0&0&0\\
\beta s_W&-\beta s_Wt_W&\sqrt{1-\beta^2t_W^2}&0&0\\
\gamma s_W&-\gamma s_Wt_W&-\frac{\beta\gamma t_W^2}{\sqrt{1-\beta^2t^2_W}}&\fr {1}{\sqrt{1+\gamma^2t_X^2}}&0\\
\frac{s_W}{t_X}&-\frac{s_Wt_W}{t_X}&-\frac{\beta t_W^2}{t_X\sqrt{1-\beta^2 t_W^2}}&-\frac{\gamma t_X}{\sqrt{1+\gamma^2t_X^2}}&0\\0&0&0&0&1
\end{array} \right).
\eea
The mass matrix $M^2$ is correspondingly changed to 
\bea
 M'^{2} =U_1^T M^2 U_1 =\left(%
\begin{array}{ccc} 
0& 0 \\
0& M_s^{\prime 2}
\end{array} \right), \hs M'^2_s \equiv
\left(
\begin{array}{cccc}
m^2_{Z} & m^2_{ZZ'_2}& m^2_{ZZ'_3} & m^2_{ZC'}\\
m^2_{ZZ'_2} & m^2_{Z'_2} & m^2_{Z'_2Z'_3} & m^2_{Z'_2C'}\\
m^2_{ZZ'_3} & m^2_{Z'_2Z'_3} & m^2_{Z'_3} & m^2_{Z'_3C'}\\
m^2_{ZC'} & m^2_{Z'_2C'} & m^2_{Z'_3C'}& m^2_{C'}
\end{array}\right). 
\eea where
 \bea
 m^2_Z&=&\fr {g^2}{4c^2_W} (u^2 + v^2),\hs m^2_{ZZ'_2} =\fr{g^2\sqrt{1-(\beta^2+\gamma^2)t_W^2}}{4\sqrt3 c_W\sqrt{1+\gamma^2 t^2_X}}\{\beta_2u^2+(2\sqrt3 \beta t_X^2-\beta_2)v^2\},\crn 
 m^2_{ZZ'_3}& =&\fr{g^2}{4\sqrt6 c_W\sqrt{1+\gamma^2 t^2_X}}\{(1+\gamma\beta_1t_X^2)u^2+[\gamma(2\sqrt6-\beta_1)t_X^2-1]v^2\},\crn 
m^2_{ZC'}&=& \fr{g^2}{4\sqrt6 c_W \sqrt{1-\delta^2}}\{[\delta\beta_1t_X-(\sqrt2 b+c)t_N]u^2+[\delta(2\sqrt6-\beta_1)t_X+(\sqrt2 b+c)t_N]v^2\},\crn 
 m^2_{Z'_2}&= & \fr{g^2[1-(\beta^2+\gamma^2)t_W^2]}{12(1+\gamma^2t_X^2)}\{\beta_2^2u^2+(2\sqrt3\beta t_X^2-\beta_2)^2v^2+4[1+(\beta^2+\gamma^2)t_X^2]^2w^2\} ,\crn
m^2_{Z'_2Z'_3}&= & \fr{g^2\sqrt{1-(\beta^2+\gamma^2)t_W^2}}{12\sqrt2(1+\gamma^2t_X^2)}\{(1+\gamma\beta_1t_X^2)\beta_2u^2+[\gamma(2\sqrt6-\beta_1)t_X^2-1](2\sqrt3\beta t_X^2-\beta_2)v^2\crn
&&+2[\gamma(2\sqrt2 \beta-\gamma)t_X^2-1][1+(\beta^2+\gamma^2)t_X^2]w^2\},\crn
m^2_{Z'_2C'}&= & \fr{g^2\sqrt{1-(\beta^2+\gamma^2)t_W^2}}{12\sqrt2\sqrt{(1-\delta^2)(1+\gamma^2t_X^2)}}\{[\delta\beta_1t_X-(\sqrt2 b+c)t_N]\beta_2u^2+[\delta(2\sqrt6-\beta_1)t_X+(\sqrt2 b+c)t_N]\crn
&&\times(2\sqrt3\beta t_X^2-\beta_2)v^2+2[\delta(2\sqrt2\beta-\gamma)t_X-(2\sqrt2 b-c)t_N][1+(\beta^2+\gamma^2)t_X^2]w^2\},\crn
m^2_{Z'_3}&= &\fr{g^2}{24(1+\gamma^2t_X^2)}\{(1+\gamma\beta_1t_X^2)^2u^2+[\gamma(2\sqrt6-\beta_1)t_X^2-1]^2v^2+[\gamma(2\sqrt2 \beta-\gamma)t_X^2-1]^2w^2\crn
&&+9(1+\gamma^2t_X^2)^2V^2\},\crn
m^2_{Z'_3C'}&= &\fr{g^2}{24\sqrt{(1-\delta^2)(1+\gamma^2t_X^2)}}\{[\delta\beta_1t_X-(\sqrt2 b+c)t_N](1+\gamma\beta_1t_X^2)u^2+[\delta(2\sqrt6-\beta_1)t_X+(\sqrt2 b+c)t_N]\crn
&&\times[\gamma(2\sqrt6-\beta_1)t_X^2-1]v^2+[\delta(2\sqrt2 \beta-\gamma)t_X-(2\sqrt2 b-c)t_N][\gamma(2\sqrt2 \beta-\gamma)t_X^2-1]w^2\crn
&&+9(\delta\gamma t_X-ct_N)(1+\gamma^2t_X^2)V^2\},\crn
  m^2_{C'}&= & m^2_{55}, \nn
 \eea where we have defined $\beta_2=1+(\sqrt3 \beta+\beta^2+\gamma^2)t_X^2$.

Since $u, v \ll w, V, \Lambda$, the first row and first column of $M'^2_s$ consist of the elements much smaller than those of the remaining entries. We diagonalize $M'^2_s$ using the seesaw formula \cite{seesaw} that separates $Z$ from the heavy fields, given by   
 \bea
\left(Z \, Z'_2 \, Z'_3 \, C' \right)^T = U_2 \left(Z_1 \, \mathcal{Z}'_2 \, \mathcal{Z}'_3 \, \mathcal{C}' \right)^T, \hs M''^2=U^{T}_2M'^2_sU_2= \left(\begin{array}{cc}
m^{2}_{Z_{1}} &0 \\
0&M''^2_{s}\end{array}\right),\label{tran2}
\eea
where $Z_1$ is physical as decoupled, while $\mathcal{Z}'_2$, $\mathcal{Z}'_3$ and $\mathcal{C}'$ mix via $M''^2_s$, such that  
\bea
U_2 &\simeq &  \left(\begin{array}{cccc}
1 & \epsilon_1 & \epsilon_2 & \epsilon_3  \\
-\epsilon_1 &  1 & 0 & 0\\
-\epsilon_2 & 0 & 1 & 0 \\
-\epsilon_3 & 0 & 0 & 1 \\
\end{array}
\right), \hs  M''^2_s \simeq 
\left(
\begin{array}{ccc}
 m^2_{Z'_2} & m^2_{Z'_2Z'_3} & m^2_{Z'_2C'}\\
 m^2_{Z'_2Z'_3} & m^2_{Z'_3} & m^2_{Z'_3C'}\\
 m^2_{Z'_2C'} & m^2_{Z'_3C'}& m^2_{C'}
\end{array}\right),\label{tranu2}\\
m^{2}_{Z_{1}}&\simeq & m^2_Z - \epsilon_1 m^2_{ZZ'_2} -  \epsilon_2 m^2_{ZZ'_3} -  \epsilon_3 m^2_{ZC'}.
\eea
We further separate $\epsilon_{1,2,3}\equiv\epsilon_{1,2,3}^0+\epsilon_{1,2,3}^\delta$, where $\epsilon^0_{1,2,3}$ are the mixing of $Z$ with $Z'_2, Z'_3,$ and $C'$ due to the symmetry breaking, whereas $\epsilon^{\delta}_{1,2,3}$ determine those mixings due to the kinetic mixing,  
\bea
\epsilon_1^0&=&\frac{1}{4c_W\sqrt{1+\gamma^2 t_X^2}[1+(\beta^2+\gamma^2)t_X^2]^{3/2}}\left\{\frac{\sqrt3(1+\gamma^2 t_X^2)[\beta_2u^2+(2\sqrt3 \beta t_X^2-\beta_2)v^2]}{w^2}\right.\crn
&&+\frac{(\beta+2\sqrt2\gamma)[1+\gamma(\gamma-2\sqrt2\beta)t_X^2]t_X^2(u^2+v^2)+\sqrt3[1-\gamma(2\sqrt2\beta-\gamma)t_X^2][1+(\beta^2+\gamma^2)t_X^2](u^2-v^2)}{3V^2}\crn
&&\left.+\frac{(b\beta+c\gamma)[b-\gamma(c\beta-b\gamma)t_X^2]t_X^2(u^2+v^2)}{4\Lambda^2}\right\},\\
\epsilon_2^0&=&\frac{1}{c_W\sqrt{1+\gamma^2 t_X^2}[1+(\beta^2+\gamma^2)t_X^2]}\left\{\frac{(\beta+2\sqrt2\gamma)t_X^2(u^2+v^2)+\sqrt3[1+(\beta^2+\gamma^2)t_X^2](u^2-v^2)}{3\sqrt2V^2}\right.\crn
&&\left.+\frac{c(b\beta+c\gamma)t_X^2(u^2+v^2)}{16\Lambda^2}\right\},\\
\epsilon_3^0&=&\frac{(b\beta+c\gamma)t_X^2(u^2+v^2)}{16c_W[1+(\beta^2+\gamma^2)t_X^2]t_N\Lambda^2},\\
\epsilon_1^\delta&=&\frac{\delta\{[b(1+\gamma^2 t_X^2)-\beta(b\beta+2c\gamma)t_X^2]t_N-\delta\beta t_X\}t_X(u^2+v^2)}{16c_W\sqrt{1+\gamma^2 t_X^2}[1+(\beta^2+\gamma^2)t_X^2]^{3/2}t_N^2\Lambda^2},\\
\epsilon_2^\delta&=&\frac{\delta[(ct_N-\delta\gamma t_X)-\gamma(b\beta+c\gamma)t_X^2t_N]t_X(u^2+v^2)}{16 c_W\sqrt{1+\gamma^2 t_X^2}[1+(\beta^2+\gamma^2)t_X^2]t_N^2\Lambda^2},\\
\epsilon_3^\delta&=&\frac{\delta t_X(u^2+v^2)}{16c_W[1+(\beta^2+\gamma^2)t_X^2]t_N\Lambda^2}\left\{\fr{\sqrt{1-\delta^2}}{t_N}-\fr{\delta(b\beta+c\gamma)t_X}{1+\sqrt{1-\delta^2}}\right\}.
\eea
Because $\epsilon_{1,2,3}\sim (u^2, v^2)/ (w^2, V^2, \Lambda^2)$, the mixings are very small. 

Next, the symmetry breaking is done through three possible ways, corresponding to the assumptions: $w, V\ll \Lambda$, $w\ll V, \Lambda$, or $w, \Lambda\ll V$. Let us consider the first case, $w, V\ll \Lambda$. We have the element $m^2_{C'}$ much larger than the remainders. The mass matrix $M''^2_s$ can be diagonalized by using the seesaw formula, which yields
\bea
 (\mathcal{Z}'_2\, \mathcal{Z}'_3\, \mathcal{C}')^T=U_3 (\mathcal{Z}_2\,\mathcal{Z}_3\,Z_4)^T, \hs M'''^2 =U_3^{T}M''^2_sU_3= \left(\begin{array}{cc}
M^2_{2\times 2} &0 \\
0&m^{2}_{Z_{4}}\end{array}\right),
\eea
where $Z_4$ is physical as decoupled, while $\mathcal{Z}_2,\mathcal{Z}_3$ mix via $M^2_{2\times 2}$.  We obtain
\bea
U_3 &\simeq &  \left(\begin{array}{ccc}
1 & 0& \zeta _1  \\
0 &  1 & \zeta_2 \\
-\zeta_1  & -\zeta_2  & 1\\
\end{array}
\right), \hs  M^2_{2\times 2} =
\left(
\begin{array}{cc}
 m^2_{11} & m^2_{12} \\
 m^2_{12} & m^2_{22}
\end{array}\right), \hs m^{2}_{Z_{4}}\simeq m^2_{C'},
\eea
where $\zeta_{1,2}\equiv \zeta_{1,2}^0+\zeta_{1,2}^\delta$, 
\bea
\zeta_1^0&=&-\fr{(2\sqrt2 b-c) w^2}{24\sqrt2 \sqrt{1-\beta^2t^2_W}t_N \Lambda^2},\\
\zeta_2^0&=&-\fr{9V^2(1+\gamma^2t_X^2)c-w^2[1-\gamma(2\sqrt2 \beta-\gamma)t_X^2](2\sqrt2 b-c)}{96\sqrt{1+\gamma^2t_X^2}t_N\Lambda^2},\\
\zeta_1^\delta&=&\fr{\delta w^2[(1-\delta^2+\sqrt{1-\delta^2})(2\sqrt2 \beta-\gamma)t_X+\delta(2\sqrt2 b-c)t_N]}{24\sqrt2\sqrt{1-\beta^2t_W^2}(1+\sqrt{1-\delta^2})t_N^2\Lambda^2},\\
\zeta_2^\delta&=&\fr{\delta}{96\sqrt{1+\gamma^2t_X^2}(1+\sqrt{1-\delta^2})t_N^2\Lambda^2}\left\{9V^2(1+\gamma^2t_X^2)[\gamma(1-\delta^2+\sqrt{1-\delta^2})t_X+\delta c t_N]\right.\\
&&\left.-w^2[1-\gamma(2\sqrt2 \beta-\gamma)t_X^2][(1-\delta^2+\sqrt{1-\delta^2})(2\sqrt2 \beta-\gamma)t_X+\delta(2\sqrt2 b-c)t_N] \right\},
\eea
which are very small, and 
\bea
m^2_{11}&\simeq&m^2_{Z'_2} - \zeta_1 m^2_{Z'_2C'}\simeq m^2_{Z'_2},\\
m^2_{12}&\simeq&m^2_{Z'_2Z'_3}- \zeta_1 m^2_{Z'_3C'}\simeq m^2_{Z'_2Z'_3},\\
m^2_{22}&\simeq&m^2_{Z'_3}- \zeta_2 m^2_{Z'_3C'}\simeq m^2_{Z'_3}.
\eea
Last, we diagonalize $M^2_{2\times 2}$ to yield two remaining physical gauge bosons, 
\bea
Z_2 = c_\varphi \mathcal{Z}_2 - s_\varphi \mathcal{Z}_3, \hs Z_3 = s_\varphi \mathcal{Z}_2+ c_\varphi \mathcal{Z}_3.
\eea
The $\mathcal{Z}_2-\mathcal{Z}_3$ mixing angle and $Z_2, Z_3$ masses are given by
 \bea 
t_{2\varphi }&\simeq& \fr{4\sqrt2 w^2[1-\gamma(2\sqrt2 \beta-\gamma)t_X^2]\sqrt{1+(\beta^2+\gamma^2)t_X^2}}{w^2[7-\gamma^2(2\sqrt2 \beta-\gamma)^2t_X^4+(8\beta^2+4\sqrt2 \beta\gamma+6\gamma^2)t_X^2]-9V^2(1+\gamma^2t_X^2)^2},\\
m^2_ {Z_2,Z_3}&=& \fr 1 2 [m^2_{11}  + m^2_{22} \mp \sqrt{(m^2_{11}  - m^2_{22})^2 + 4 m^4_{12}}]. 
 \eea

Now we consider two other cases, $w\ll V, \Lambda$ and $w,\Lambda\ll V$. Because $m^2_{Z'_2}$, $ m^2_{Z'_2Z'_3}$, $ m^2_{Z'_2C'}\ll $ $ m^2_{Z'_3}$, $ m^2_{Z'_3C'}$, $ m^2_{C'}$, the mass matrix $M''^2_s$ can be diagonalized, obeying
\bea
 (\mathcal{Z}'_2\, \mathcal{Z}'_3\, \mathcal{C}')^T=U_3' (Z_2\,\mathcal{Z}_3\,\mathcal{C})^T, \hs M'''^2 =U_3'^{T}M''^2_sU_3'= \left(\begin{array}{cc}
m^{2}_{Z_{2}} &0 \\
0&M'^2_{2\times 2}\end{array}\right),
\eea
where $Z_2$ is physical as decoupled, while $\mathcal{Z}_3$ and $\mathcal{C}$ mix via $M'^2_{2\times 2}$, and
\bea
U_3' &\simeq &  \left(\begin{array}{ccc}
1 & \mathcal{E}_1 & \mathcal{E}_2  \\
-\mathcal{E}_1 &  1 & 0 \\
-\mathcal{E}_2 & 0 & 1\\
\end{array}
\right), \hs  M'^2_{2\times 2} \simeq 
\left(
\begin{array}{cc}
 m^2_{Z'_3} & m^2_{Z'_3C'}\\
 m^2_{Z'_3C'}& m^2_{C'}
\end{array}\right),\\
m^{2}_{Z_{2}}&\simeq & m^2_{Z'_2} - \mathcal{E}_1 m^2_{Z'_2Z'_3} -\mathcal{E}_2 m^2_{Z'_2C'}.
\eea
Further for the case $w\ll V, \Lambda$, we achieve $\mathcal{E}_{1,2}\equiv\mathcal{E}^0_{1,2}+\mathcal{E}^\delta_{1,2}$, where  
\bea
\mathcal{E}_1^0&=&\fr{\sqrt{1+(\beta^2+\gamma^2)t_X^2}w^2}{3(1+\gamma^2t_X^2)^2}\left\{\fr{4[\gamma(2\sqrt2\beta-\gamma)t_X^2-1]}{3\sqrt2V^2}-\fr{[b(1+\gamma^2t_X^2)-c\beta\gamma t_X^2]c}{4\Lambda^2}\right\},\\
\mathcal{E}_2^0&=&\fr{\sqrt{1+(\beta^2+\gamma^2)t_X^2}[\gamma(c\beta-b\gamma)t_X^2-b]w^2}{12(1+\gamma^2t_X^2)^{3/2}t_N\Lambda^2},\\
\mathcal{E}_1^\delta&=&\fr{\delta\sqrt{1+(\beta^2+\gamma^2)t_X^2}t_X[b\gamma(1+\gamma^2t_X^2)t_N+c\beta(1-\gamma^2t_X^2) t_N-\delta\beta\gamma t_X]w^2}{12(1+\gamma^2t_X^2)^2t_N^2\Lambda^2},\\
\mathcal{E}_2^\delta&=&\fr{\delta\sqrt{1+(\beta^2+\gamma^2)t_X^2}w^2}{12(1+\gamma^2t_X^2)^{3/2}t_N\Lambda^2}\left\{\fr{\delta[b+\gamma(b\gamma-c\beta)t_X^2]}{1+\sqrt{1-\delta^2}}+\fr{\sqrt{1-\delta^2}\beta t_X}{t_N}\right\},
\eea
which are very small. Otherwise, for the case $w, \Lambda \ll V$, we have 
\bea
\mathcal{E}_1&=&-\fr{\sqrt{1+(\beta^2+\gamma^2)t_X^2}(\delta\gamma t_X-ct_N)[\beta(\delta+c\gamma t_Xt_N)t_X-b(1+\gamma^2t_X^2)t_N]w^2}{[\beta(\delta+c\gamma t_Xt_N)t_X-b(1+\gamma^2t_X^2)t_N]^2w^2+12(1+\gamma^2t_X^2)^2t_N^2\Lambda^2},\\
\mathcal{E}_2&=&\fr{\sqrt{(1-\delta^2)(1+\gamma^2t_X^2)[1+(\beta^2+\gamma^2)t_X^2]}[\beta(\delta+c\gamma t_Xt_N)t_X-b(1+\gamma^2t_X^2)t_N]w^2}{[\beta(\delta+c\gamma t_Xt_N)t_X-b(1+\gamma^2t_X^2)t_N]^2w^2+12(1+\gamma^2t_X^2)^2t_N^2\Lambda^2},
\eea
which may be large.

We diagonalize the mass matrix $M'^2_{2\times 2}$ to get two remaining physical gauge bosons, such that  
\bea
Z_3 = c_\xi \mathcal{Z}_3 - s_\xi \mathcal{C}, \hs Z_4 = s_\xi \mathcal{Z}_3+ c_\xi \mathcal{C}.
\eea
The $\mathcal{Z}_3-\mathcal{C}$ mixing angle for the case $w\ll V,\Lambda$ is given by
 \bea 
t_{2\xi}\simeq \frac{6\sqrt{1-\delta^2}\sqrt{1+\gamma^2t_X^2}(\delta\gamma t_X-ct_N)V^2}{3[(\delta\gamma t_X-ct_N)^2-(1-\delta^2)(1+\gamma^2t_X^2)]V^2+32t_N^2\Lambda^2},\label{mixingangle}
 \eea
which may be large. For the case $w,\Lambda\ll V$, the $\mathcal{Z}_3-\mathcal{C}$ mixing angle is defined similarly to \eqref{mixingangle}, but the term associated to $\Lambda$ should be omitted. In particular, all the two cases imply $\xi =0$ when $\delta=ct_N/\gamma t_X$, the condition by which the kinetic mixing and symmetry breaking effects cancels out. Besides, the $Z_3, Z_4$ masses are given by
 \bea 
m^2_ {Z_3,Z_4}= \fr1 2 [m^2_ {Z'_3} +m^2_{C'} \mp \sqrt{(m^2_ {Z'_3} - m^2_{C'})^2 + 4m^4_ {Z'_3 C'} }]. 
 \eea

In summary, the original fields are related to the mass eigenstates by $(A_3\, A_8\, A_{15}\, B\, C)^T=U(A\, Z_1\, Z_2\,Z_3\,Z_4)^T$. For the first case, $w,V\ll\Lambda$, we have $U= U_\delta U_1U_2U_3U_\varphi\simeq U_\delta U_1U_2U_\varphi$. For the second case, $w\ll V, \Lambda$, we obtain $U= U_\delta U_1U_2U'_3U_\xi\simeq U_\delta U_1U_2U_\xi$. For the last case, $w, \Lambda\ll V$, the mixing matrix is $U= U_\delta U_1U_2U'_3U_\xi$. Here we define
\bea
U_\delta= \left(%
\begin{array}{ccccc}
1&0&0&0&0\\
0&1&0&0&0\\
0&0&1&0&0\\
0&0&0&1&-\fr{\delta}{\sqrt{1-\delta^2}}\\
0&0&0&0&\fr{1}{\sqrt{1-\delta^2}}
\end{array} \right),  \hs
U_\varphi= \left(%
\begin{array}{ccccc}
1&0&0&0&0\\
0&1&0&0&0\\
0&0&c_\varphi&s_\varphi&0\\
0&0&-s_\varphi&c_\varphi&0\\
0&0&0&0&1
\end{array} \right),  \hs U_\xi= \left(%
\begin{array}{ccccc}
1&0&0&0&0\\
0&1&0&0&0\\
0&0&1&0&0\\
0&0&0&c_\xi&s_\xi\\
0&0&0&-s_\xi&c_\xi
\end{array} \right).
\eea
The fields $A, Z_1$ are identical to the standard model, whereas $Z_2, Z_3$ and $Z_4$ are new, heavy gauge bosons. The mixings of the standard model gauge bosons with the new gauge bosons are very small, while the mixing within the new gauge bosons may be large.

\section{\label{pheno}Electroweak precision test}

\subsection{$\rho$ parameter}

The new physics that contributes to the $\rho$-parameter starts from the tree-level. This is caused by the mixing of the $Z$ boson with the new neutral gauge bosons. We evaluate 
\bea \Delta \rho &=&\fr{m_W^2}{c_W^2m_{Z_1}^2}-1=\fr{m_Z^2}{m^2_Z - \epsilon_1 m^2_{ZZ'_2} -  \epsilon_2 m^2_{ZZ'_3} -  \epsilon_3 m^2_{ZC'}}-1\simeq\fr{\epsilon_1 m^2_{ZZ'_2} +  \epsilon_2 m^2_{ZZ'_3} +  \epsilon_3 m^2_{ZC'}}{m_Z^2}\crn
&\equiv& (\Delta \rho)^0+(\Delta \rho)^\delta,
\eea
where
\bea
(\Delta \rho)^0&\simeq&\fr{1}{4[1+(\beta^2+\gamma^2)t_X^2]^2}\left\{\fr{[\beta_2u^2+(2\sqrt3\beta t_X^2-\beta_2)v^2]^2}{(u^2+v^2)w^2}\right.\crn
&&\left.+\fr{\{(\beta+2\sqrt2\gamma)t_X^2(u^2+v^2)+\sqrt3[1+(\beta^2+\gamma^2)t_X^2](u^2-v^2)\}^2}{3(u^2+v^2)V^2}+\fr{(b\beta+c\gamma)^2t_X^4(u^2+v^2)}{4\Lambda^2}\right\},\\
(\Delta \rho)^\delta&\simeq&\fr{\delta[\delta+2(b\beta+c\gamma)t_Xt_N]t_X^2(u^2+v^2)}{16[1+(\beta^2+\gamma^2)t_X^2]^2t_N^2\Lambda^2}.
\eea This tree-level contribution is appropriately suppressed due to $u,v\ll w, V,\La$. The $\rho$ deviation may receive one-loop corrections by non-degenerate vector multiplets, such as $(W_{13},W_{23})$ and $(W_{14},W_{24},W_{34})$, similar to the 3-3-1 model~\cite{sturho}. However, this source can be neglected if the new gauge bosons are heavy at TeV. In this analysis, we consider only the tree-level contribution.     

Let us note that $\beta^2+\gamma^2<1/s^2_W-1\simeq 3.329$, which fixes $|\ga|,|\beta|<1.82456$. The condition $q=-(1+\sqrt3\beta)/2$ leads to $-2.08012<q<1.08012$. Considering $q$ to be integer implies $q=-2,-1,0,$ and $1$. When $q=-2$, i.e. $\beta=\sqrt3$, we obtain $|\gamma|<0.57359$. The condition $p=-(1+\sqrt6\gamma-q)/3$ provides $-1.46833<p<-0.53167$, thus $p=-1$, i.e. $\gamma=0$, given that $p$ is integer. When $q=-1$, i.e. $\beta=1/\sqrt3$, we get $p=-2, -1, 0$, thus $\gamma=4/\sqrt6,1/\sqrt6, -\sqrt2/\sqrt3$, respectively. When $q=0$, i.e. $\beta=-1/\sqrt3$, we gain $p=-1, 0, 1$, thus $\gamma=\sqrt2/\sqrt3,-1/\sqrt6, -4/\sqrt6$, respectively. When $q=1$, i.e. $\beta=-\sqrt3$, we have $p=0$, thus $\gamma=0$. 

However, we are interested in the four models for dark matter, such that $q=p=-1$ (or $\beta=1/\sqrt3,\gamma=1/\sqrt6$), $q=-1,p=0$ (or $\beta=1/\sqrt3,\gamma=-\sqrt2/\sqrt3$), $q=0,p=-1$ (or $\beta=-1/\sqrt3,\gamma=\sqrt2/\sqrt3$), and $q=p=0$ (or $\beta=-1/\sqrt3,\gamma=-1/\sqrt6$).\footnote{In these cases, the Landau pole is high enough, such that the new physics is viable \cite{landau}.} Besides, we take $n=m=0$, thus $b=-2/\sqrt3$, $c=-\sqrt2/\sqrt3$, for brevity. This case implies that the wrong particles are old. On the other hand, the $W$ mass, $m_W^2=\fr{g^2}{4}(u^2+v^2)$, implies $u^2+v^2=(246\ \mathrm{GeV})^2$. We will take $u$ in the range $(0,246)\ \mathrm{GeV}$, while $v$ is related to $u$.

The $\rho$ deviation is given from the global fit by $0.0002 < \Delta \rho < 0.00058$ \cite{pdg2018}. 
For the cases, $w, V\ll\La$ and $w\ll V, \Lambda$, $\Delta\rho$ is independent of $\delta$. Additionally, in the latter case ($w\ll V, \Lambda$), $\Delta\rho$ is independent of $\gamma$. However, in the case $w,\La\ll V$, all the parameters contribute to $\Delta\rho$, except for $V$. Without loss of generality, we impose $V=2w$ for the case $w, V\ll\La$ while $\Lambda=2w$ for the case $w,\La\ll V$. Besides, we put $t_N=0.5$. 

\begin{figure}[!h]
\begin{center}
\includegraphics[scale=0.31]{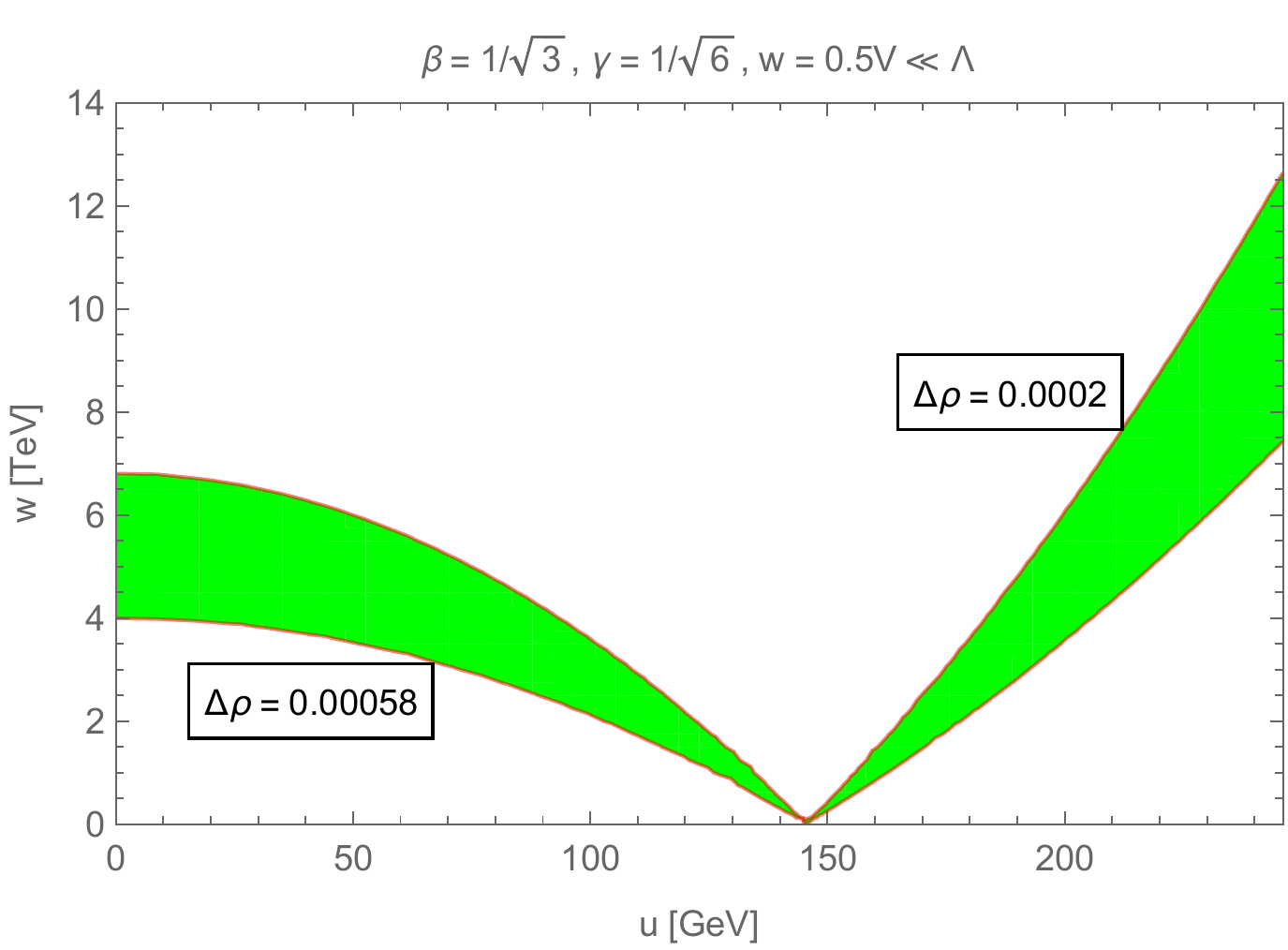}
\includegraphics[scale=0.31]{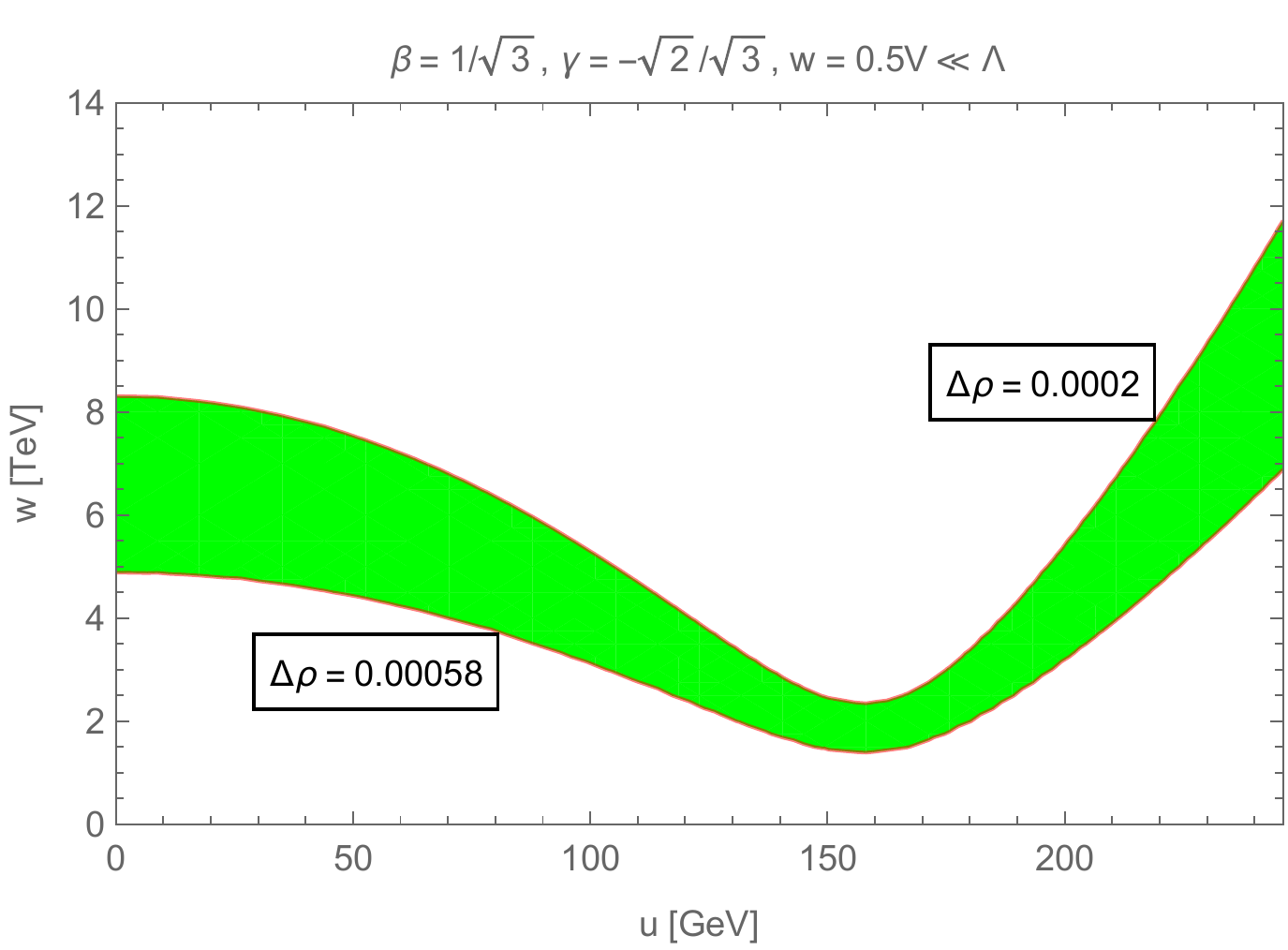}
\includegraphics[scale=0.31]{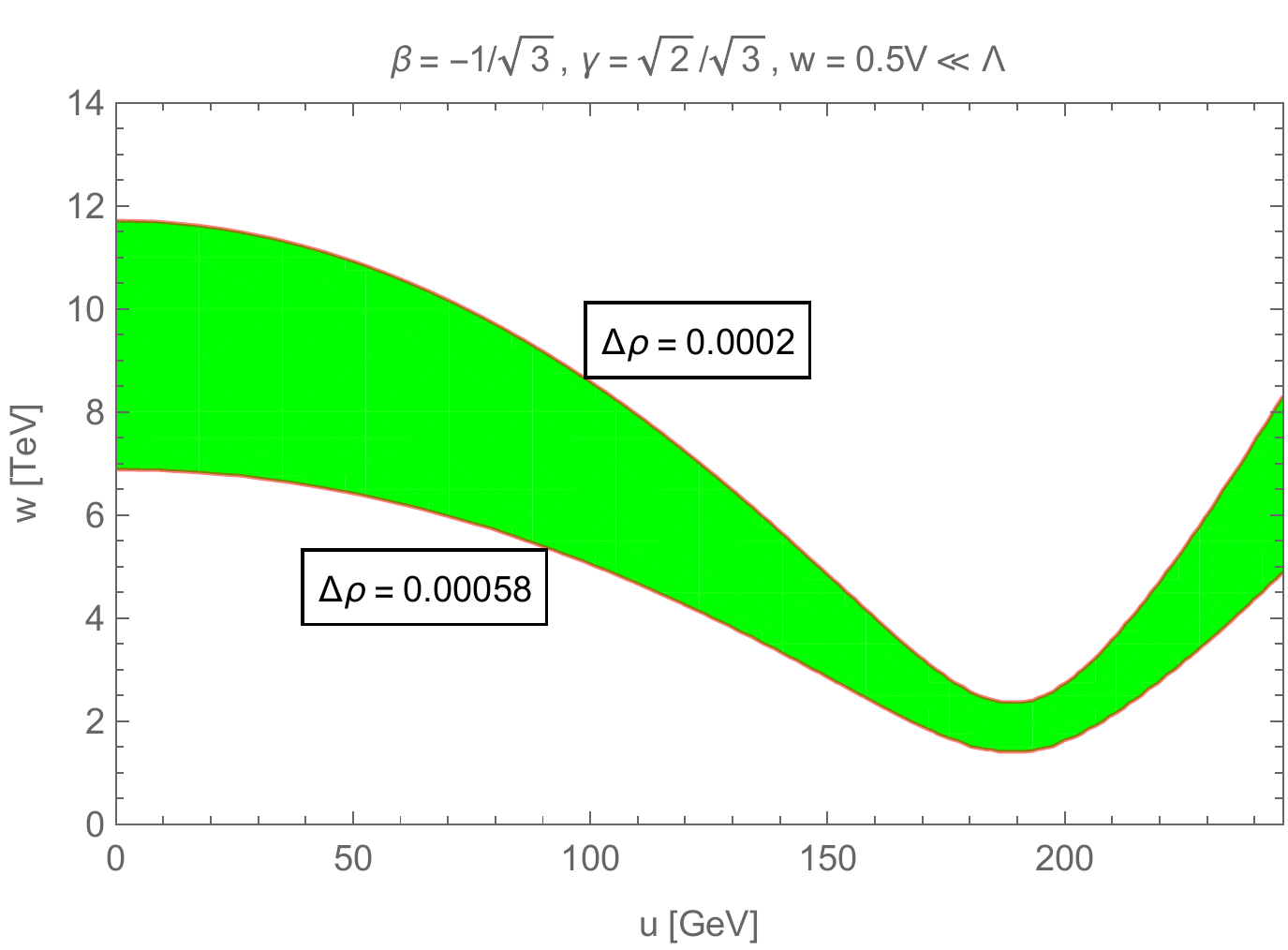}
\includegraphics[scale=0.31]{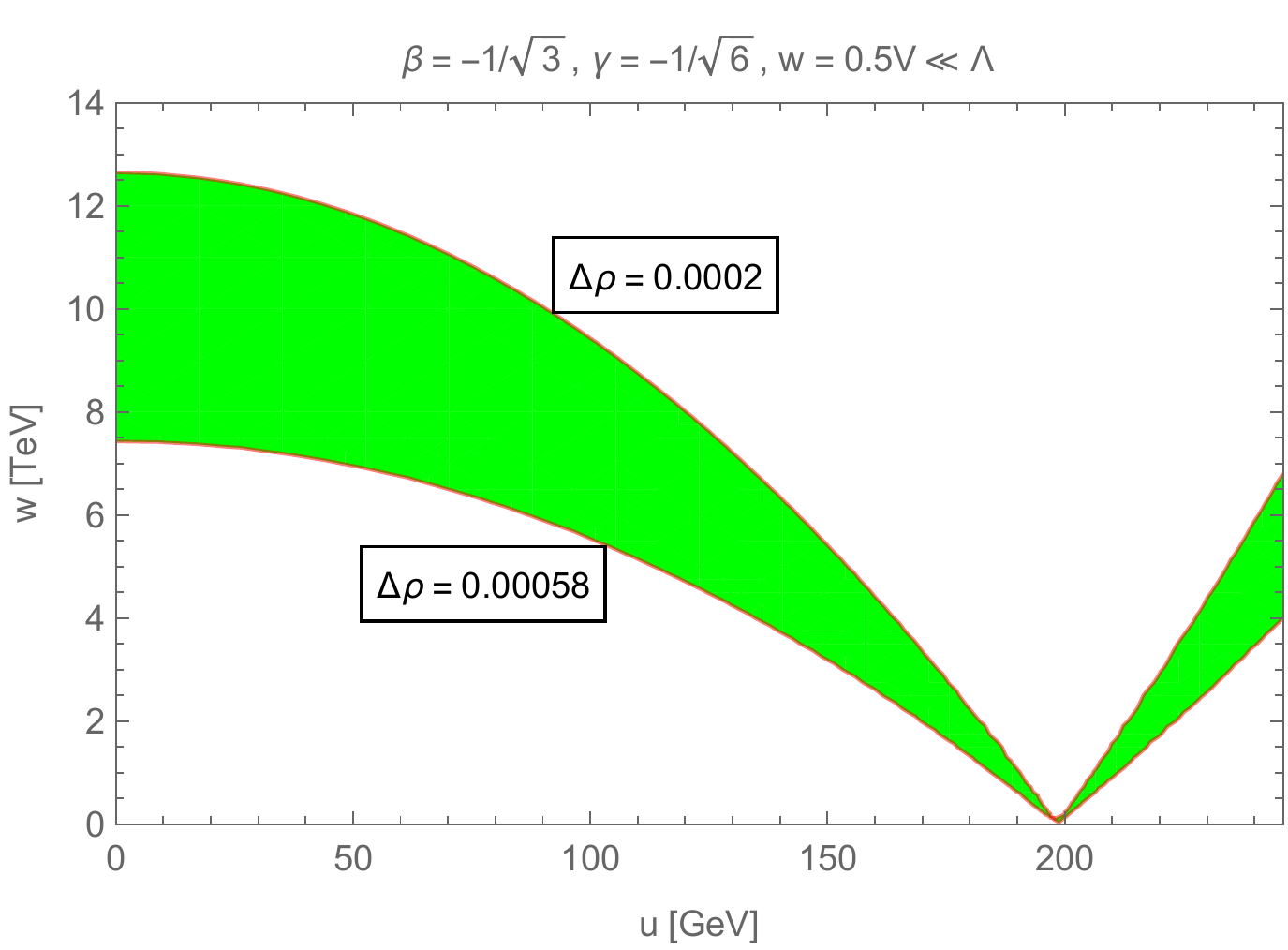}
\caption[]{\label{DT1} The $(u,w)$ regime that is bounded by the $\rho$ parameter for $w=0.5V\ll\La$, where the panels from left to right correspond to the four dark matter models.}
\end{center}
\end{figure}

In Fig. \ref{DT1}, we make a contour of $\Delta\rho$ as the function of ($u,w$) concerning the first case of VEV arrangement. Here, the panels arranging from left to right correspond to the four dark matter models such as ($\beta=1/\sqrt3, \gamma=1/\sqrt6$), ($\beta=1/\sqrt3, \gamma=-\sqrt2/\sqrt3$), ($\beta=-1/\sqrt3, \gamma=\sqrt2/\sqrt3$), and ($\beta=-1/\sqrt3, \gamma=-1/\sqrt6$), respectively.

\begin{figure}[!h]
\begin{center}
\includegraphics[scale=0.35]{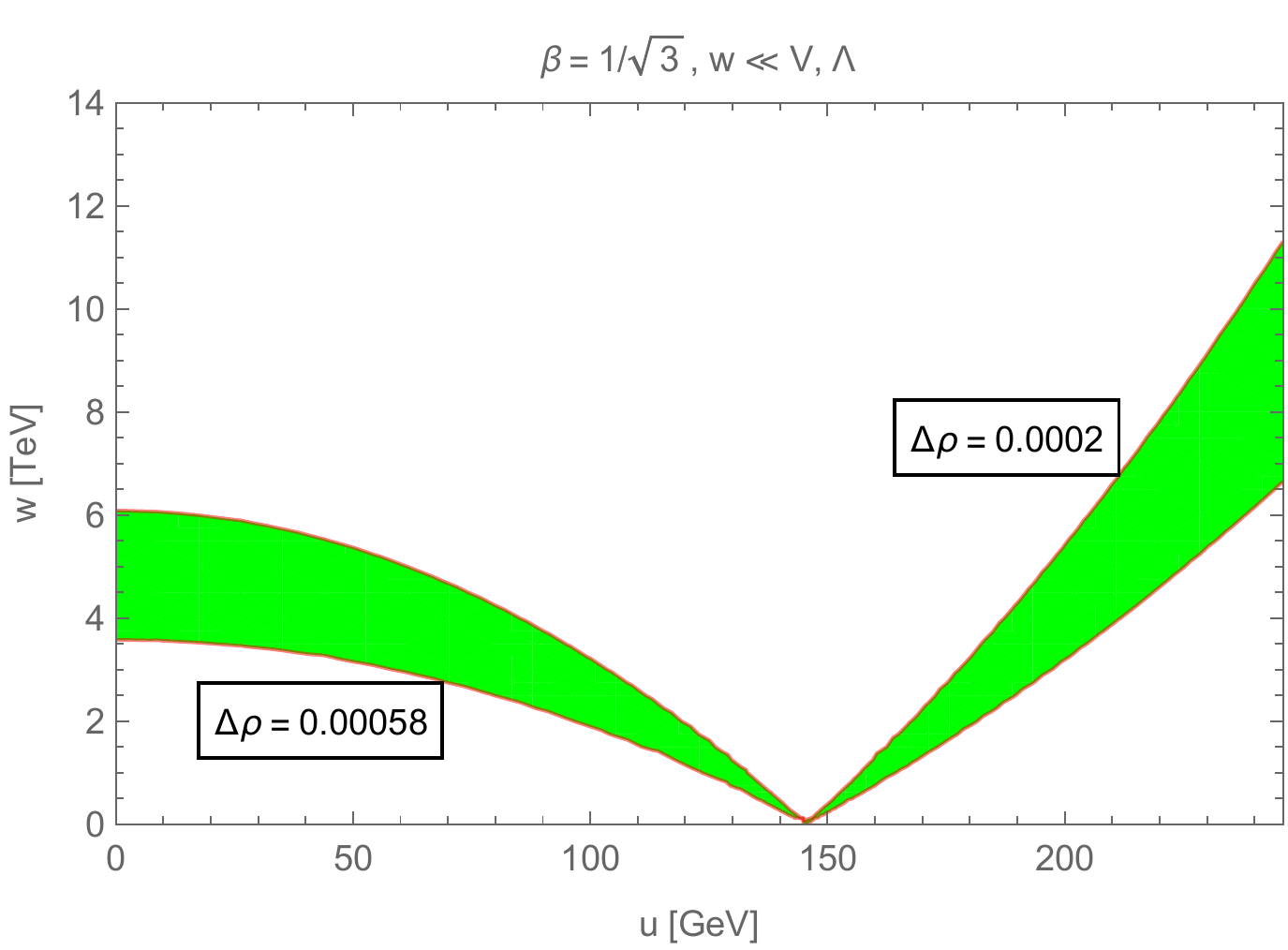}
\includegraphics[scale=0.35]{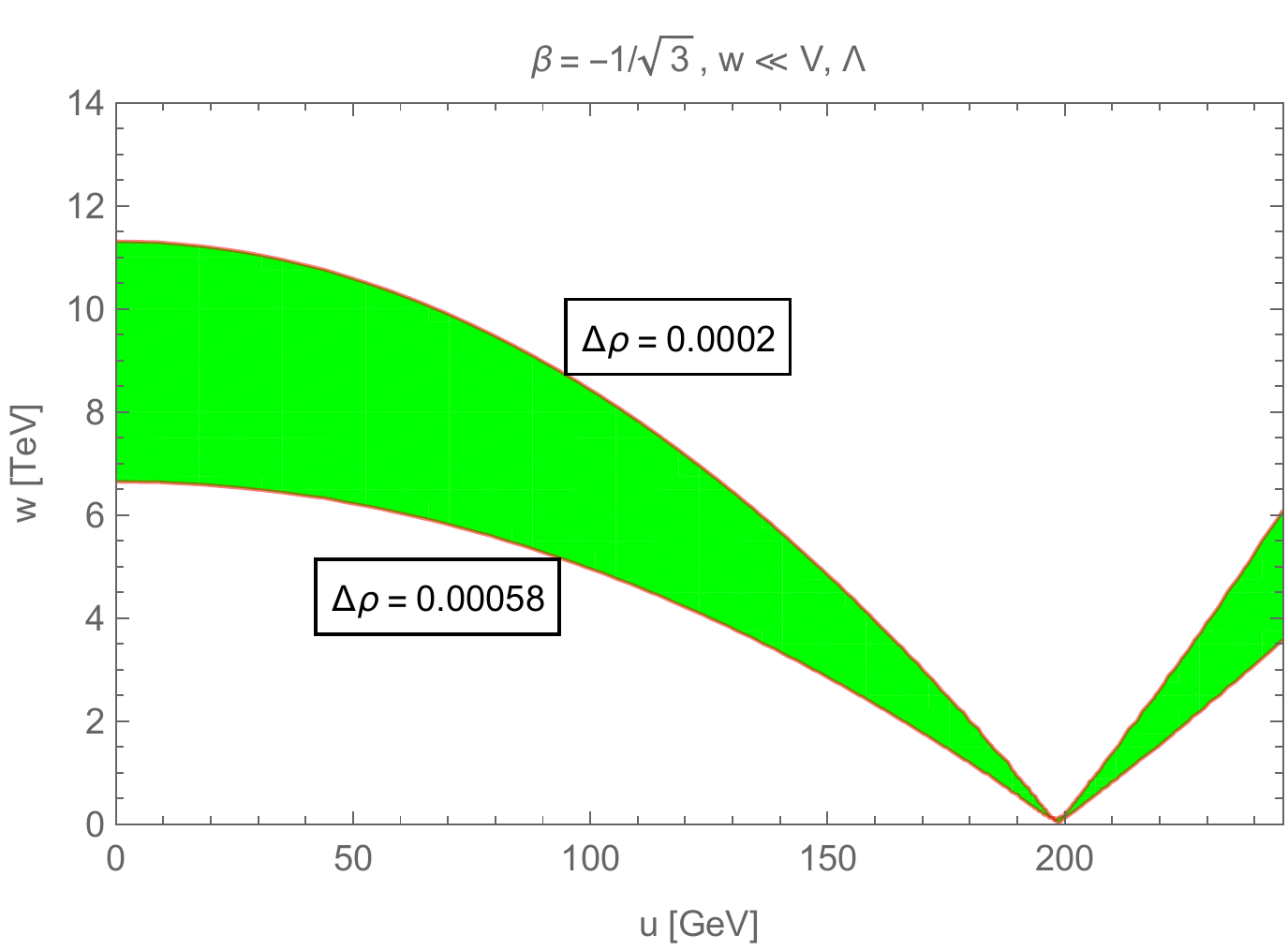}
\caption[]{\label{DT2} The $(u,w)$ regime that is bounded by the $\rho$ parameter for $w\ll V, \La$, where the left and right panels correspond to $\beta=1/\sqrt3$ and $\beta=-1/\sqrt3$.}
\end{center}
\end{figure}

In Fig. \ref{DT2}, we make a contour of $\Delta\rho$ as the function of ($u,w$) for the second case of VEV arrangement. Here, we have only two viable cases, the left panel for $\beta=1/\sqrt3$ and the right panel for $\beta=-1/\sqrt3$.

\begin{figure}[!h]
\begin{center}
\includegraphics[scale=0.31]{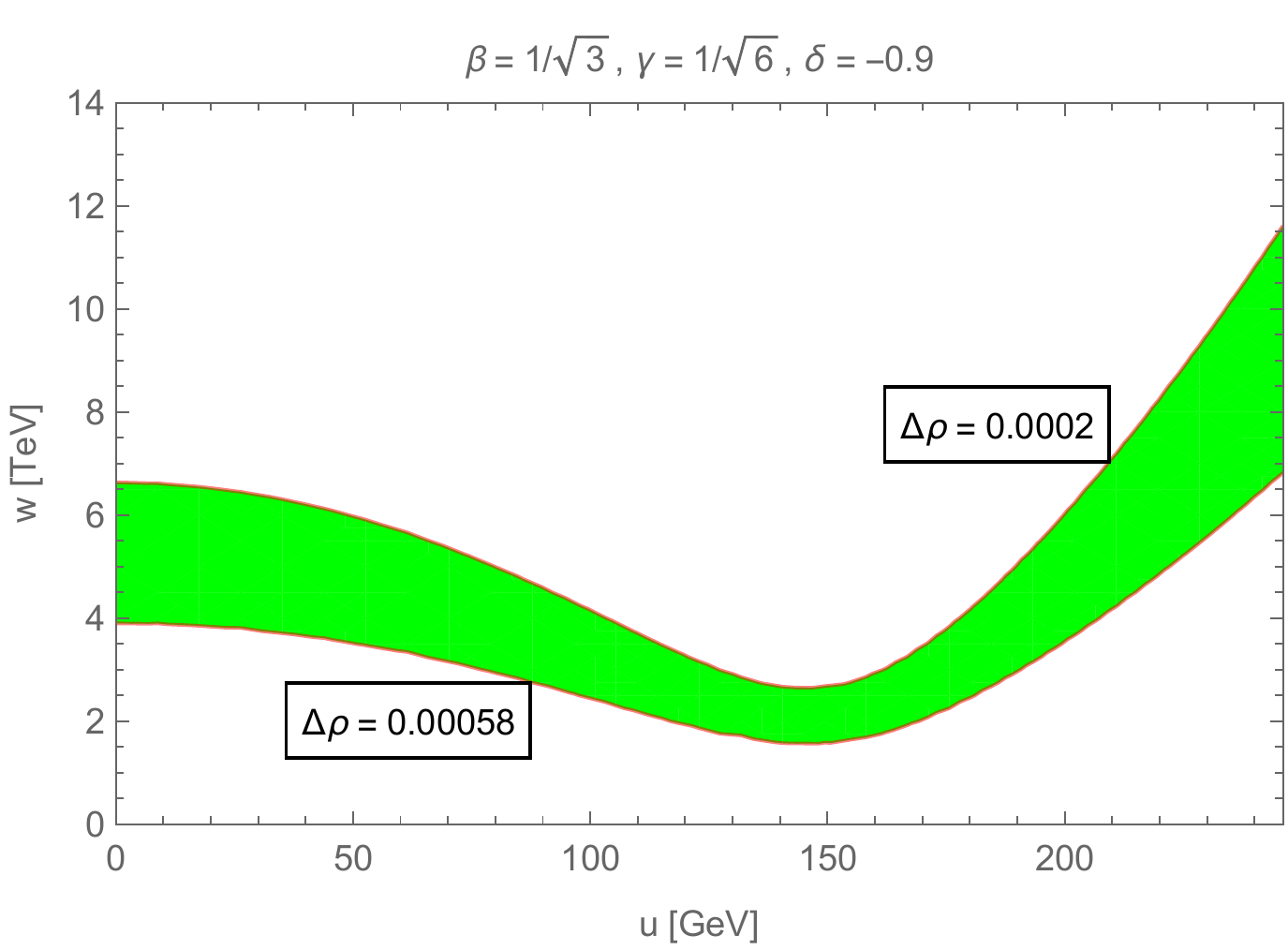}
\includegraphics[scale=0.31]{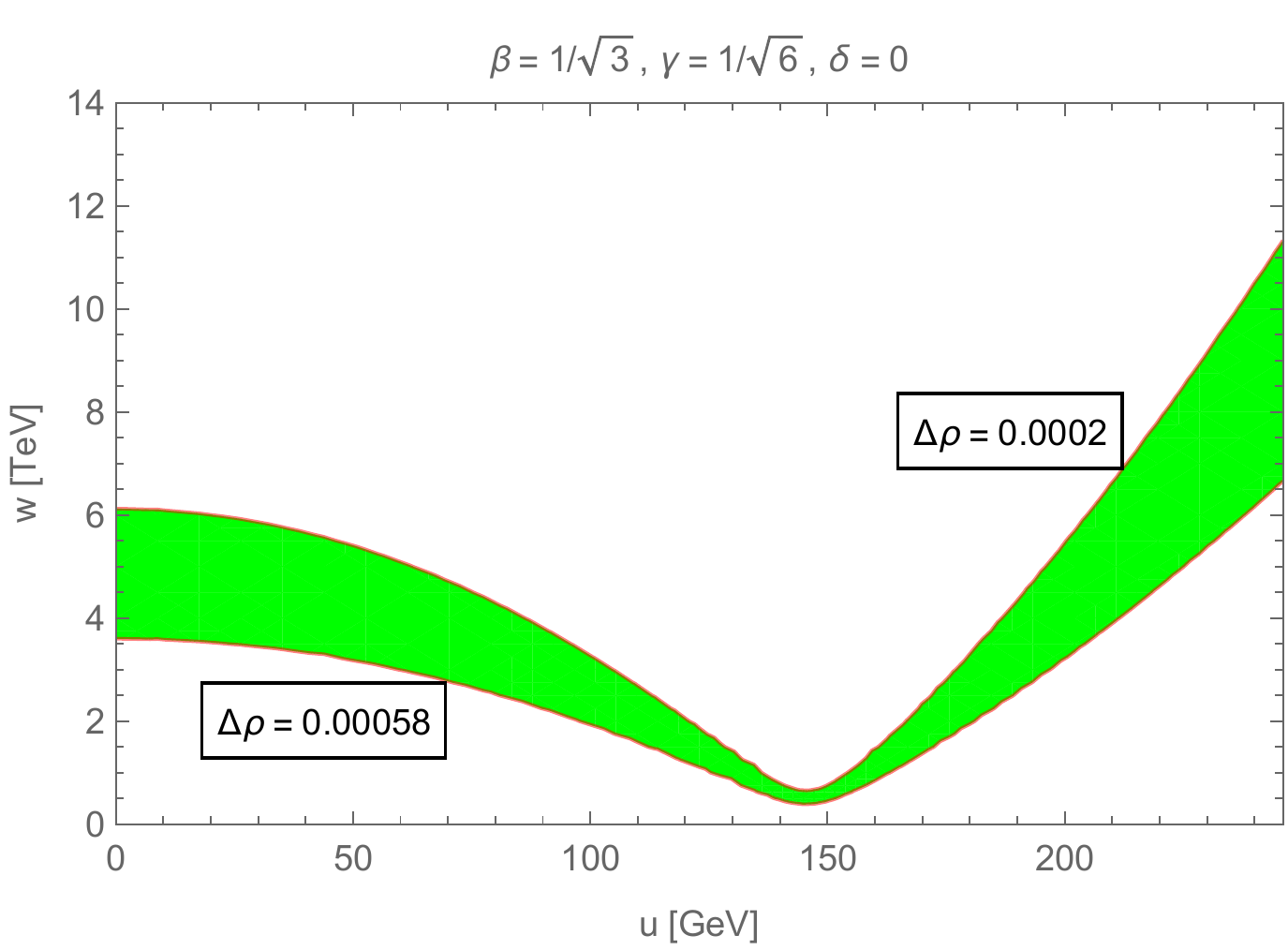}
\includegraphics[scale=0.31]{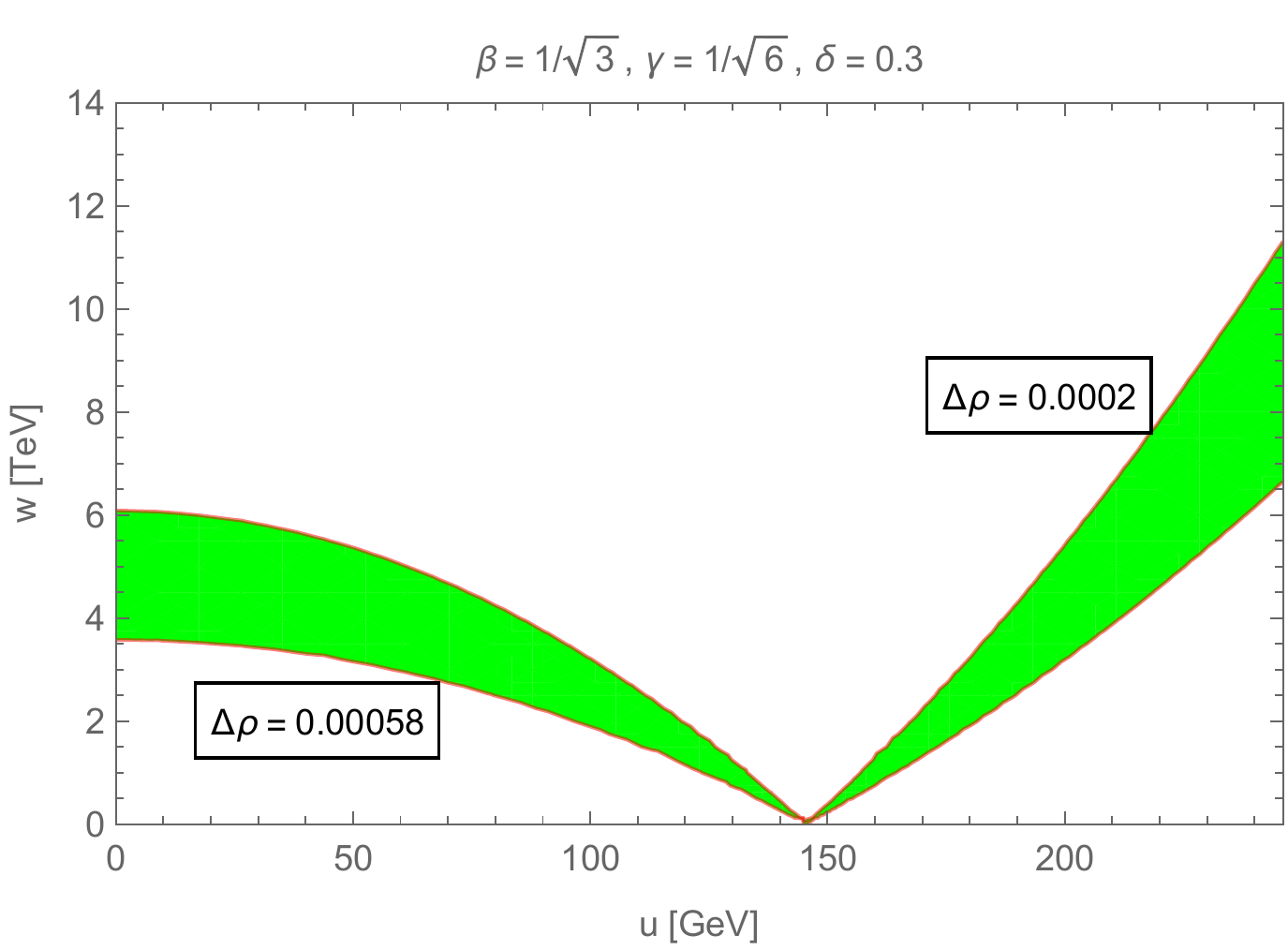}
\includegraphics[scale=0.31]{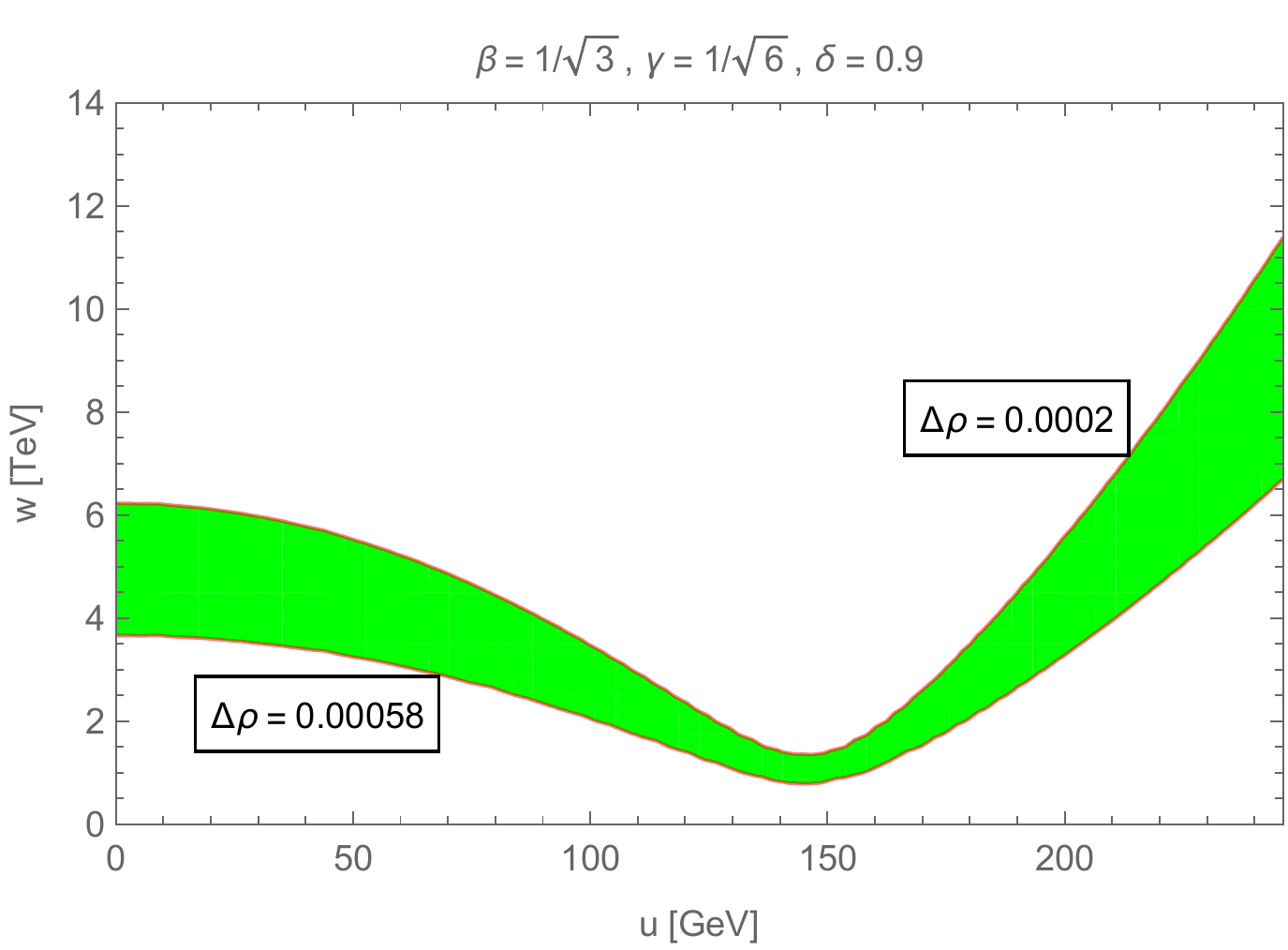}
\caption[]{\label{DT3} The $(u,w)$ regime that is bounded by the $\rho$ parameter for ($\beta=1/\sqrt3, \gamma=1/\sqrt6, b=-2/\sqrt3$, $c=-\sqrt2/\sqrt3$) and $w=0.5\Lambda\ll V$, where the panels from left to right correspond to $\delta=-0.9,\ 0,\ 0.3$, and $0.9$.}
\end{center}
\end{figure}

\begin{figure}[!h]
\begin{center}
\includegraphics[scale=0.35]{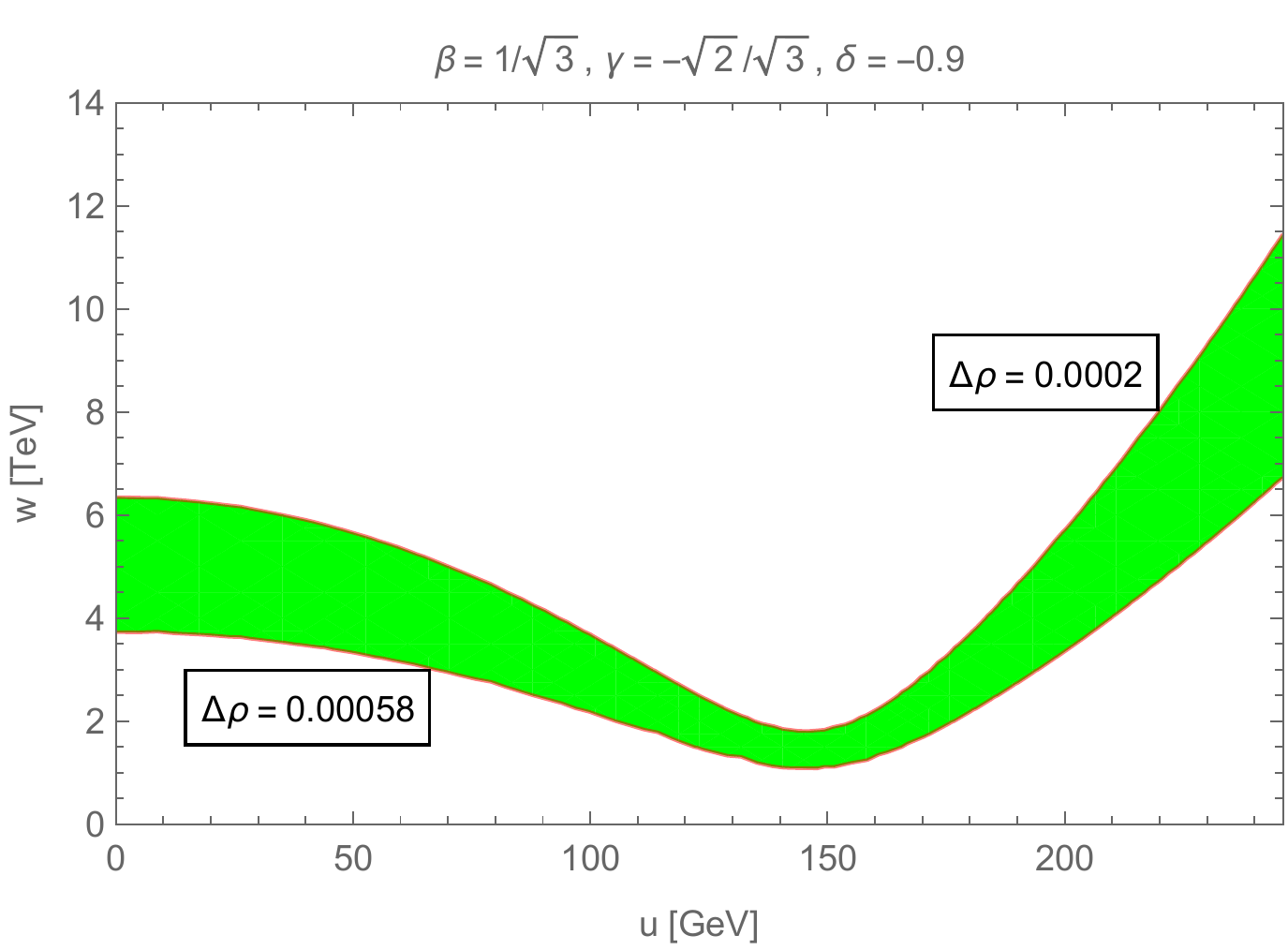}
\includegraphics[scale=0.35]{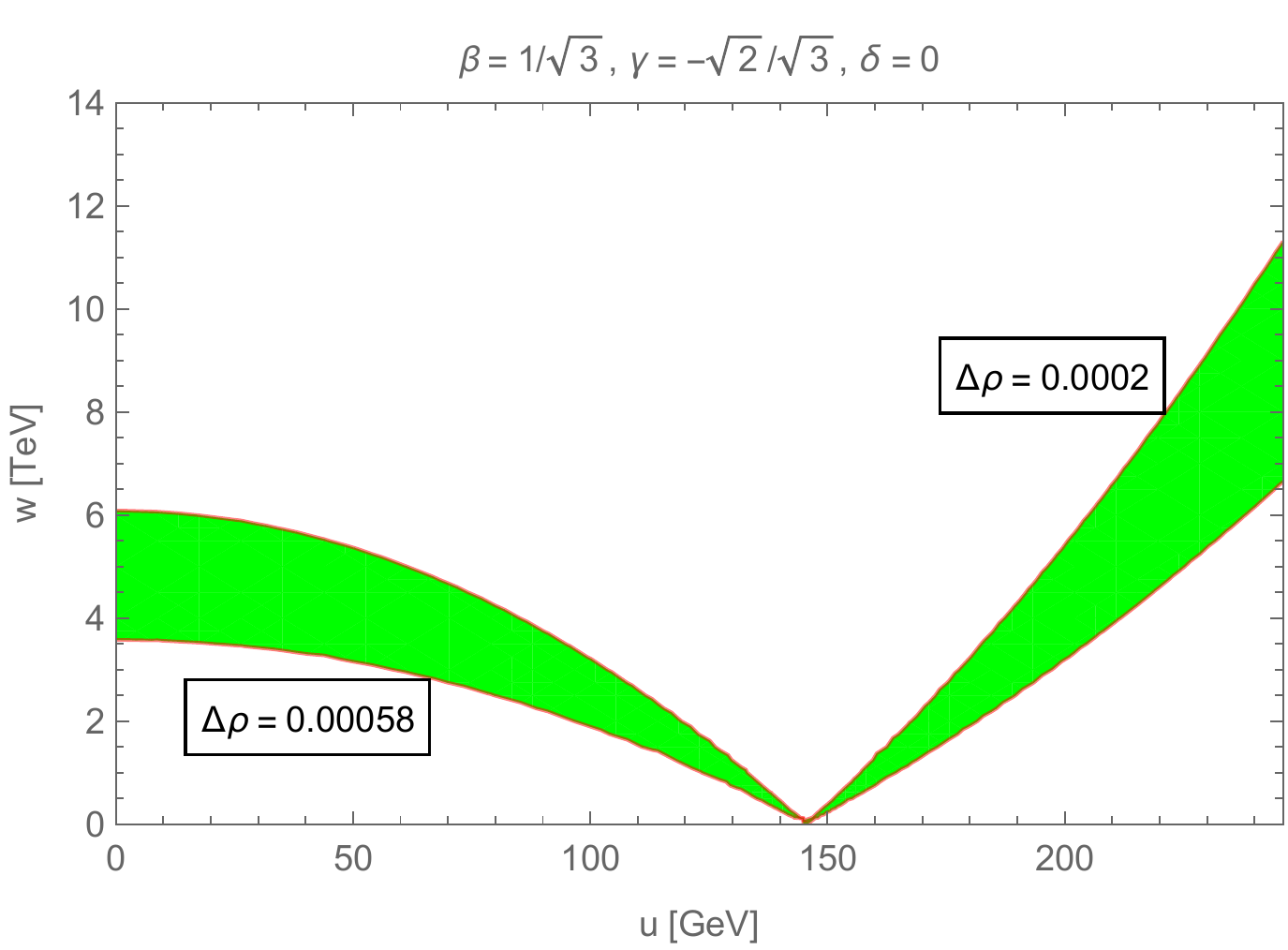}
\includegraphics[scale=0.35]{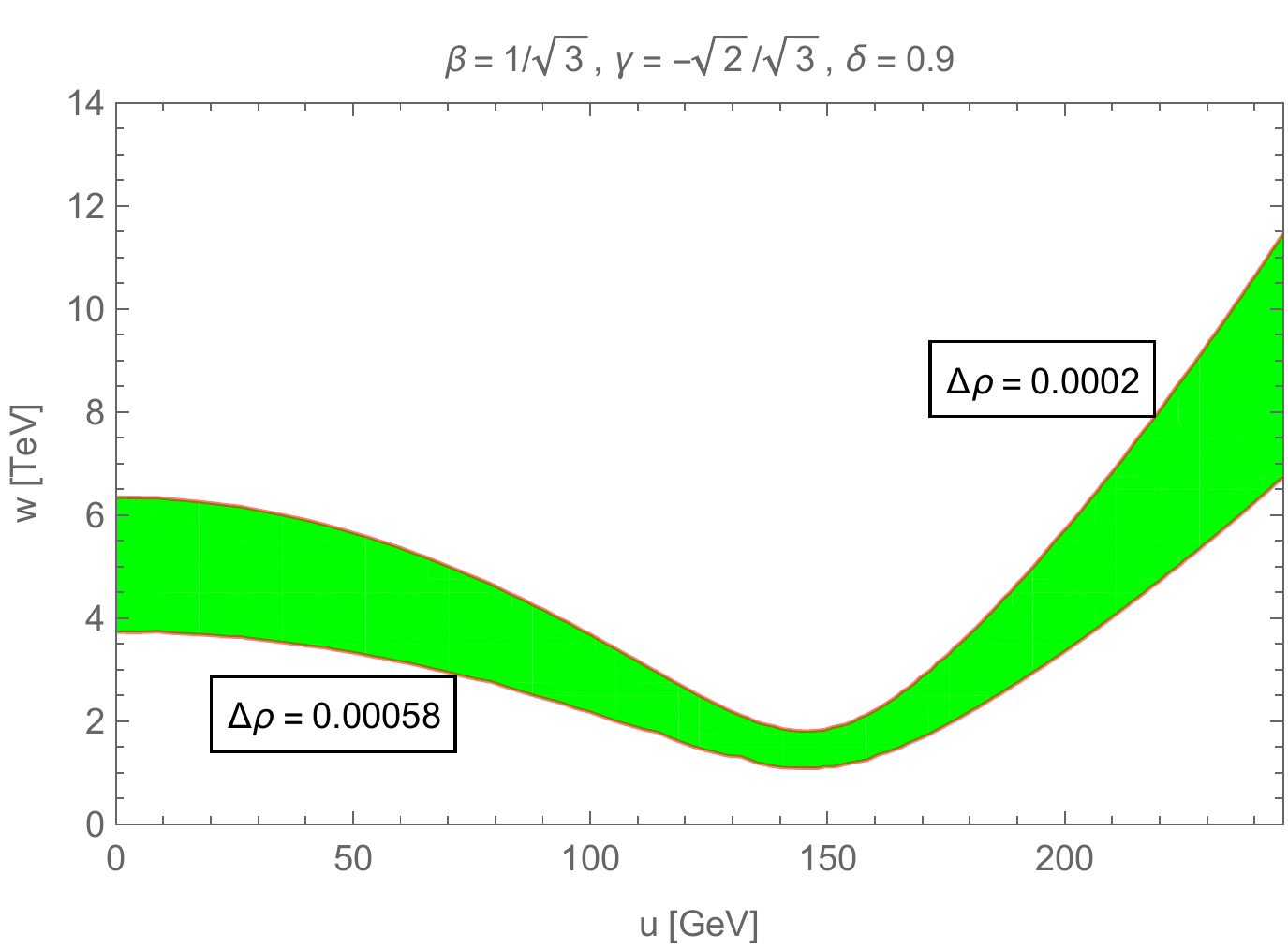}
\caption[]{\label{DT4} The $(u,w)$ regime that is bounded by the $\rho$ parameter for ($\beta=1/\sqrt3, \gamma=-\sqrt2/\sqrt3, b=-2/\sqrt3$, $c=-\sqrt2/\sqrt3$) and $w=0.5\Lambda\ll V$, where the panels from left to right are for $\delta=-0.9,\ 0$, and $0.9$, respectively.}
\end{center}
\end{figure}

\begin{figure}[!h]
\begin{center}
\includegraphics[scale=0.35]{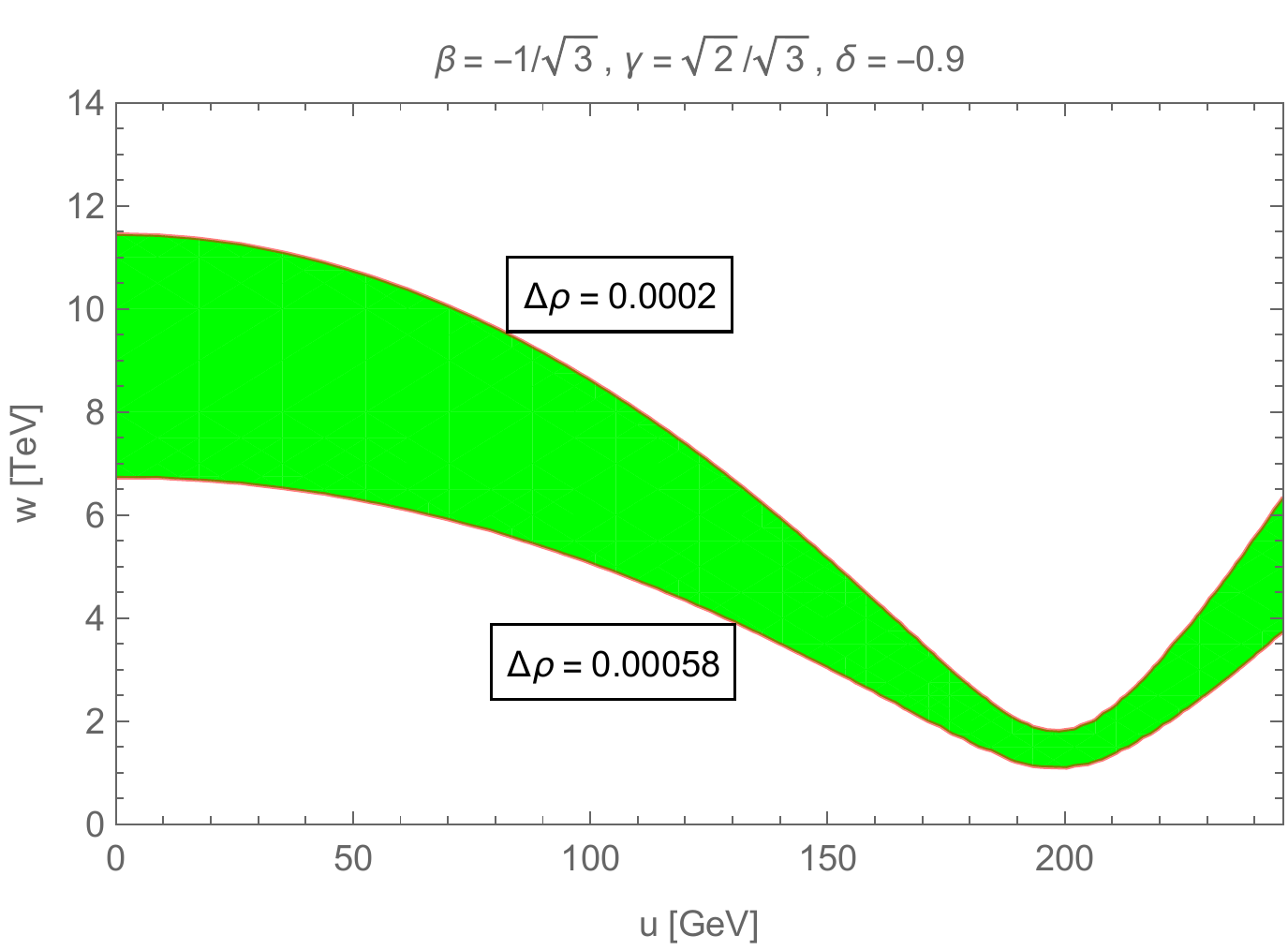}
\includegraphics[scale=0.35]{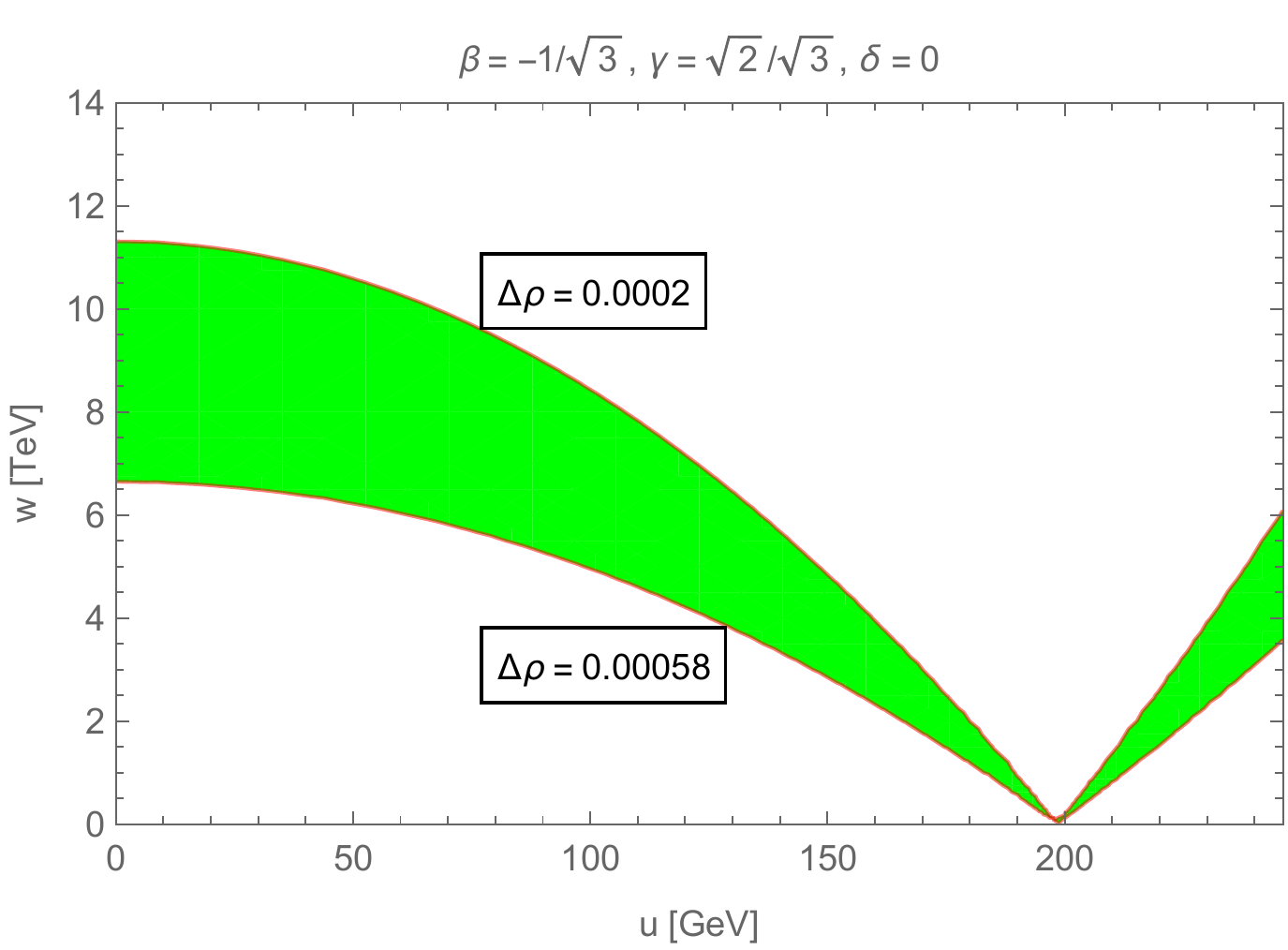}
\includegraphics[scale=0.35]{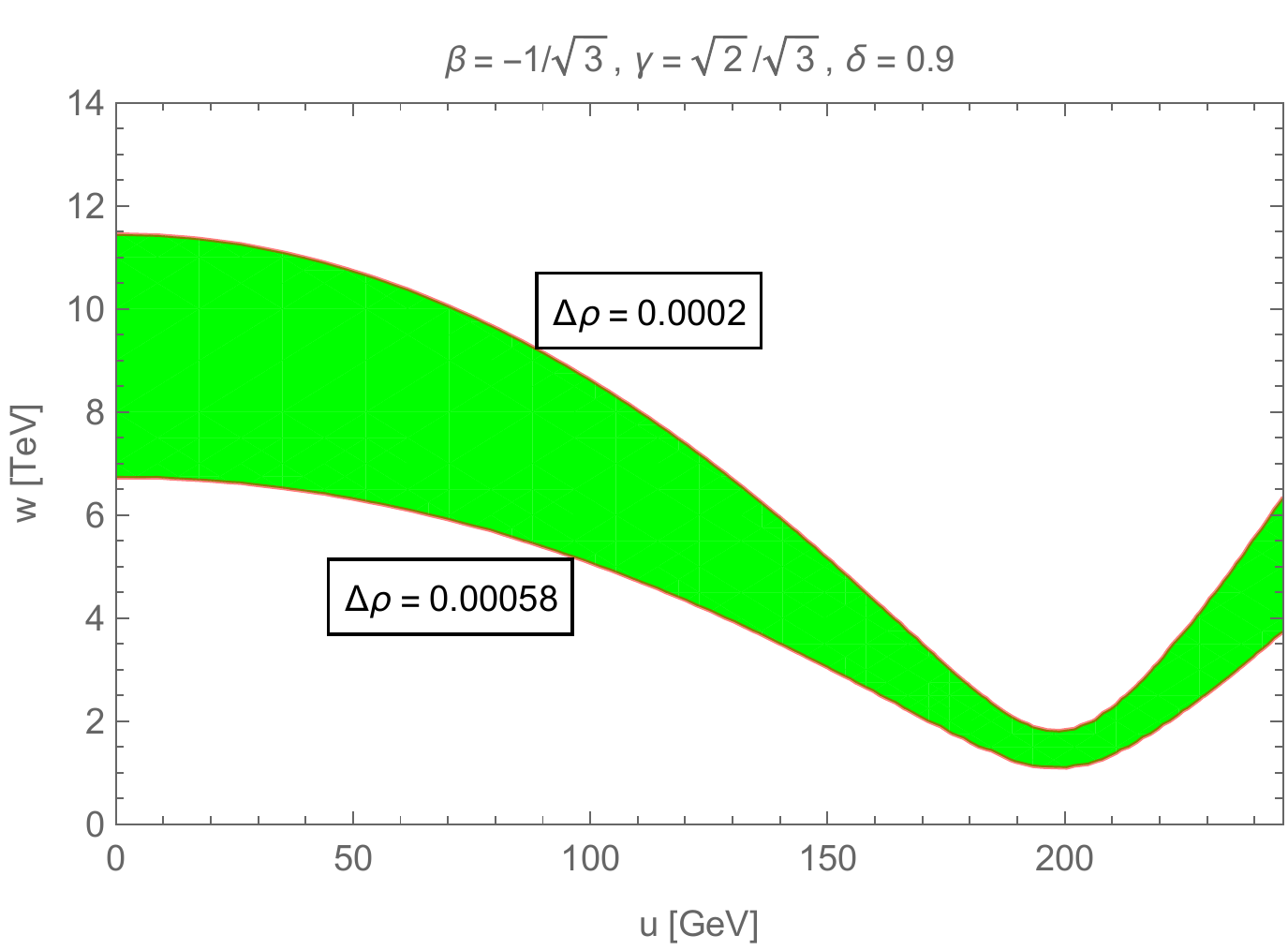}
\caption[]{\label{DT5} The $(u,w)$ regime that is bounded by the $\rho$ parameter for ($\beta=-1/\sqrt3, \gamma=\sqrt2/\sqrt3, b=-2/\sqrt3$, $c=-\sqrt2/\sqrt3$) and $w=0.5\Lambda\ll V$, where the panels from left to right are for $\delta=-0.9,\ 0$, and $0.9$, respectively.}
\end{center}
\end{figure}

\begin{figure}[!h]
\begin{center}
\includegraphics[scale=0.31]{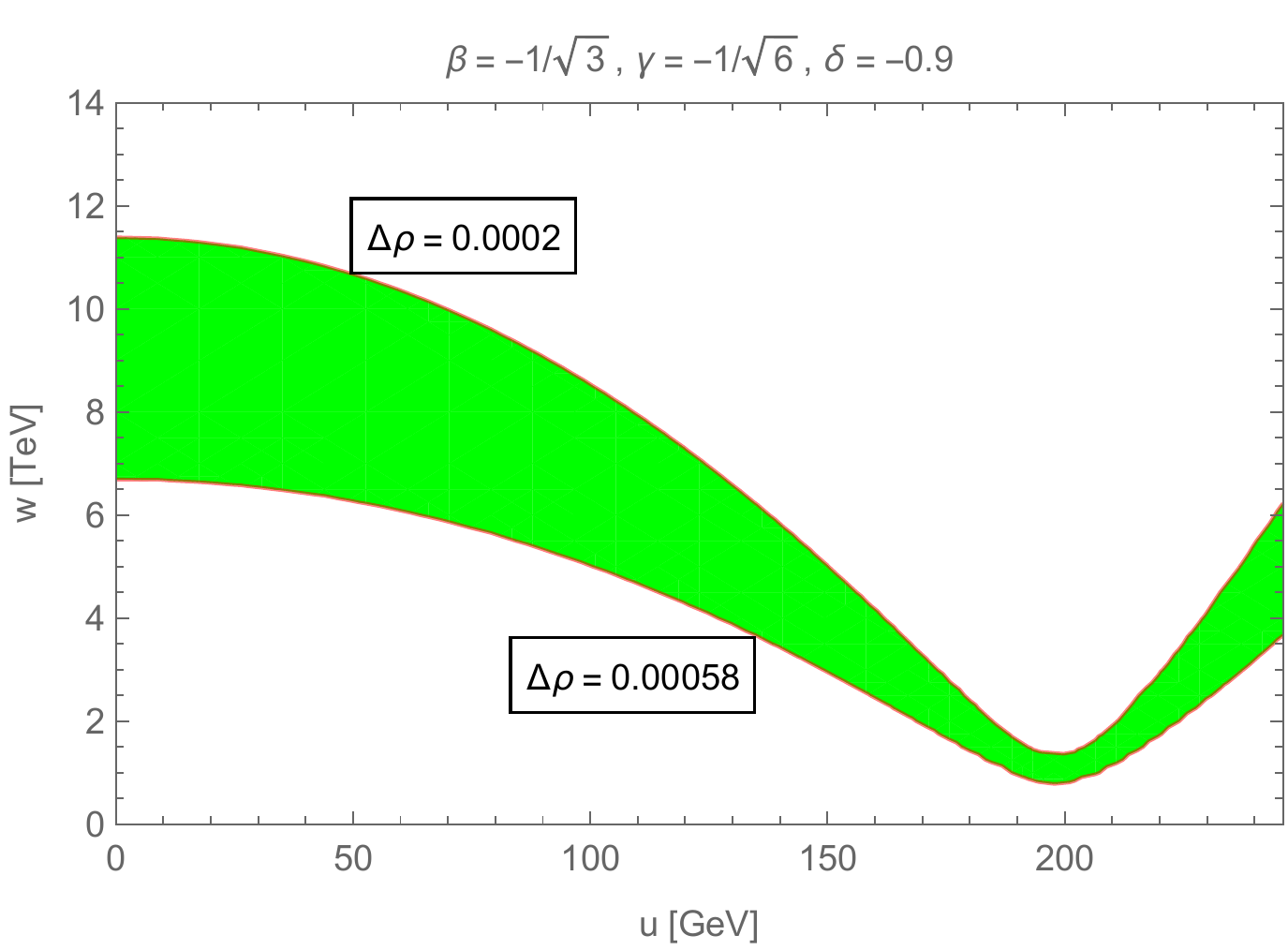}
\includegraphics[scale=0.31]{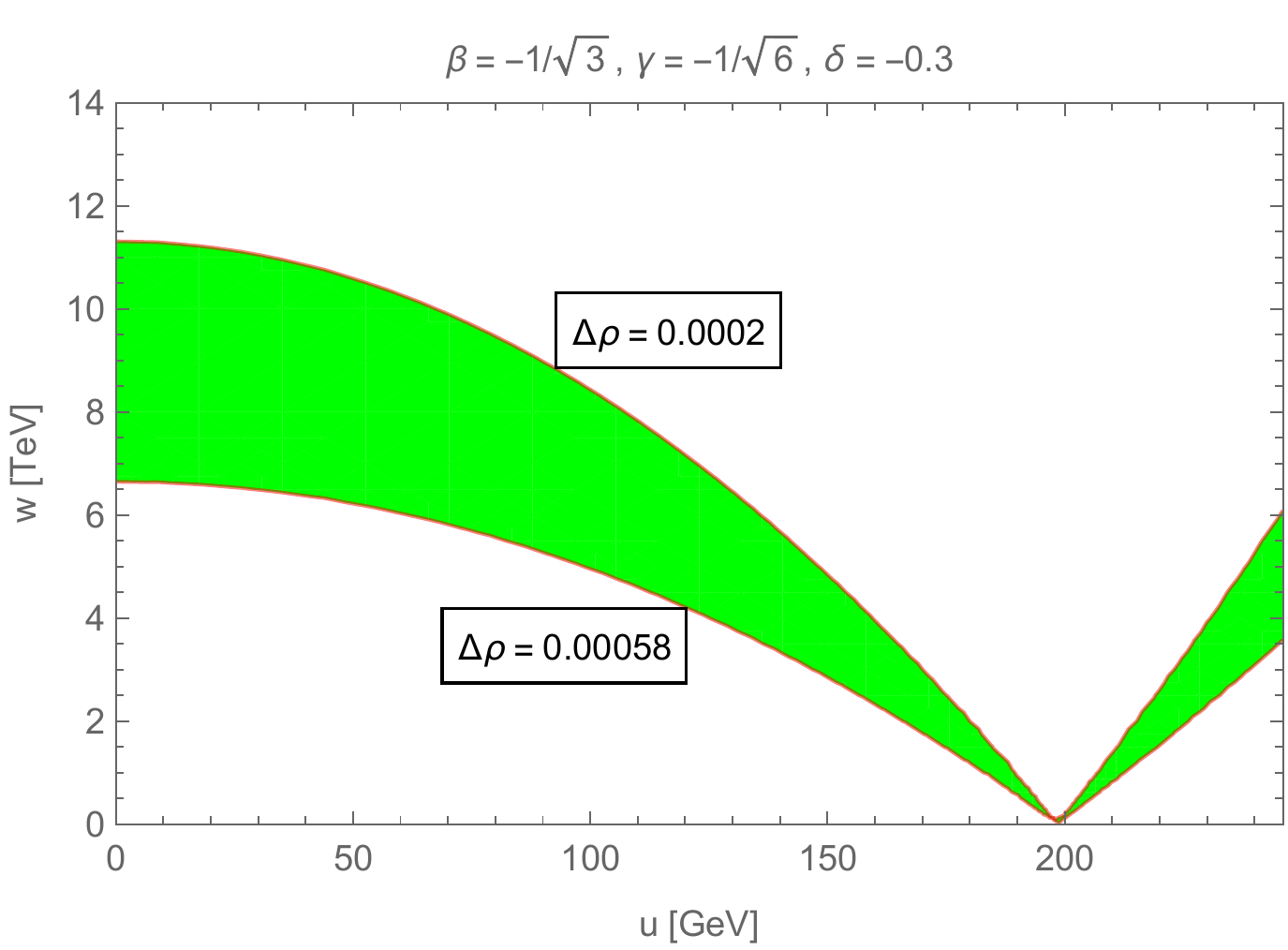}
\includegraphics[scale=0.31]{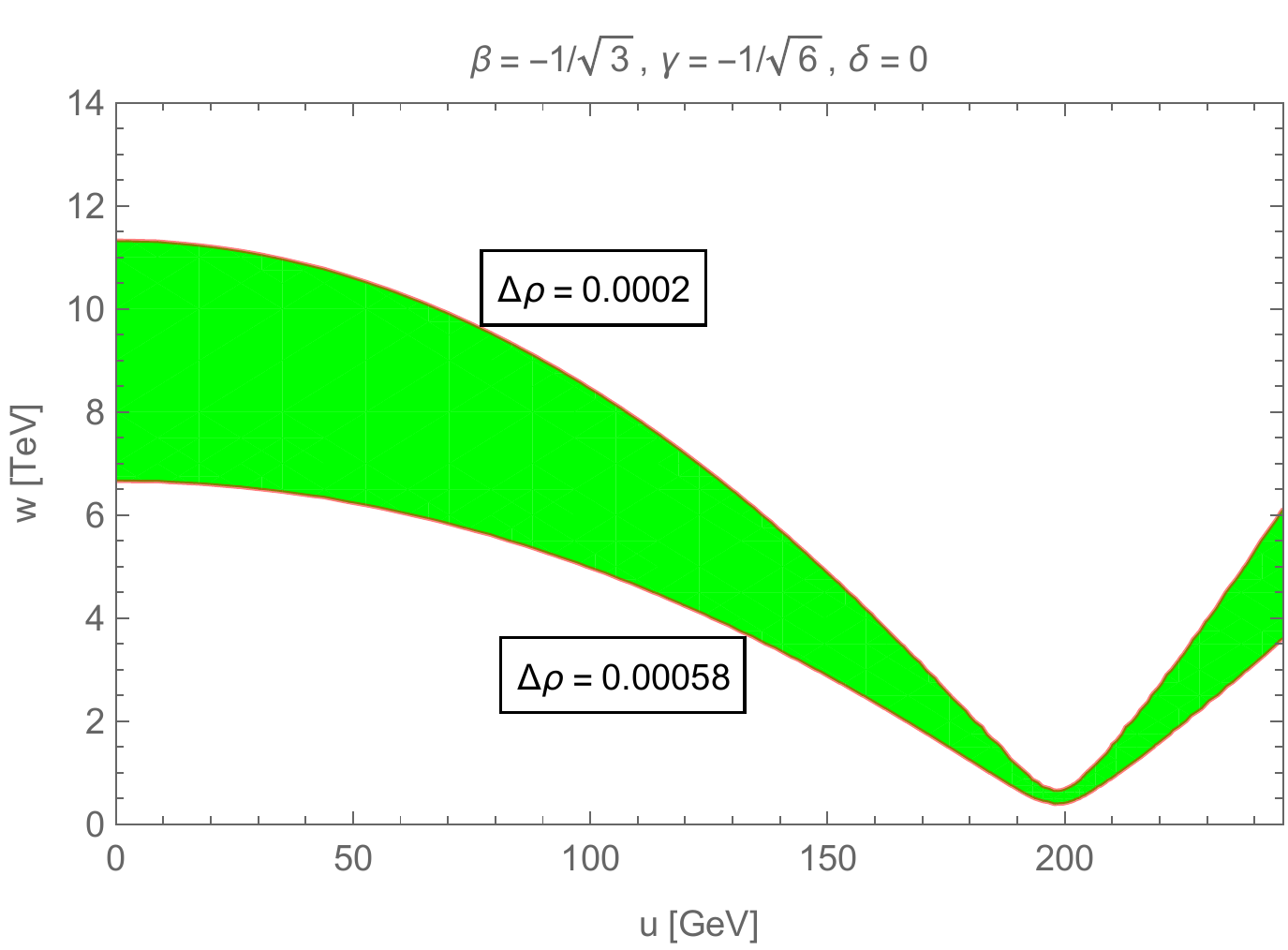}
\includegraphics[scale=0.31]{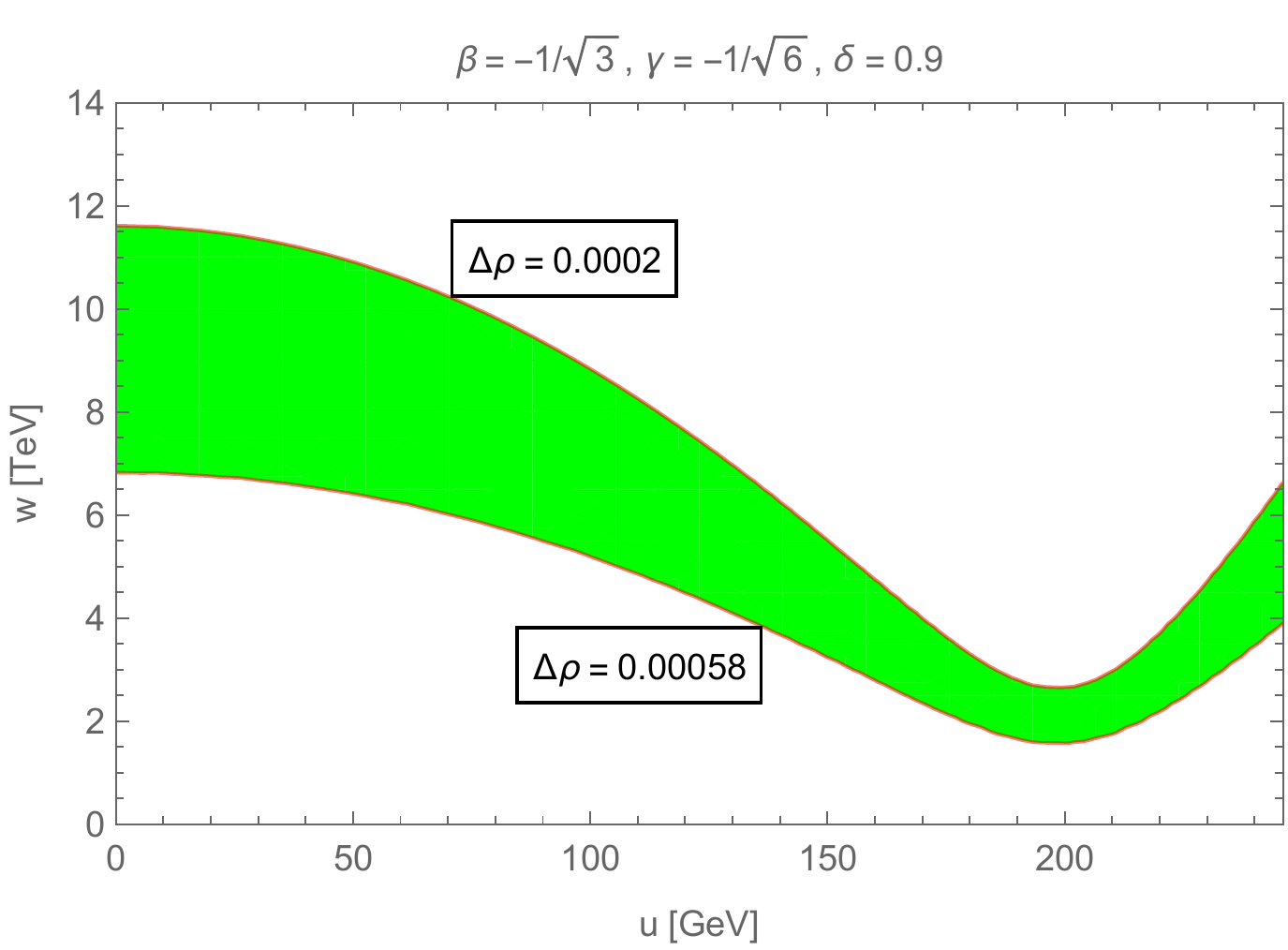}
\caption[]{\label{DT6} The $(u,w)$ regime that is bounded by the $\rho$ parameter for ($\beta=-1/\sqrt3, \gamma=-1/\sqrt6, b=-2/\sqrt3$, $c=-\sqrt2/\sqrt3$) and $w=0.5\Lambda\ll V$, where the panels ordered correspond to $\delta=-0.9,\ -0.3,\ 0$, and $0.9$, respectively.}
\end{center}
\end{figure}

The third case depending on the kinetic mixing parameter is given in Figs. \ref{DT3}, \ref{DT4}, \ref{DT5}, and \ref{DT6} according to the dark matter models ($\beta=1/\sqrt3, \gamma=1/\sqrt6$), ($\beta=1/\sqrt3, \gamma=-\sqrt2/\sqrt3$), ($\beta=-1/\sqrt3, \gamma=\sqrt2/\sqrt3$), and ($\beta=-1/\sqrt3, \gamma=-1/\sqrt6$), respectively. It is clear that the new physics scale bound is increased, when $|\delta|$ increases. The effect of $\delta$ is strong, when $u$ reaches values near $145$ GeV for the first dark matter model. By contrast, when $u$ approaches $0$ or $246$ GeV, the effect is negligible. In summary, the kinetic mixing effect is important when the new physics is considered.

\subsection{$Z_1 \bar{f}f $ couplings}
As stated, the considering model has the mixing of the $Z$ boson with the new neutral gauge bosons. From \eqref{tran2} and \eqref{tranu2}, we get $Z=Z_1+\epsilon_1\mathcal{Z}'_2+\epsilon_2\mathcal{Z}'_3+\epsilon_3\mathcal{C}'$, $Z'_2=-\epsilon_1Z_1+\mathcal{Z}'_2$, $Z'_3=-\epsilon_2Z_1+\mathcal{Z}'_3$, and $C'=-\epsilon_3Z_1+\mathcal{C}'$. Hence, the couplings of $Z_1$ to fermions are modified by the mixing parameters $\epsilon_{1,2,3}$. Fitting the standard model precision test, the room for the mixing parameters is only $10^{-3}$ order. Hence, we impose the bound $|\epsilon_{1,2,3}|=10^{-3}$.

\begin{figure}[!h]
\begin{center}
\includegraphics[scale=0.31]{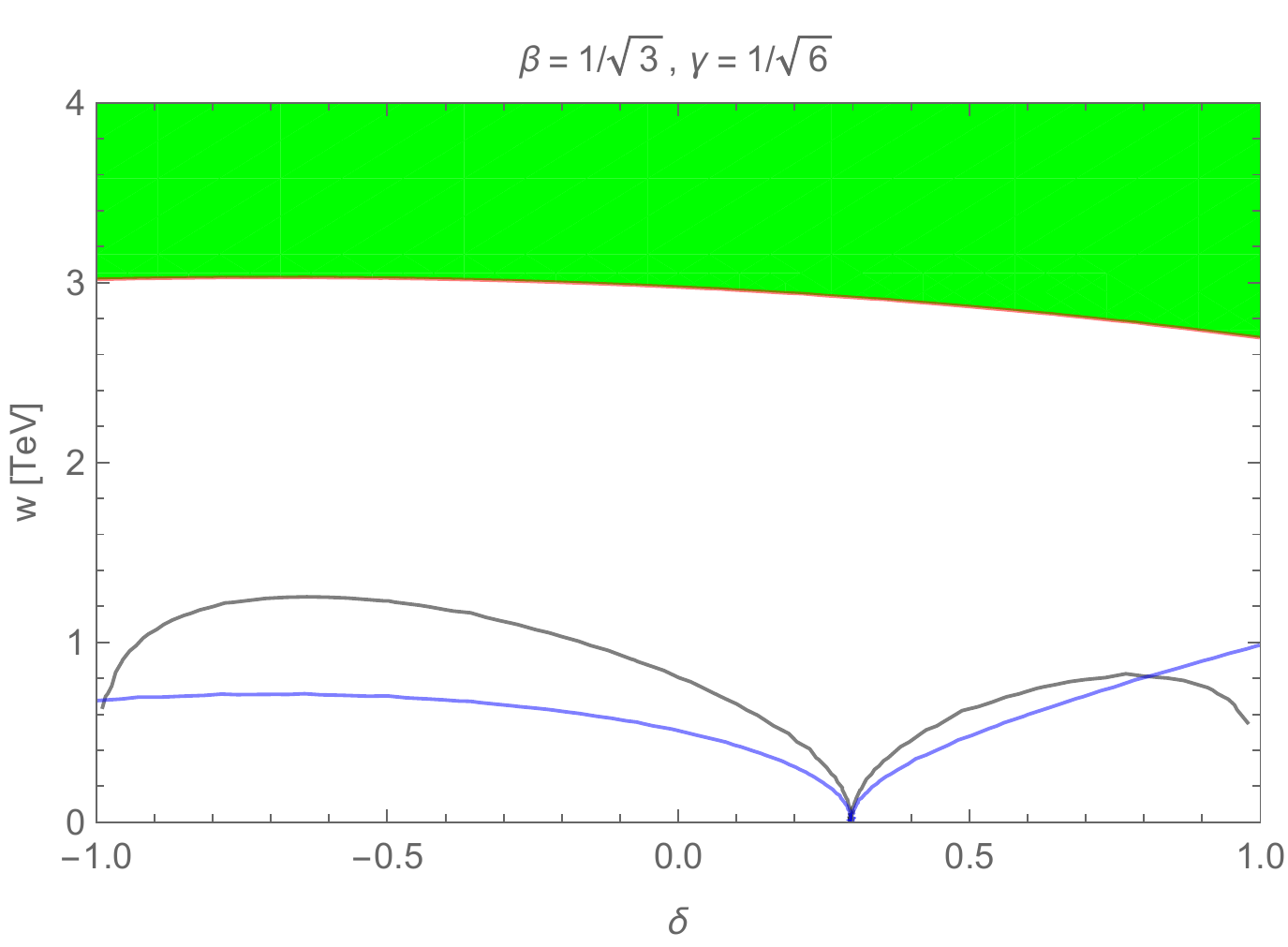}
\includegraphics[scale=0.31]{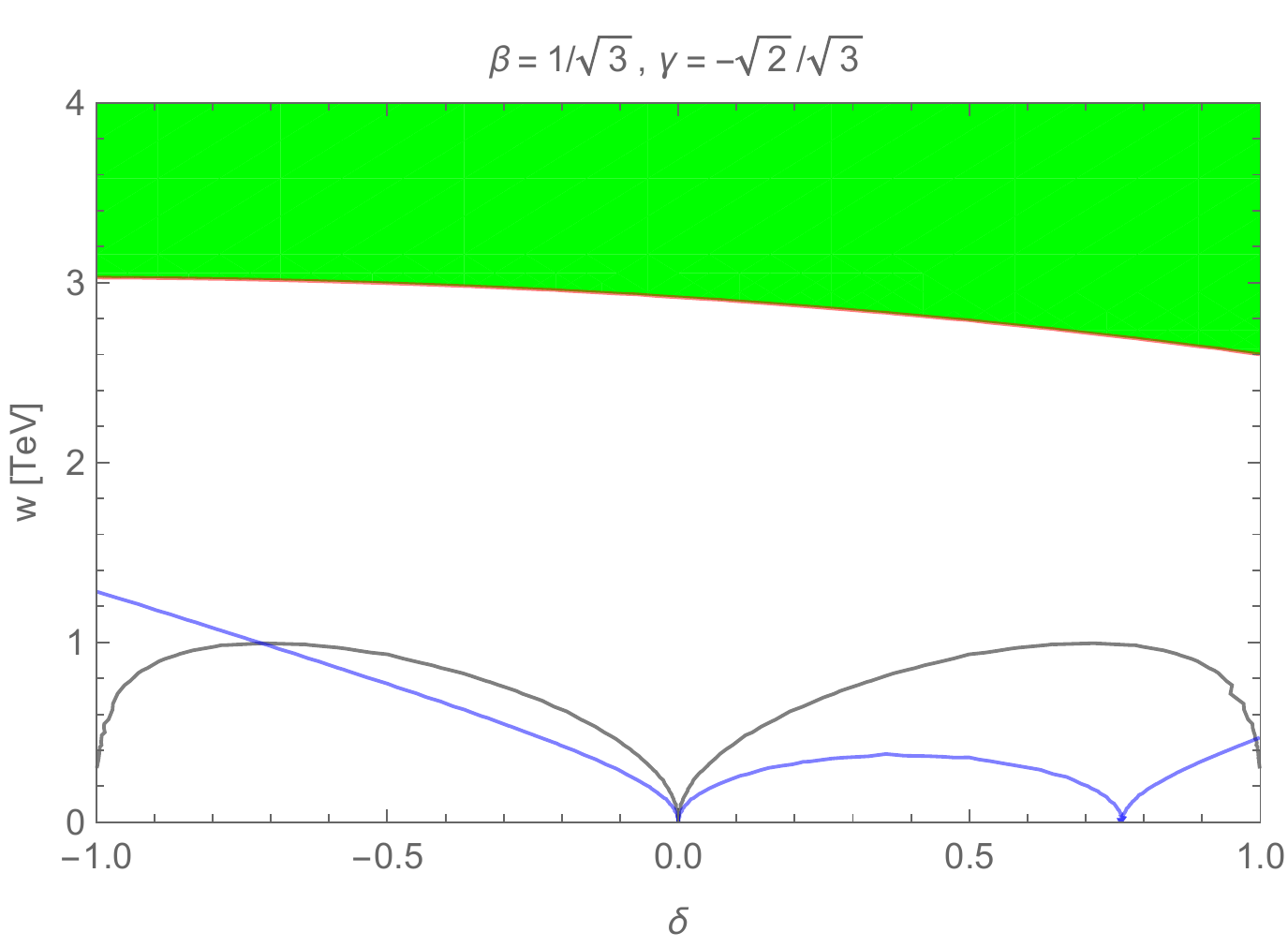}
\includegraphics[scale=0.31]{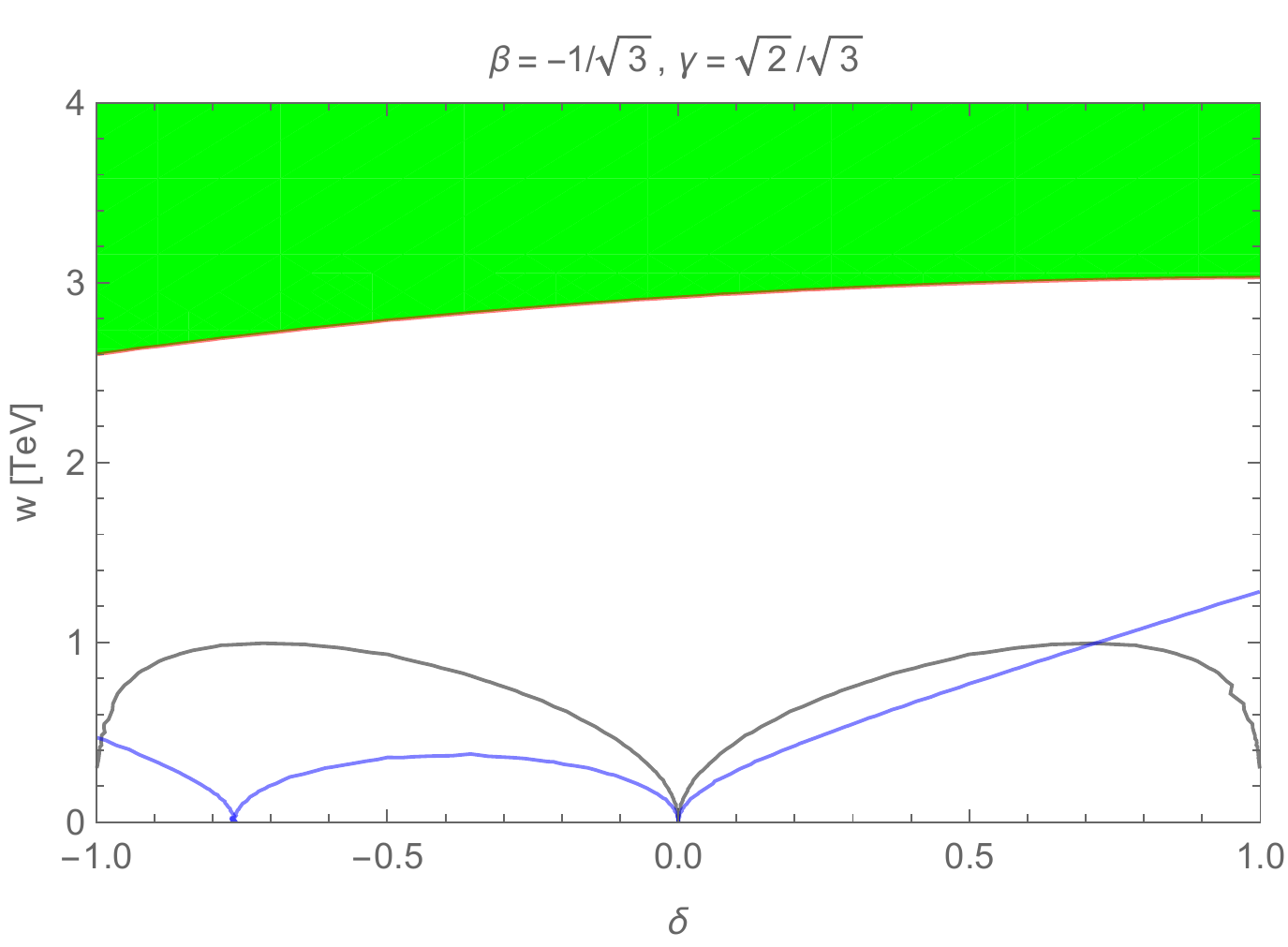}
\includegraphics[scale=0.31]{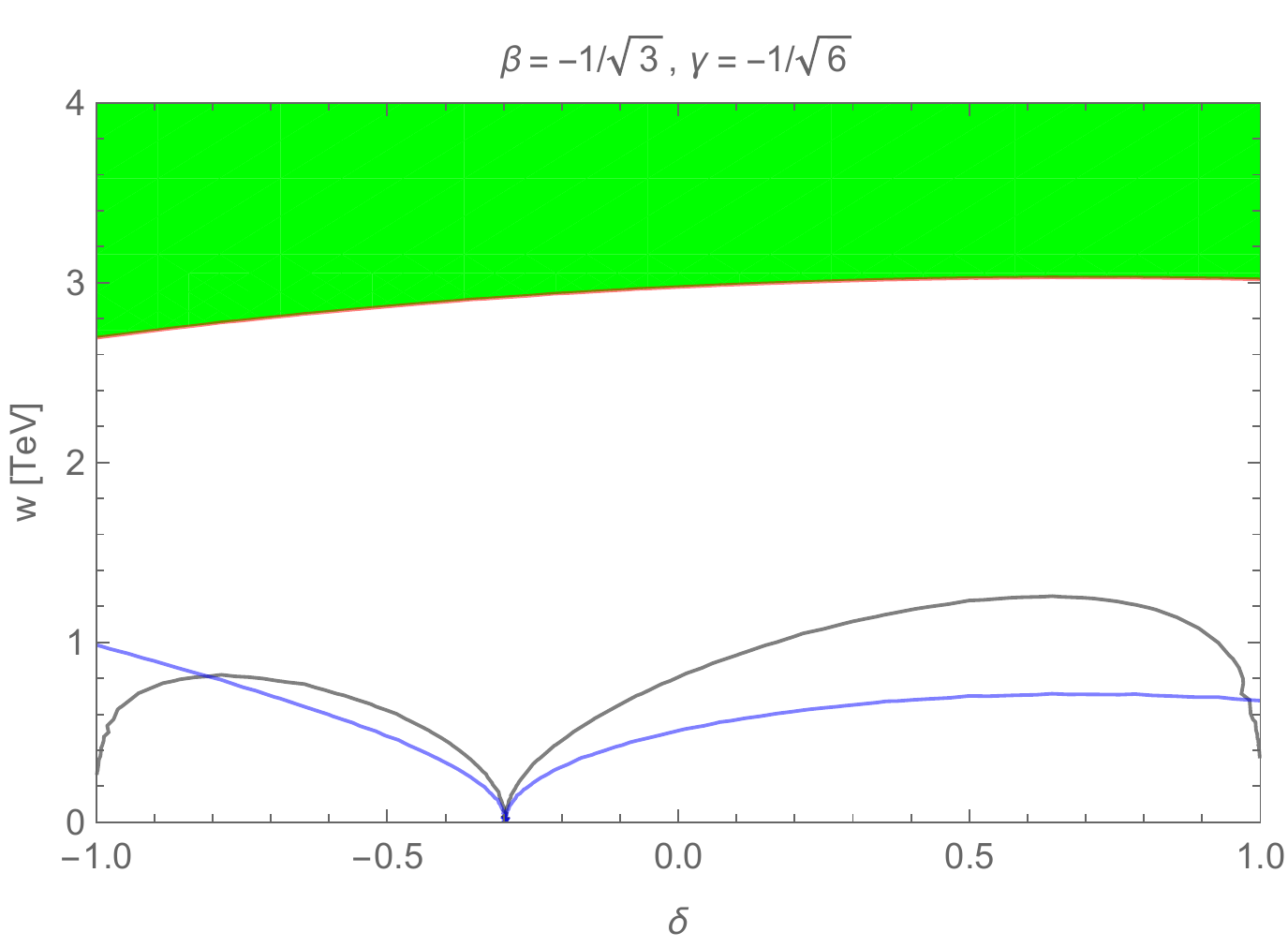}
\caption[]{\label{DT7} The bounds on new physics scales as the functions of $\delta$ for $|\epsilon_{1,2,3}|=10^{-3}$, where the red, blue, and black lines correspond to $\epsilon_1$, $\epsilon_2$, $\epsilon_3$ for the four kinds of dark matter models ($\beta=1/\sqrt3, \gamma=1/\sqrt6$), ($\beta=1/\sqrt3, \gamma=-\sqrt2/\sqrt3$), ($\beta=-1/\sqrt3, \gamma=\sqrt2/\sqrt3$), and ($\beta=-1/\sqrt3, \gamma=-1/\sqrt6$), respectively.}
\end{center}
\end{figure}

It is observed that in the first case ($w, V\ll \Lambda$), $\epsilon_3=0$ while $\epsilon_{1,2}$ are independent of $\delta, \Lambda$. In the second case ($w\ll V,\Lambda$), $\epsilon_{2,3}=0$ while $\epsilon_1$ is independent of  $\delta, V, \Lambda$. In the last case ($w,\Lambda\ll V$), all the parameters contribute to $\epsilon_{1,2,3}$, except for $V$. Hence, we consider only the sensitivity of the new physics scales in terms of the kinetic mixing parameter for the last case. Since the effect of kinetic mixing does not depend on the $u,v$ relation, we impose $u=v=246/\sqrt2$ GeV and use also the previous inputs. The results are given in Fig. \ref{DT7}. It indicates that the new physics regime changes when $\delta$ varies.

\section{\label{pheno1} FCNCs} 

Because the fermion generations transform differently under the gauge symmetry $SU(4)_L\otimes U(1)_X\otimes U(1)_N$, the tree-level FCNCs are present. Indeed, the neutral currents arise from  
\bea
\mathcal{L}_{\mathrm{NC}}&=& -g\bar{F}\ga^\mu [T_3A_{3\mu}+T_8A_{8\mu}+T_{15}A_{15\mu}+t_XXB_{\mu}+t_NNC_\mu] F \crn
&=&-g\bar{F}\ga^\mu [T_3A_{3\mu}+T_8A_{8\mu}+T_{15}A_{15\mu}+t_X(Q-T_3-\beta T_8-\gamma T_{15})B_{\mu}+t_N(B-L-bT_8-cT_{15})C_\mu] F.
\eea 

It is clear that the leptons and exotic quarks do not flavor-change. Furthermore, the terms of $T_3$, $Q$, and $B-L$ also conserve flavors. Hence, the FCNCs couple only the ordinary quarks to $T_{8,15}$, such that
\bea
\mathcal{L}_{\mathrm{NC}}\supset-g[\bar{q}_L\gamma^\mu T_8^q q_L(A_{8\mu}-\beta t_X B_\mu-bt_N C_\mu)+\bar{q}_L\gamma^\mu T_{15}^q q_L(A_{15\mu}-\gamma t_X B_\mu-ct_N C_\mu)],
\eea
where $q$ is denoted either $q=(u_1, u_2, u_3)$ or $q=(d_1, d_2, d_3)$, $T_8^q=\fr{1}{2\sqrt3}\text{diag}(-1,-1,1)$, and $T_{15}^q=\fr{1}{2\sqrt6}\text{diag}(-1,-1,1)$. Changing to the mass basis, $q_{L,R}=V_{qL,qR}q'_{L,R}$ where either $q'=(u,c,t)$ or $q'=d,s,b$, and $(A_3\, A_8\, A_{15}\, B\, C)^T=U(A\, Z_1\, Z_2\, Z_3\, Z_4)$, this yields
\bea
\mathcal{L}_{\mathrm{FCNC}}=-\bar{q}'_{iL}\gamma^\mu q'_{jL}(V^*_{qL})_{3i}(V_{qL})_{3j}(g_0A_\mu+g_1Z_{1\mu}+g_2Z_{2\mu}+g_3Z_{3\mu}+g_4Z_{4\mu})\hs (i\neq j).
\eea

It is noted that the photon always conserves flavors, $g_0=0$. In the first case ($w,V\ll \Lambda$), the couplings $g_{1,2,3,4}$ are
\bea
g_1&=&-\fr{g}{\sqrt6}\left[\fr{\sqrt2}{\sqrt{1-\beta^2t_W^2}}\epsilon_1+\fr{1+\gamma (\sqrt2 \beta+\gamma)t_X^2}{\sqrt{1+\gamma^2t_X^2}}\epsilon_2+\fr{\delta(\sqrt2 \beta+\gamma)t_X-(\sqrt2 b+c)t_N}{\sqrt{1-\delta^2}}\epsilon_3\right],\label{g1}\\
g_2&=&\fr{g}{\sqrt6}\left[\fr{\sqrt2}{\sqrt{1-\beta^2t_W^2}}c_\varphi-\fr{1+\gamma(\sqrt2 \beta+\gamma)t_X^2}{\sqrt{1+\gamma^2t_X^2}}s_\varphi\right],\\
g_3&=&g_2(c_\varphi\rightarrow s_\varphi, s_\varphi\rightarrow -c_\varphi),\\
g_4&=&\fr{g}{\sqrt6}\fr{\delta(\sqrt2 \beta+\gamma)t_X-(\sqrt2 b+c)t_N}{\sqrt{1-\delta^2}}.
\eea
In the third case ($w, \Lambda\ll V$), the coupling $g_1$ is identical to (\ref{g1}), while 
\bea
g_2&=&-\fr{g}{\sqrt6}\left[\fr{\sqrt2}{\sqrt{1-\beta^2t_W^2}}+\fr{1+\gamma (\sqrt2 \beta+\gamma)t_X^2}{\sqrt{1+\gamma^2t_X^2}}\mathcal{E}_1+\fr{\delta(\sqrt2 \beta+\gamma)t_X-(\sqrt2 b+c)t_N}{\sqrt{1-\delta^2}}\mathcal{E}_2\right],\\
g_3&=&\fr{g}{\sqrt6}\left\{c_\xi\left[\fr{\sqrt2}{\sqrt{1-\beta^2t_W^2}}\mathcal{E}_1+\fr{1+\gamma (\sqrt2 \beta+\gamma)t_X^2}{\sqrt{1+\gamma^2t_X^2}}\right]-s_\xi\left[\fr{\sqrt2}{\sqrt{1-\beta^2t_W^2}}\mathcal{E}_2+\fr{\delta(\sqrt2 \beta+\gamma)t_X-(\sqrt2 b+c)t_N}{\sqrt{1-\delta^2}}\right]\right\},\\
g_4&=&g_3(c_\xi\rightarrow s_\xi, s_\xi\rightarrow -c_\xi).
\eea
In the second case ($w\ll V, \Lambda$), the couplings can be obtained from those in the third case by $\mathcal{E}_{1,2}\rightarrow 0$.

The contribution of the new physics to the meson mixing is given after integrating $Z_{1,2,3,4}$ out,
\bea
 \mathcal{L}^{\mathrm{eff}}_{\mathrm{FCNC}}&=&(\bar{q}'_{iL}\ga^\mu q'_{jL})^2 [(V^*_{qL})_{3i}(V_{qL})_{3j}]^2\left(\fr{g^2_1}{m^2_{Z_1}} + \fr{g^2_2}{m^2_{Z_2}}+\fr{g^2_3}{m^2_{Z_3}}+\fr{g^2_4}{m^2_{Z_4}}\right)\crn
&\simeq&(\bar{q}'_{iL}\ga^\mu q'_{jL})^2 [(V^*_{qL})_{3i}(V_{qL})_{3j}]^2\left(\fr{g^2_2}{m^2_{Z_2}}+\fr{g^2_3}{m^2_{Z_3}}+\fr{g^2_4}{m^2_{Z_4}}\right),
\eea
where the $Z_1$ contribution is small and omitted. 

The strongest bound comes from $B^0_s-\bar{B}^0_s$ mixing, implying \cite{pdg2018}
\bea
[(V^*_{dL})_{32}(V_{dL})_{33}]^2\left(\fr{g^2_2}{m^2_{Z_2}}+\fr{g^2_3}{m^2_{Z_3}}+\fr{g^2_4}{m^2_{Z_4}}\right)<\fr{1}{(100 \text{ TeV})^2}.
\eea
We assume the sector of up quarks to be flavor diagonal, i.e. $V_{\text{CKM}}\equiv V^\dagger_{uL}V_{dL} =V_{dL}$. We have $|(V^*_{dL})_{32}(V_{dL})_{33}|\simeq 3.9\times 10^{-2}$ \cite{pdg2018}, which leads to
\bea
\sqrt{\fr{g^2_2}{m^2_{Z_2}}+\fr{g^2_3}{m^2_{Z_3}}+\fr{g^2_4}{m^2_{Z_4}}}<\fr{1}{3.9 \text{ TeV}}.\label{constraint}
\eea
Our remark is that since $u,v\ll w, V, \Lambda$, the l.h.s of (\ref{constraint}) depends only on the new physics scales, not on the weak scales.

In the first case ($w, V\ll \Lambda$), the $Z_4$ contribution is negligible. The l.h.s of (\ref{constraint}) is independent of $\delta$. The other inputs given previously are used, implying the bound for $w> 4.36$ TeV for all the four dark matter models.

In the second case ($w\ll V, \Lambda$), the $Z_{3,4}$ contributions are negligible. The l.h.s of (\ref{constraint}) is independent of $\beta$, $\gamma$, and $\delta$. The bound yields $w> 3.9$ TeV for all the four models.

\begin{figure}[!h]
\begin{center}
\includegraphics[scale=0.31]{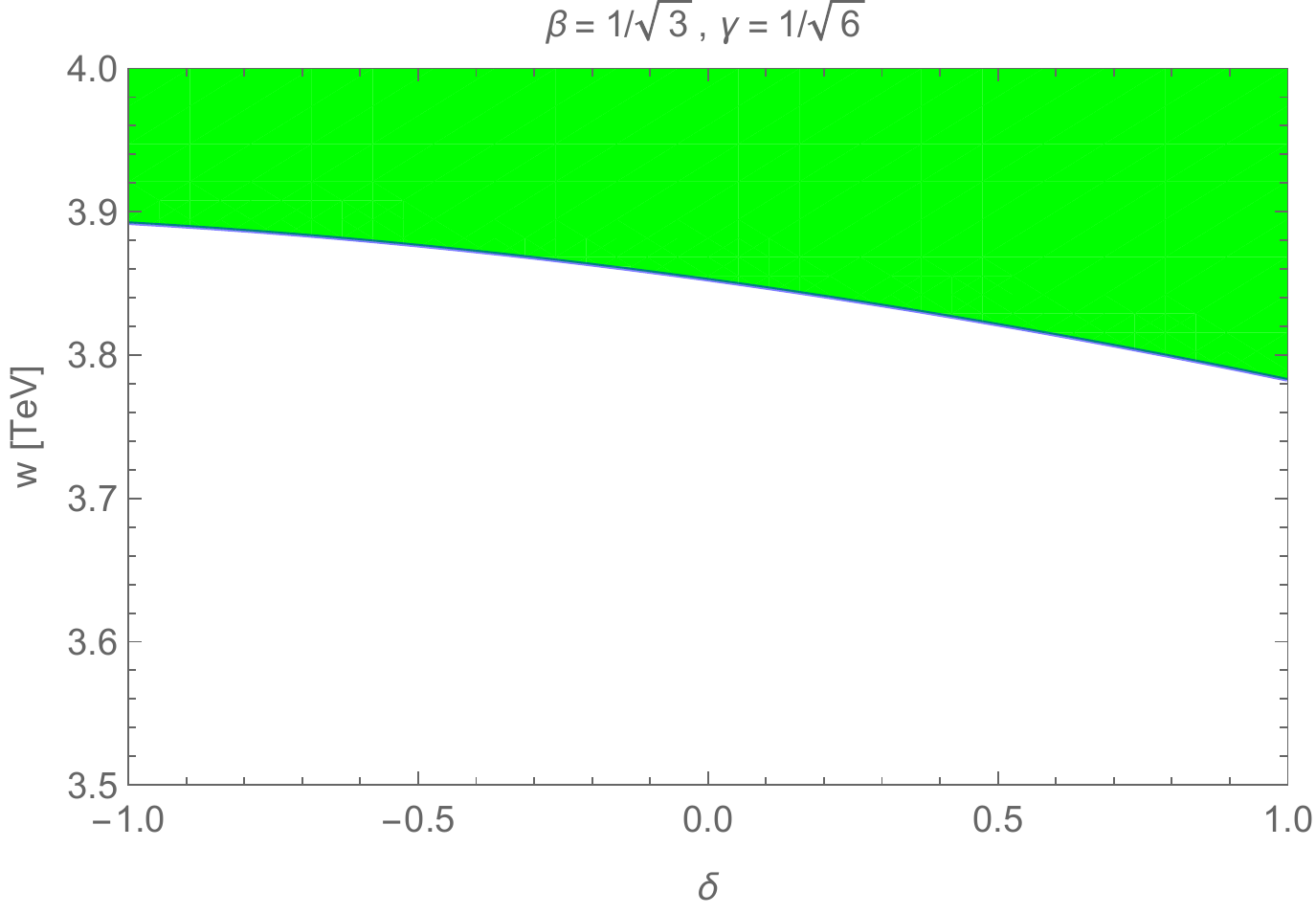}
\includegraphics[scale=0.31]{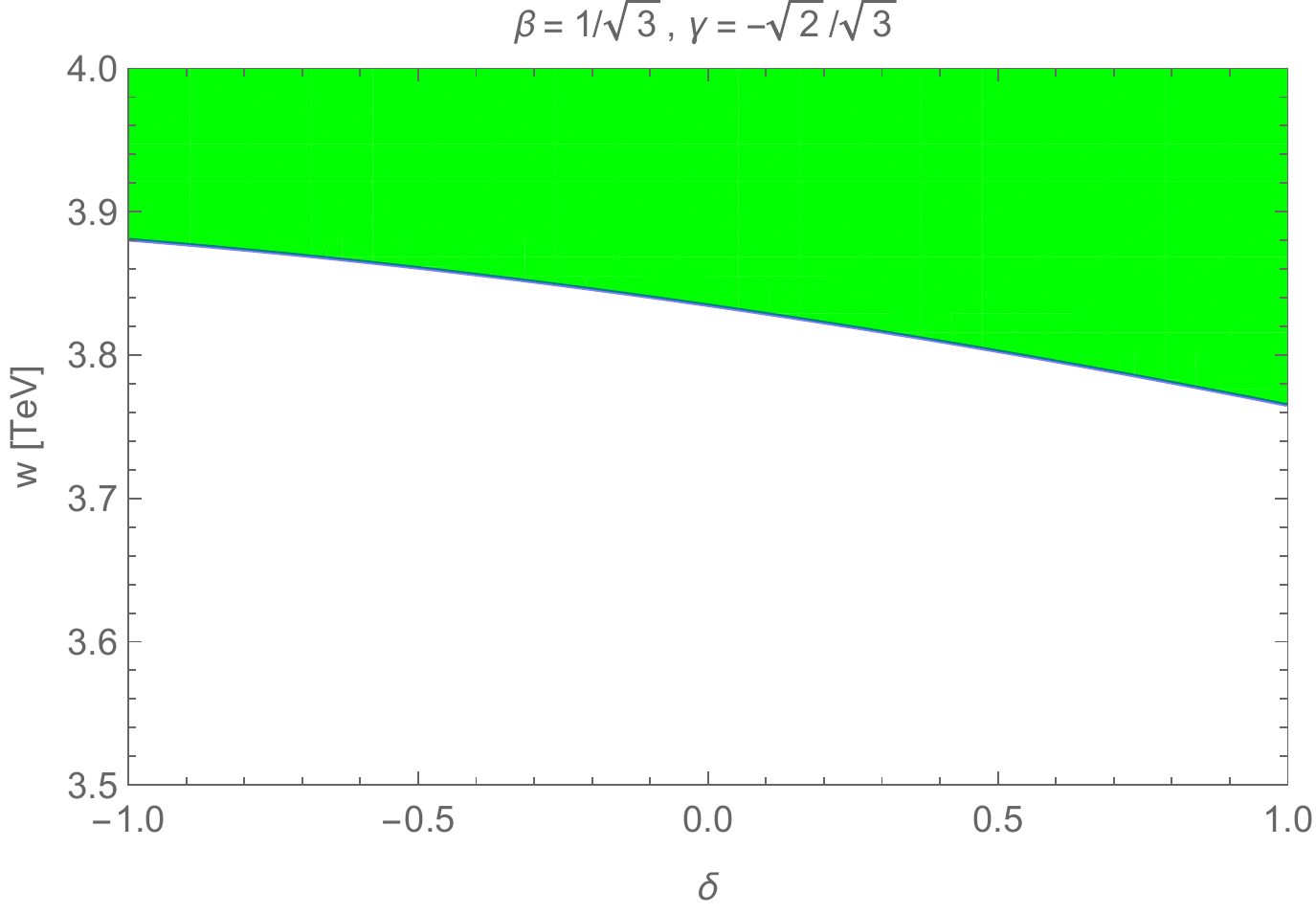}
\includegraphics[scale=0.31]{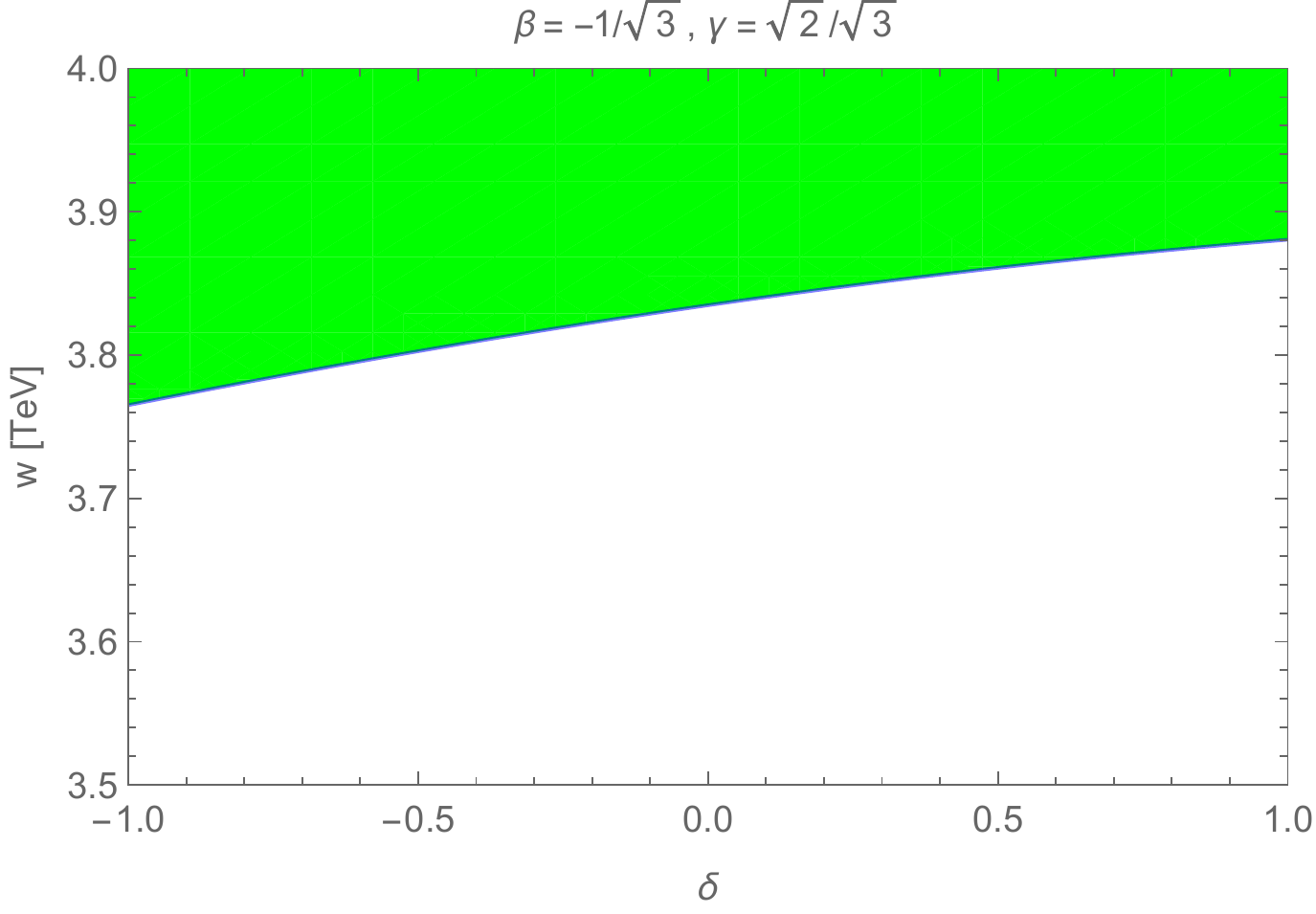}
\includegraphics[scale=0.31]{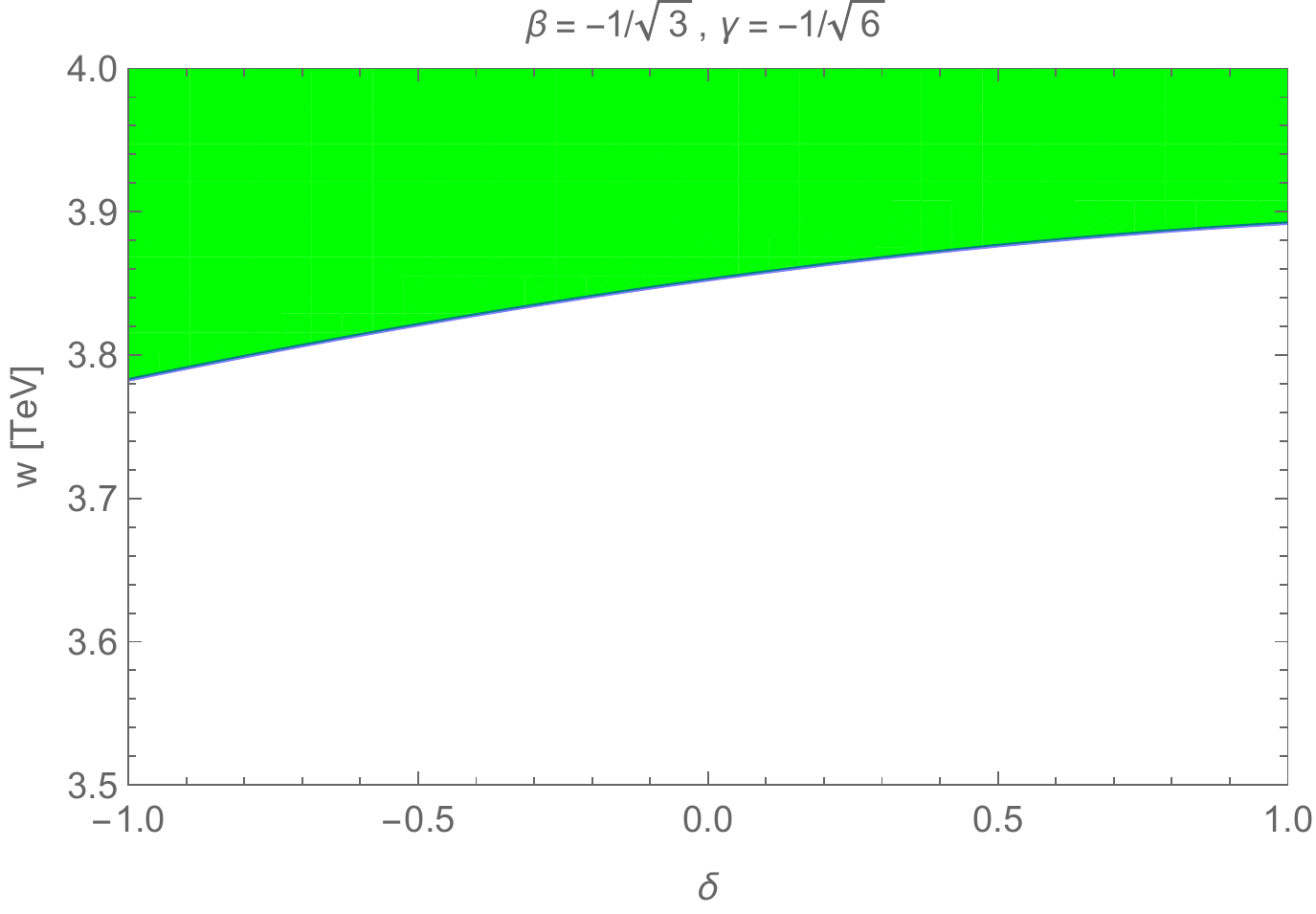}
\caption[]{\label{FCNC} The bounds on the new physics scales as functions of $\delta$ from the FCNCs for $w=0.5 \Lambda\ll V$,  where the panels left to right are for the four dark matter models, respectively.}
\end{center}
\end{figure}

In the third case ($w, \Lambda \ll V$), since the mixing angles $\mathcal{E}_{1,2}$ are finite, the l.h.s of (\ref{constraint}) depends on $\beta$, $\gamma$, and $\delta$, and is depicted in Fig. \ref{FCNC}. The figure yields that the new physics regime changes when $\delta$ varies. Furthermore, those bounds are obviously lower than that given by the two case above.

\section{\label{lhc} Collider bounds}

Since the new neutral gauge bosons couple to leptons and quarks, they contribute to the Drell-Yan and dijet processes at colliders. 

The LEPII searches for $e^+e^-\to \mu^+\mu^-$ happen similarly to the case of the 3-3-1-1 model, where all the new gauge bosons $Z_{2,3,4}$ mediate the process. Assuming that all the new physics scales are the same order, they are bounded in the TeV scale \cite{3311}.

The LHC searches for dijet and dilepton final states can be studied. Using the above condition, the new physics scales are also in TeV, similarly to \cite{331p}.       

\section{\label{3311} The 3-3-1-1 model revisited}

The 3-3-1-1 model is based upon the gauge symmetry $SU(3)_C\otimes SU(3)_L\otimes U(1)_X\otimes U(1)_N$. Thus it contains four neutral gauge bosons $A_{3,8}$, $B$, and $C$ according to the last three gauge groups, in which $B,C$ has a kinetic mixing term, $-(\delta/2)B_{\mu\nu}C^{\mu\nu}$. The kinetic mixing effect in the 3-3-1-1 model was explicitly studied in \cite{dong2016}. Here we present only new results beyond the previous investigation.    

Changing to the canonical basis, $A_{3}$, $A_8$, $B'$, and $C'$, the corresponding mass matrix $M^2=\{m^2_{ij}\}$ is given by  
\bea
m^2_{11}&= & \fr {g^2} 4 (u^2 + v^2),\hs m^2_{12}=  \fr {g^2} {4\sqrt3} (u^2-v^2),\hs m^2_{13} = -\fr{g^2t_X}{4\sqrt3}[(\sqrt3+\beta)u^2+(\sqrt3-\beta)v^2], \crn 
m^2_{14}&=& \fr{g^2}{4\sqrt{3(1-\delta^2)}}\{[\delta (\sqrt3+\beta)t_X-b t_N] u^2 +[\delta (\sqrt3-\beta)t_X+b t_N] v^2 \} ,\crn
m^2_{22}&=&\fr {g^2} {12}(u^2+v^2+4w^2),\hs
m^2_{23}= -\fr{g^2t_X}{12}[(\sqrt3+\beta)u^2-(\sqrt3-\beta)v^2+4\beta w^2], 
\crn
m^2_{24}&=& \fr{g^2} { 12 \sqrt{1-\delta^2}}\{[\delta (\sqrt3+\beta)t_X-b t_N] u^2-[\delta (\sqrt3-\beta)t_X+b t_N] v^2 +4(\delta\beta t_X-b t_N)w^2\},\crn
m^2_{33} &=&   \fr{g^2t^2_X}{12}[(\sqrt3+\beta)^2u^2 +(\sqrt3-\beta)^2v^2 + 4\beta^2w^2],\crn
m^2_{34}&=& \fr{-g^2 t_X } { 12\sqrt{1-\delta^2}}\{ (\sqrt3+\beta)[\delta (\sqrt3+\beta)t_X-b t_N] u^2+ (\sqrt3-\beta)[\delta (\sqrt3-\beta)t_X+b t_N] v^2+4\beta \left(\delta\beta t_X-b t_N\right)w^2\}, \crn
m^2_{44}&=& \fr{g^2} { 12(1-\delta^2)}\{[\delta (\sqrt3+\beta)t_X - b t_N]^2 u^2+[\delta (\sqrt3-\beta) t_X+ b  t_N]^2  v^2  +4(\delta\beta t_X-b t_N)^2w^2+48t^2_N\La^2\}.\nn
\eea This result is similar to that in \cite{dong2016}, except for the last element, $m^2_{44}$, that differs in the coefficient of $\La^2$. Note that $t_X=g_X/g$, $t_N=g_N/g$, $\beta$, $b$, $u$, $v$, and $w$ are those parameters belonging to the 3-3-1-1 model and in this case we have $s_W=e/g=t_X/\sqrt{1+(1+\beta^2)t_X^2}$.

Changing to the electroweak basis, $(A_3\, A_8 \, B' \, C')^T =U_1 (A \, Z \, Z' \, C')^T$, where $C'$ is orthogonal to $A = s_W A_3 + c_W \left(\beta t_W A_8 +\sqrt{1-\beta^2 t^2_W}B'\right)$, 
$Z=c_W A_3 - s_W \left(\beta t_W A_8 +\sqrt{1-\beta^2 t^2_W}B'\right)$, and
$Z' =\sqrt{1-\beta^2 t^2_W}\left(A_8- 
 \beta t_W B'\right)$, thus
\bea
 U_1 = \left (\begin{array}{cccc}
 s_W &c_W&0& 0\\
\beta s_W &-\beta s_W t_W &\sqrt{1-\beta^2t^2_W}&0\\
c_W\sqrt{1-\beta^2t^2_W}&- s_W\sqrt{1-\beta^2 t^2_W}&- \beta t_W&0\\
0&0&0&1
\end{array} \right ), \label{neutral2}
\eea 
the mass matrix $M^2$ changes to 
\bea M'^2 = U^T_1 M^2 U_1=
\left(
\begin{array}{cc}
0 & 0 \\
0 & M'^2_{s}\end{array} \right),\hs M'^2_s \equiv
\left(
\begin{array}{ccc}
m^2_{Z} & m^2_{ZZ'} & m^2_{ZC'}\\
m^2_{ZZ'} & m^2_{Z'} & m^2_{Z'C'}\\
m^2_{ZC'} & m^2_{Z'C'} & m^2_{C'}
\end{array}\right), 
\eea which has the elements as given in \cite{dong2016}, in which $m^2_{C'}=  m^2_{44}$. 

The light state $Z$ can be separated by using the seesaw approximation,    
 \bea
\left(Z \, Z' \, C' \right)^T = U_2 \left(Z_1 \, \mathcal{Z}' \, \mathcal{C}'\right)^T, \hs M''^2=U^{T}_2M'^2_sU_2= \left(\begin{array}{cc}
m^{2}_{Z_{1}} & 0\\
0&M^2_{2\times 2}\end{array}\right),
\eea
where 
\bea
&& U_2 \simeq  \left(\begin{array}{ccc}
1 & \epsilon_1 & \epsilon_2  \\
-\epsilon_1 &  1 &0 \\
-\epsilon_2 &0 & 1 \\
\end{array}
\right), \hs M^2_{2\times 2}\simeq    \left(\begin{array}{cc}
m^2_{Z'}&m^2_{Z'C'}\\
m^2_{Z'C'}&m^2_{C'} \end{array}\right). \\
&& m^2_{Z_1}\simeq  m^2_{Z}- \epsilon_1m^2_{ZZ'}-\epsilon_2m^2_{ZC'}.
\eea
We separate $\epsilon_{1,2}\equiv\epsilon_{1,2}^0+\epsilon_{1,2}^\delta$, where $\epsilon^0_{1,2}$ are the mixing parameters due to the symmetry breaking \cite{3311,dong2015}, while $\epsilon^{\delta}_{1,2}$ determine the kinetic mixing effect,  
\bea 
\epsilon_1^0&=&\fr{\sqrt{1-\beta^2t_W^2}}{4c_W}\left\{\fr{\sqrt3[u^2-v^2+\sqrt3\beta t_W^2(u^2+v^2)]}{w^2}+\fr{b^2\beta t_W^2(u^2+v^2)}{4\Lambda^2}\right\},\\
\epsilon_2^0&=&\fr{b\beta t_W^2(u^2+v^2)}{16c_Wt_N\Lambda^2},\\
\epsilon_1^\delta&=&\fr{\delta t_W[b(1-2\beta^2t_W^2)t_N-\delta\beta t_W\sqrt{1-\beta^2t_W^2}](u^2+v^2)}{16c_Wt_N^2\Lambda^2},\\\epsilon_2^\delta&=&\fr{\delta t_W(u^2+v^2)}{16c_Wt_N\Lambda^2}\left(\fr{\sqrt{1-\delta^2}\sqrt{1-\beta^2t_W^2}}{t_N}-\fr{\delta b \beta t_W}{1+\sqrt{1-\delta^2}}\right),
 \eea
where $\epsilon_{1,2}^\delta$ differ from those in \cite{dong2016}. 

We diagonalize $M^2_{2\times 2}$ to obtain mass eigenstates,  
\bea
Z_2 = c_\xi \mathcal{Z}' - s_\xi \mathcal{C}', \hs Z_3 = s_\xi \mathcal{Z}'+ c_\xi \mathcal{C}',
\eea in which the $\mathcal{Z}'-\mathcal{C}'$ mixing angle and masses are
 \bea 
t_{2\xi}&\simeq& \frac{2\sqrt{1-\delta^2}(\delta\beta t_W-bt_N \sqrt{1-\beta^2t_W^2})w^2}{[(\delta\beta t_W-bt_N\sqrt{1-\beta^2t_W^2})^2-(1-\delta^2)]w^2+12(1-\beta^2t_W^2)t_N^2\Lambda^2},\\
m^2_ {Z_2,Z_3}&=& \fr1 2 [m^2_ {Z'} +m^2_{C'} \mp \sqrt{(m^2_ {Z'} - m^2_{C'})^2 + 4m^4_ {Z' C'} }]. 
 \eea
Generally, $\xi$ is finite if $w\sim \La$. The kinetic mixing and symmetry breaking effects cancel out if $\delta=bt_N/\beta t_X$, which takes place between $\delta$ and $b/\beta$---the embedding coefficients of $T_8$. Whereas, in the 3-4-1-1 model, it happens between $\delta$ and $c/\gamma$---the embedding coefficients of $T_{15}$.   

Hence, the gauge states are connected to the physical states by $(A_3\, A_8\, B\, C)^T=U_\delta U_1U_2U_\xi (A\, Z_1\, Z_2\,Z_3)^T$, where  
\bea 
U_\delta= \left(%
\begin{array}{cccc}
1&0&0&0\\
0&1&0&0\\
0&0&1&-\fr{\delta}{\sqrt{1-\delta^2}}\\
0&0&0&\fr{1}{\sqrt{1-\delta^2}}
\end{array} \right),  \hs
U_\xi= \left(%
\begin{array}{cccc}
1&0&0&0\\
0&1&0&0\\
0&0&c_\xi&s_\xi\\
0&0&-s_\xi&c_\xi
\end{array} \right).
\eea

The $\rho$ deviation starts from the tree-level contribution,  
  \bea 
(\Delta\rho)_{\mathrm{tree}} &=& \fr{m^2_W}{c^2_Wm^2_{Z_1}}-1=\fr{m^2_Z}{m^2_Z-\epsilon_1m^2_{ZZ'}-\epsilon_2 m^2_{ZC'}}-1\simeq \fr{\epsilon_1m^2_{ZZ'}+\epsilon_2 m^2_{ZC'}}{m^2_Z}\equiv (\Delta \rho)_{\mathrm{tree}}^0+(\Delta \rho)_{\mathrm{tree}}^\delta,
\eea
where
\bea
(\Delta \rho)_{\mathrm{tree}}^0&\simeq&\fr{[u^2-v^2+\sqrt3\beta t_W^2(u^2+v^2)]^2}{4(u^2+v^2)w^2}+\fr{b^2\beta^2t_W^4(u^2+v^2)}{16\Lambda^2},\\
(\Delta \rho)_{\mathrm{tree}}^\delta&\simeq&\fr{\delta\sqrt{1-\beta^2t_W^2}(\delta\sqrt{1-\beta^2t_W^2}+2b\beta t_Wt_N)t_W^2(u^2+v^2)}{16t_N^2\Lambda^2}.\label{rhodelta}
\eea
In this computation, we also include one-loop contributions by the gauge vector doublet $(X,Y)$, as supplied in \cite{dong2016}.

If $\La\gg w$, $\Delta\rho$ does not depend on $\La$, $t_N$, $b$, and $\delta$. If $\La \sim w$, all the parameters modify $\Delta\rho$. Comparing to \cite{dong2016}, the difference is only expressions related to $\delta$. Hence, the first case is not investigated in this work. To finalize the result, we use the parameter values similar to those in \cite{dong2016}, namely $\La=2w$, $t_N=0.5$, $n=0$ (thus $b=-2/\sqrt{3}$), and $q=-1,0, 1$ (thus $\beta=1/\sqrt{3},-1/\sqrt{3},-\sqrt{3}$, respectively).  

\begin{figure}[!h]
\begin{center}
\includegraphics[scale=0.31]{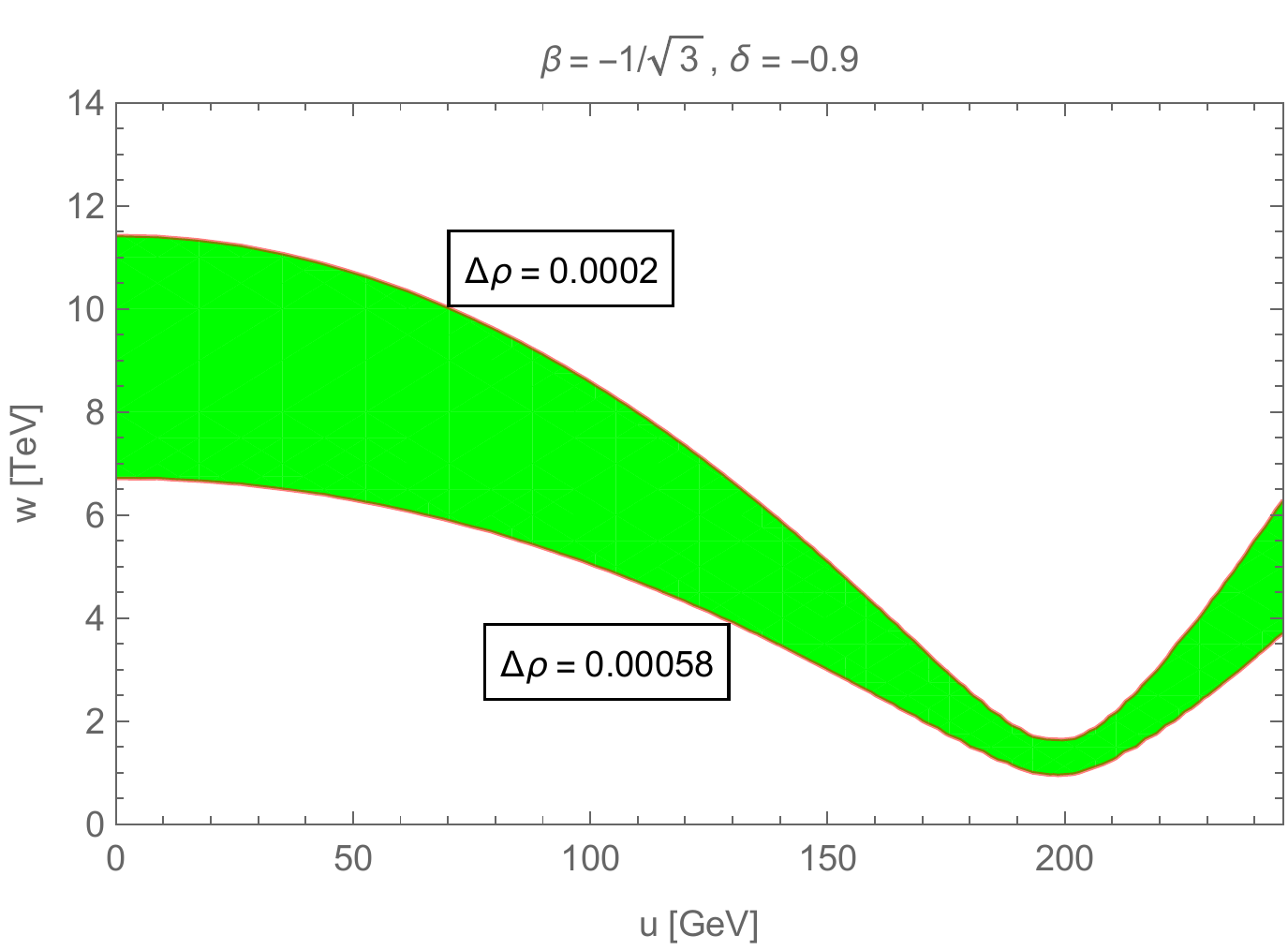}
\includegraphics[scale=0.31]{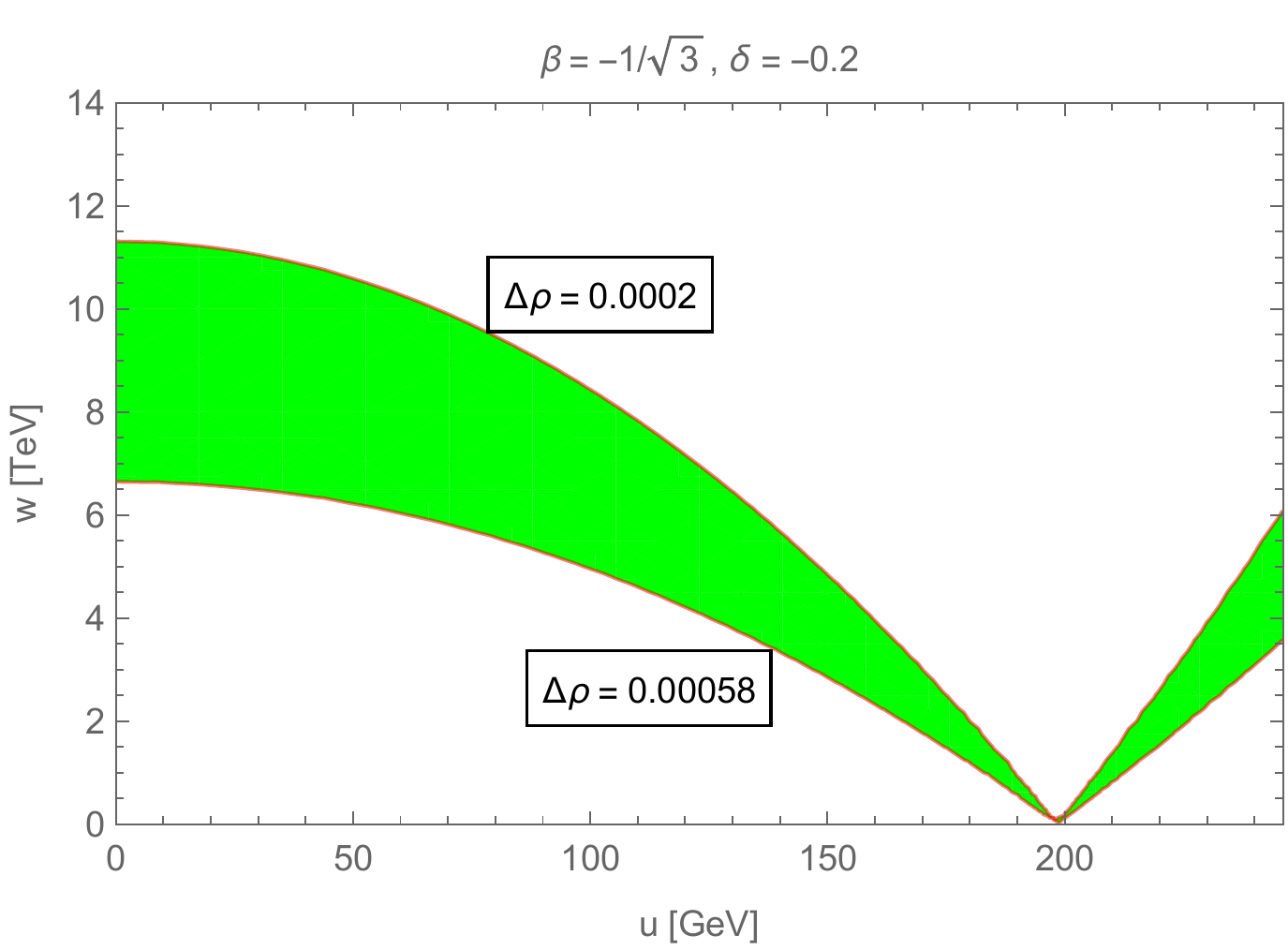}
\includegraphics[scale=0.31]{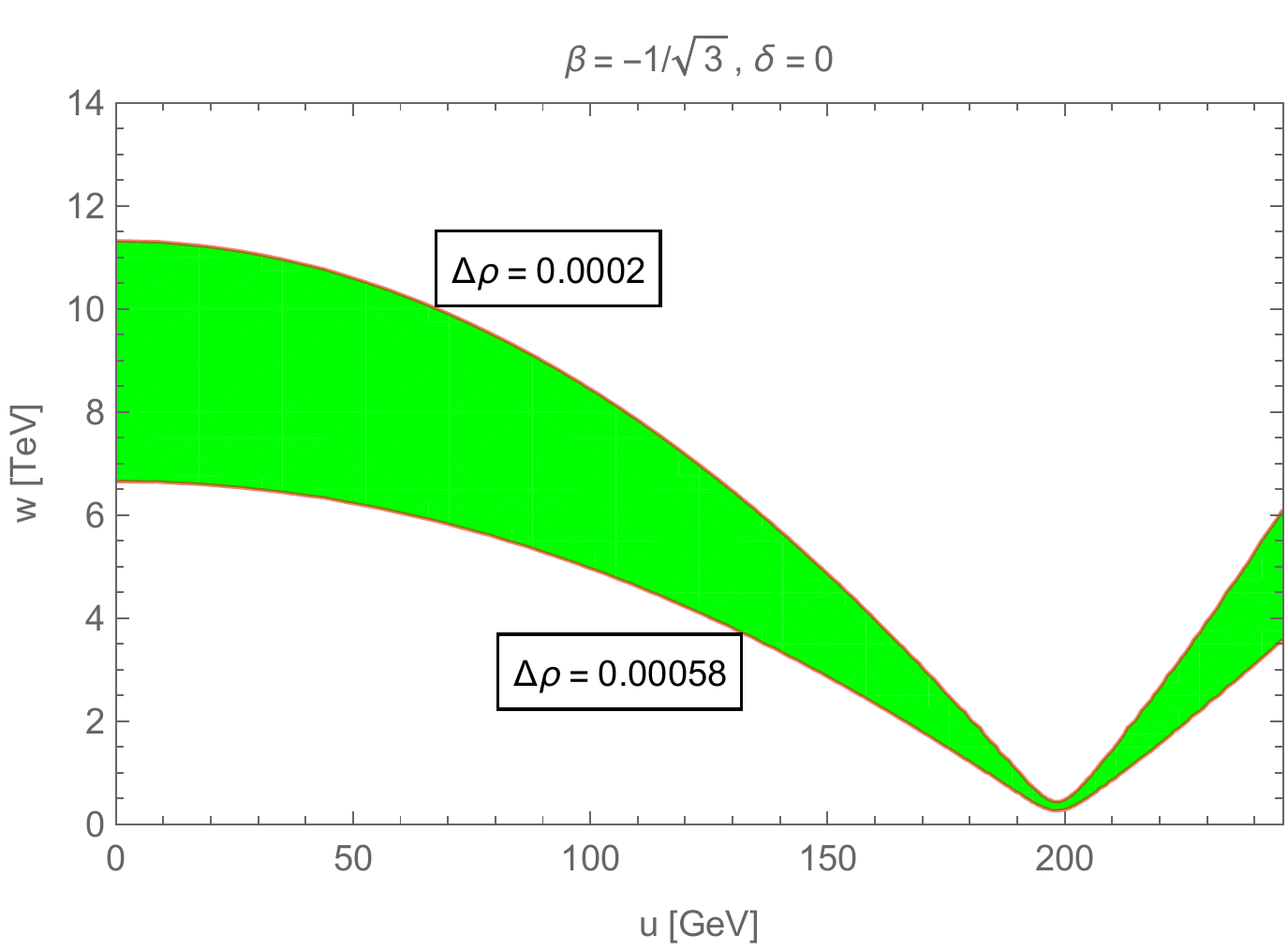}
\includegraphics[scale=0.31]{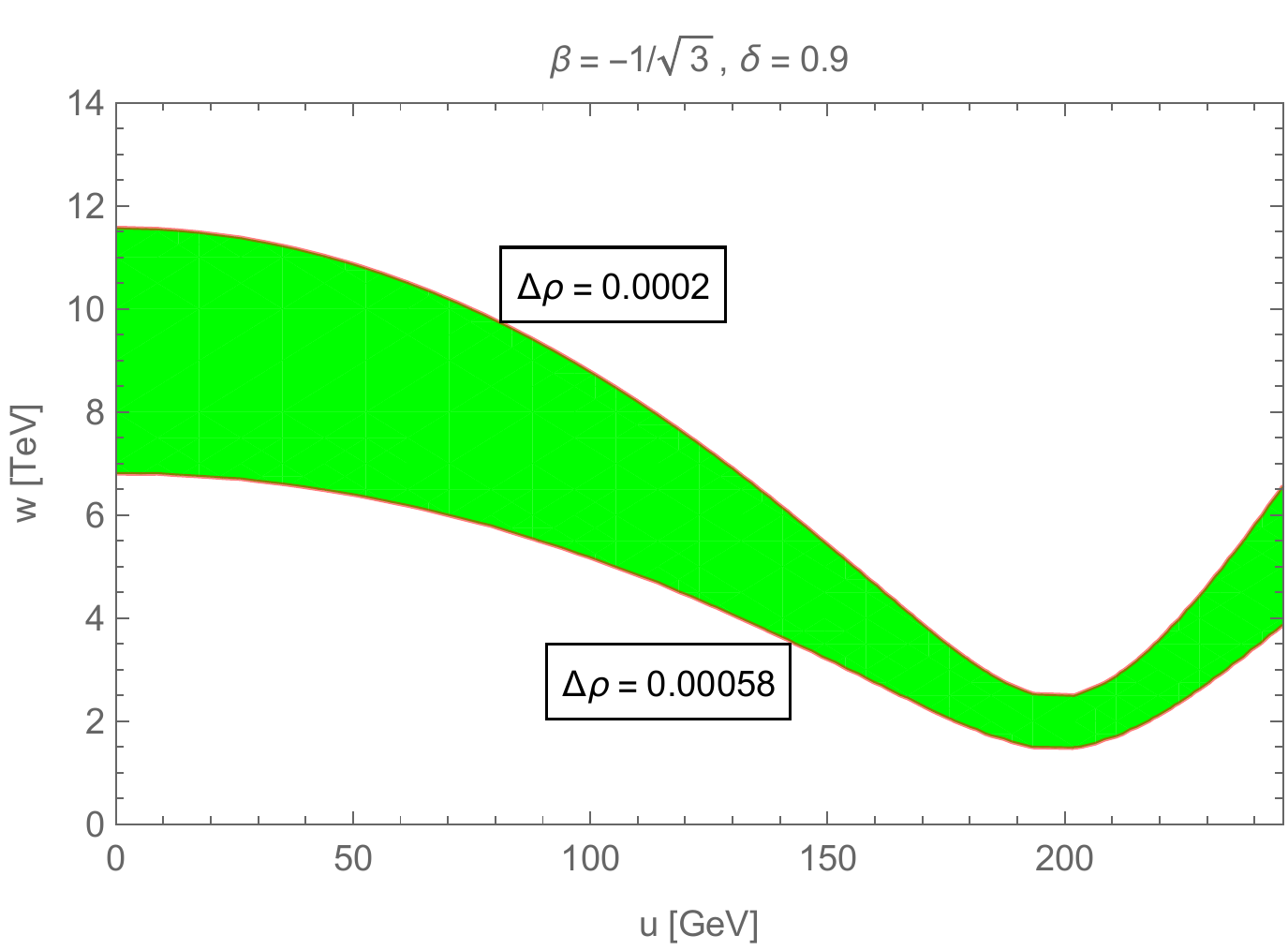}
\caption[]{\label{rho3311r} The $(u,w)$ regime constrained by $\Delta\rho$ for $\beta=-1/\sqrt{3}$, $b=-2/\sqrt{3}$, $t_N=0.5$, and $\La=2w$, where the panels correspond to $\delta=-0.9,\ -0.2$, and $0.9$.}
\end{center}
\end{figure}

\begin{figure}[!h]
\begin{center}
\includegraphics[scale=0.31]{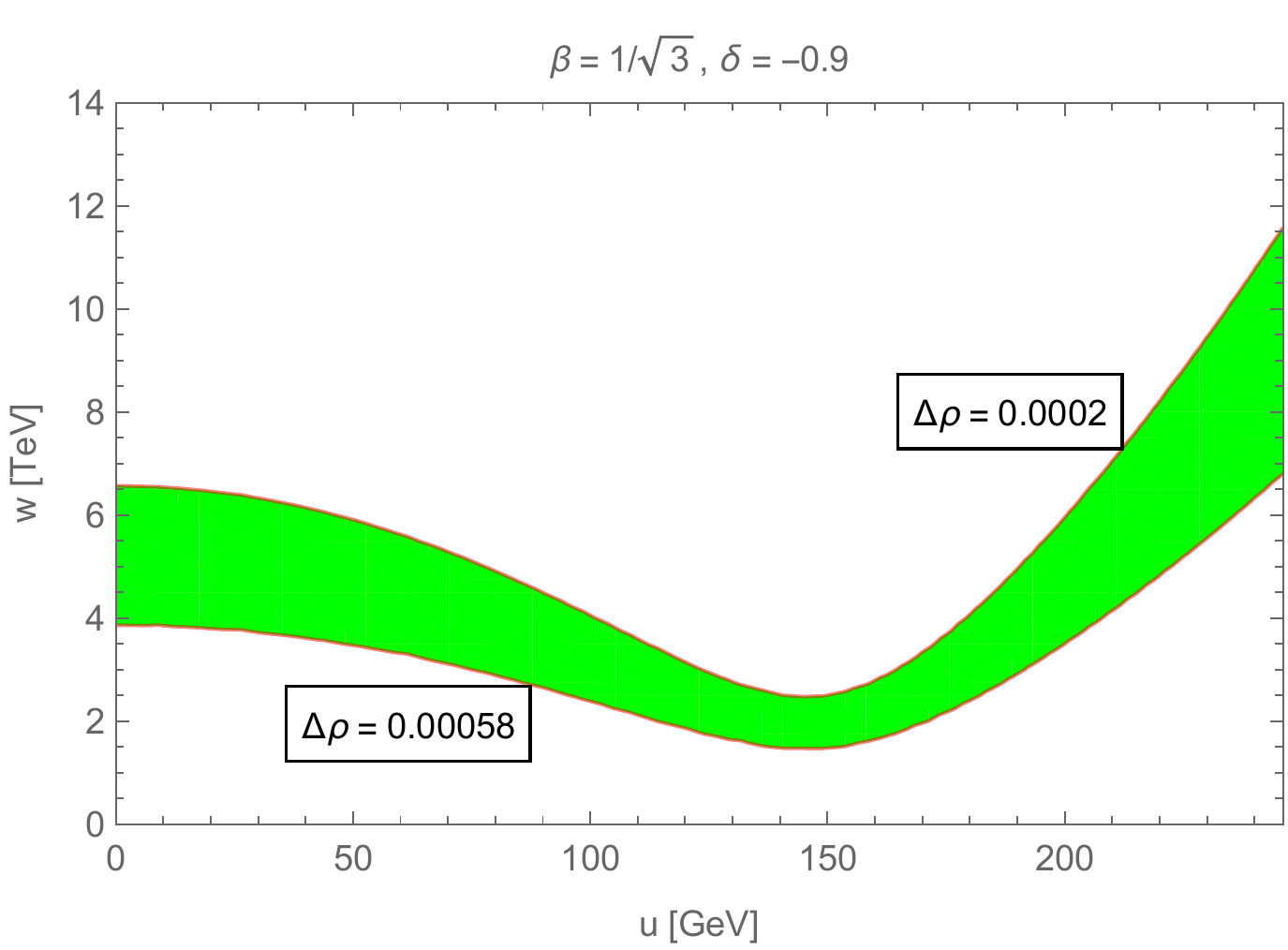}
\includegraphics[scale=0.31]{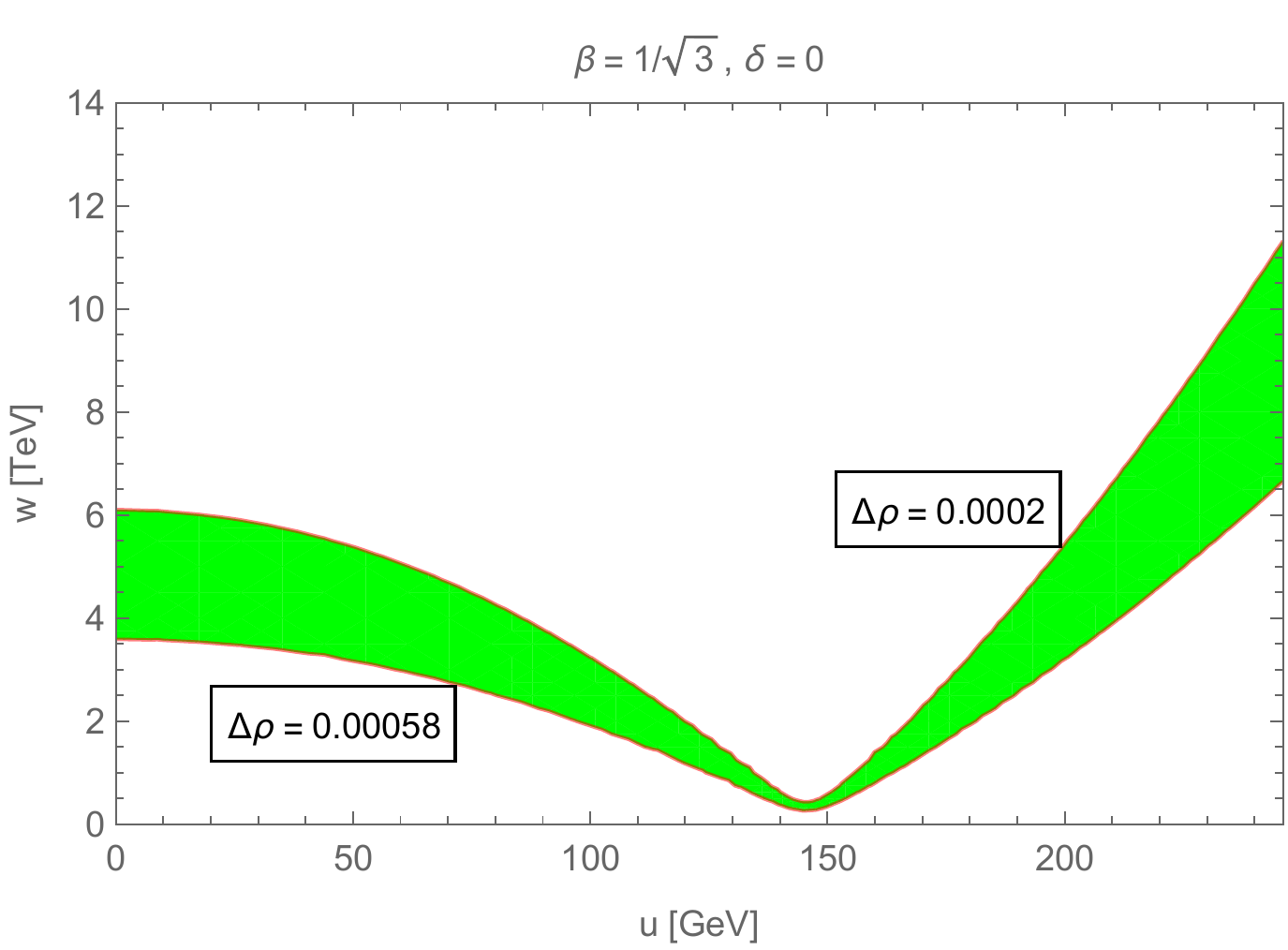}
\includegraphics[scale=0.31]{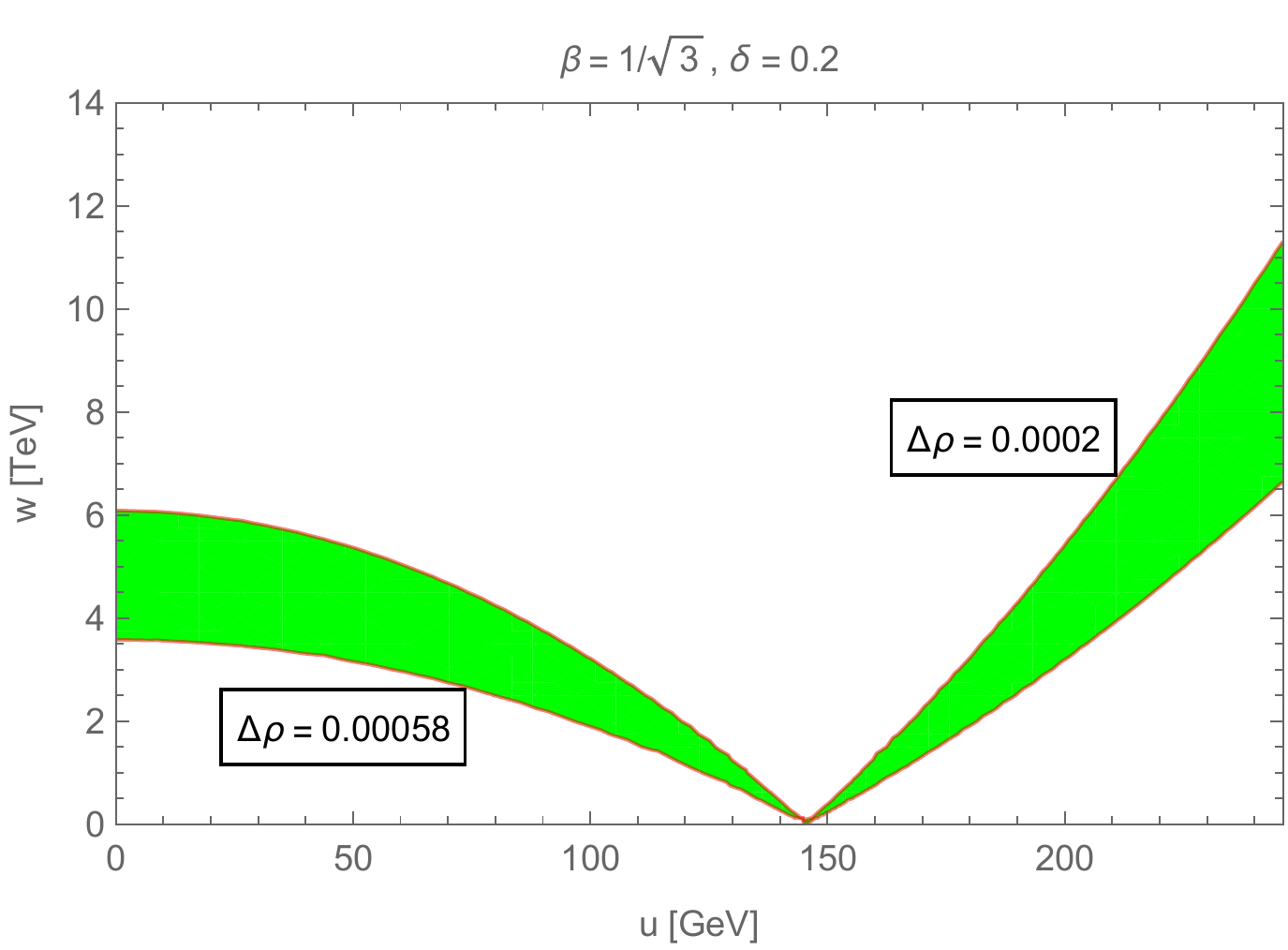}
\includegraphics[scale=0.31]{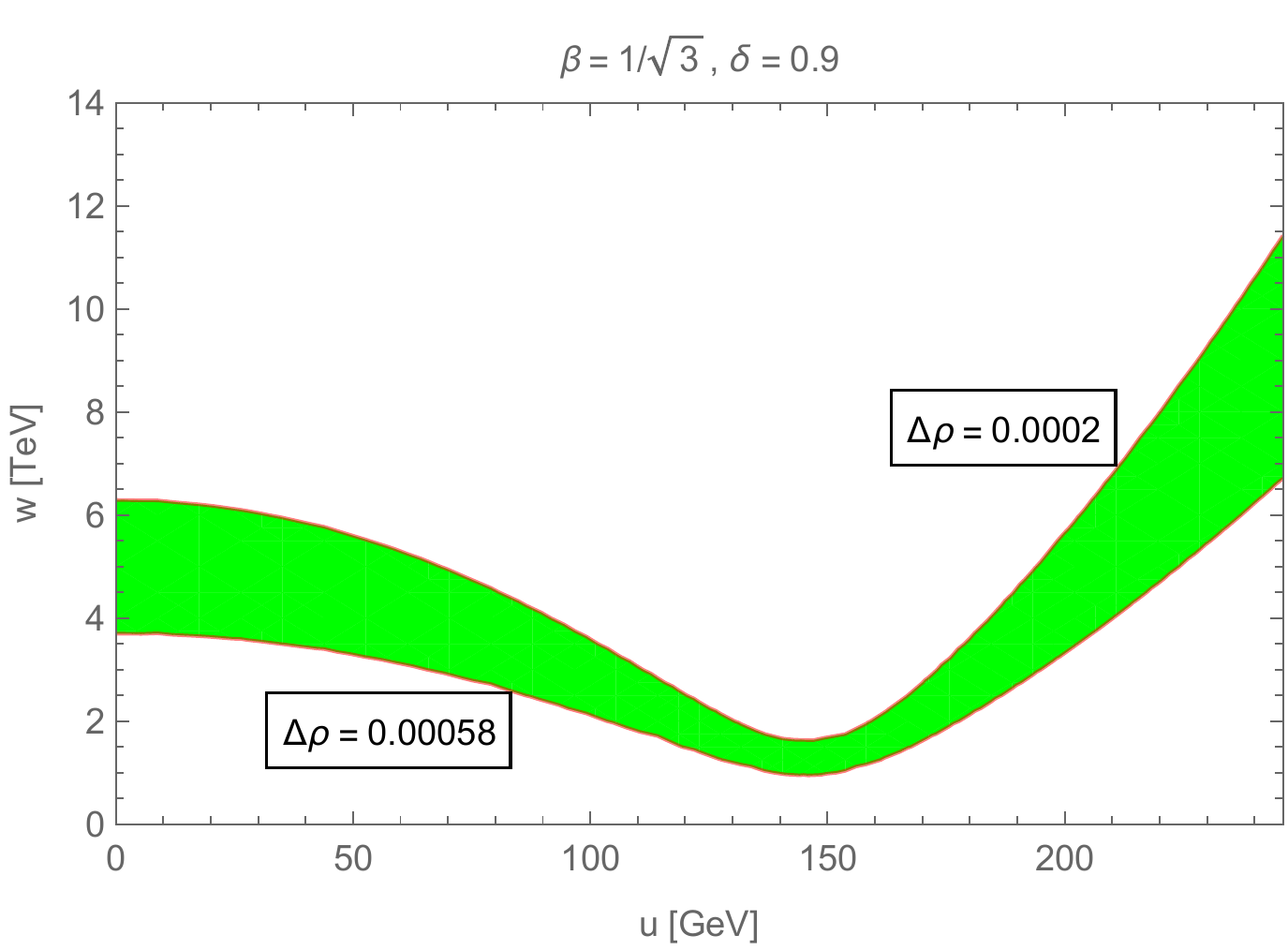}
\caption[]{\label{rho3311s} The $(u,w)$ regime constrained by $\Delta\rho$ for $\beta=1/\sqrt{3}$, $b=-2/\sqrt{3}$, $t_N=0.5$, and $\La=2w$, where the panels correspond to $\delta=-0.9,\ 0.2$, and $0.9$.}
\end{center}
\end{figure}

\begin{figure}[!h]
\begin{center}
\includegraphics[scale=0.35]{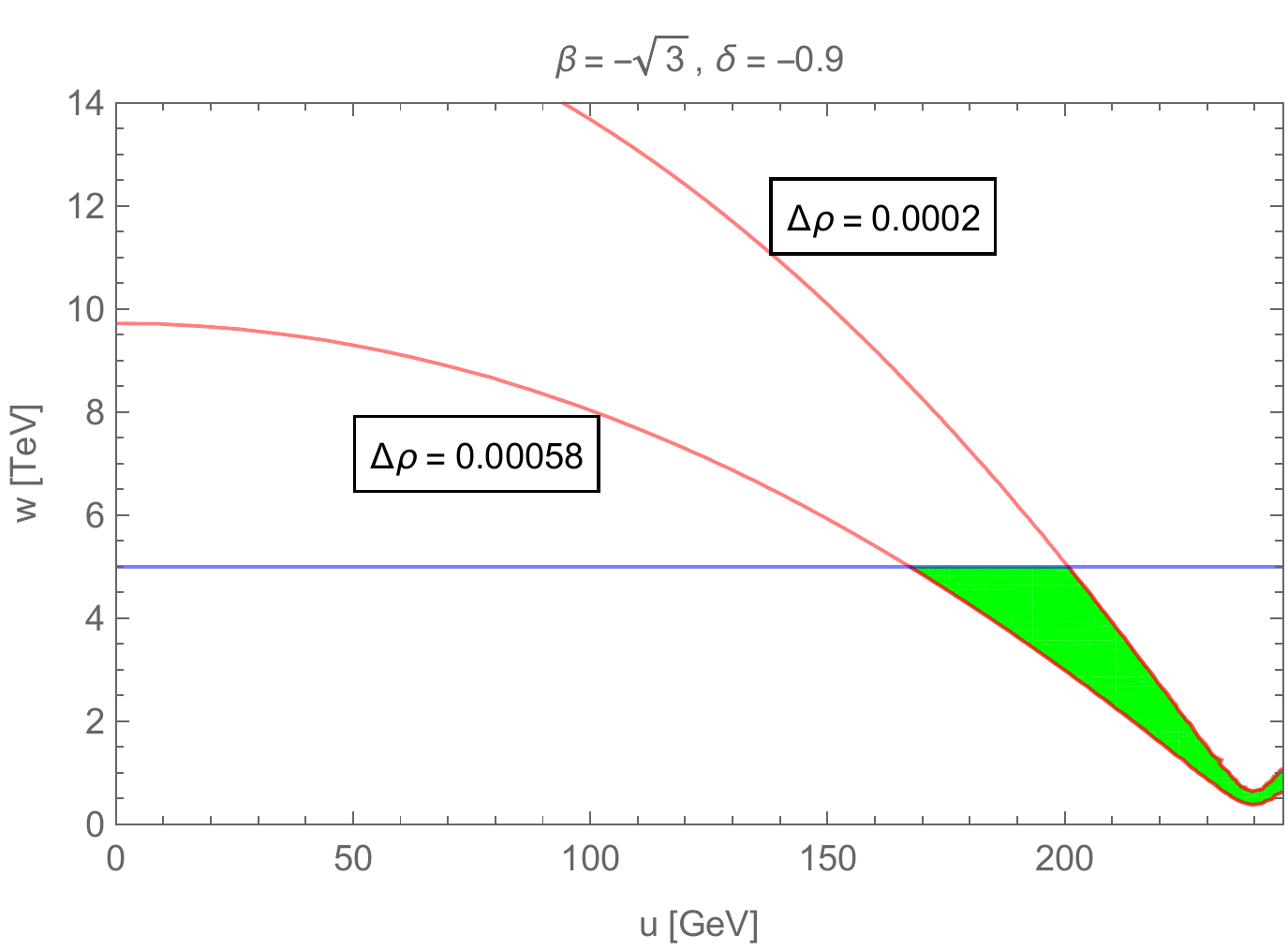}
\includegraphics[scale=0.35]{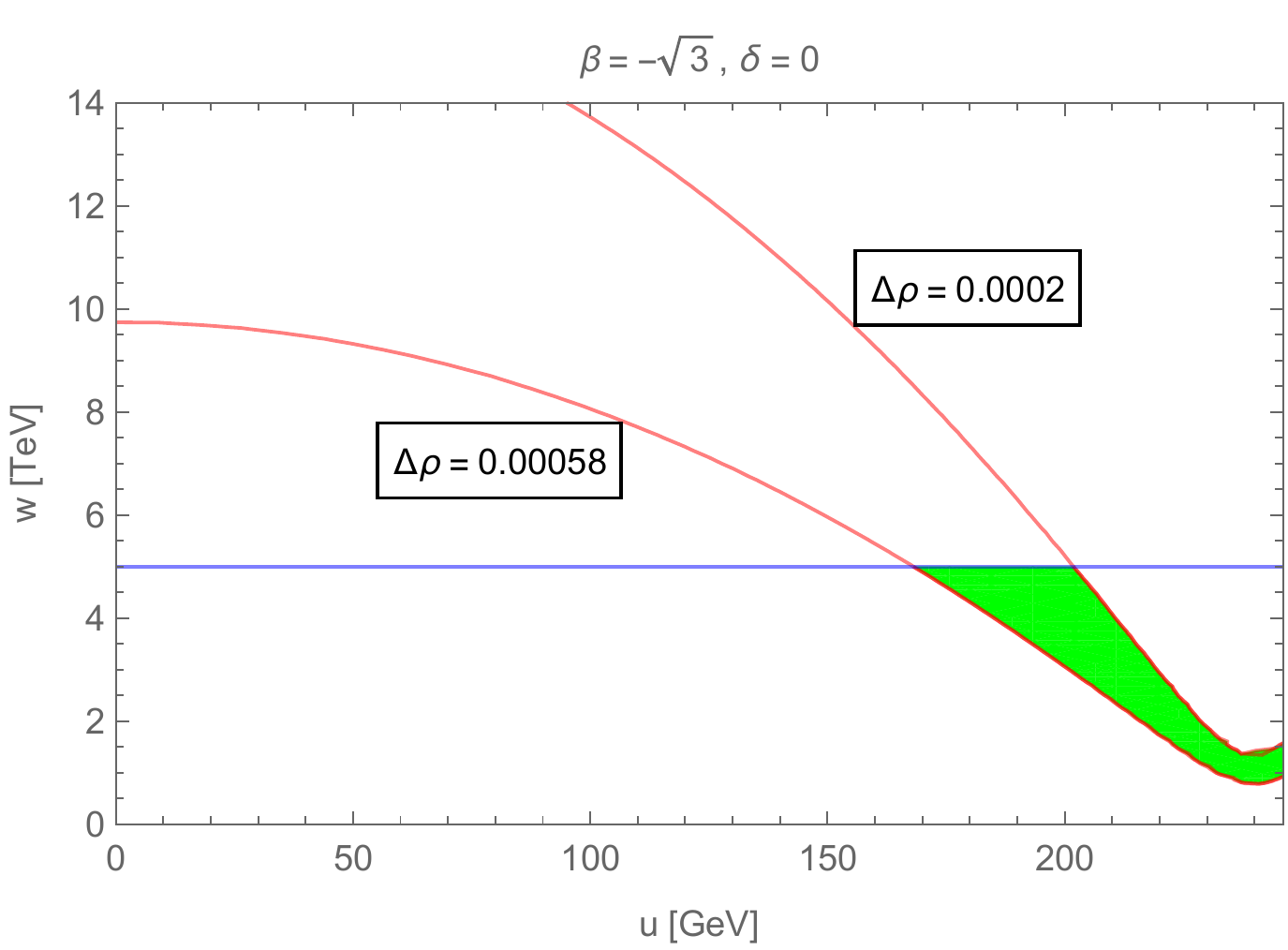}
\includegraphics[scale=0.35]{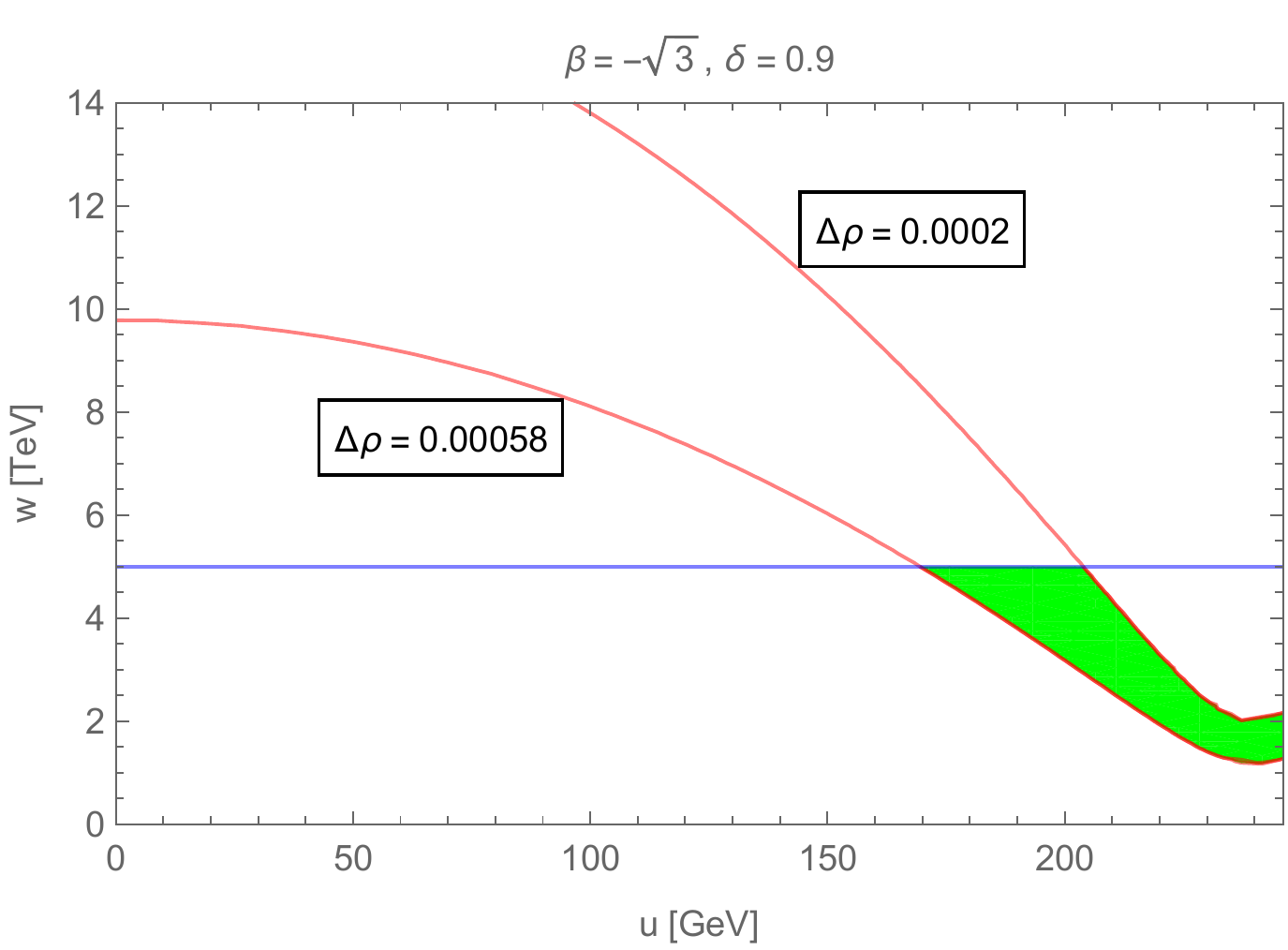}
\caption[]{\label{rho3311m} The $(u,w)$ regime that is bounded by the $\rho$ parameter for $\beta=-\sqrt{3}$, $b=-2/\sqrt{3}$, $t_N=0.5$, and $\La=2w$, where the panels, ordering from left to right, correspond to $\delta=-0.9,\ 0$, and $0.9$, respectively. In this case, the Landau pole, which is roundly $w=5$ TeV, is imposed.}
\end{center}
\end{figure}

We make a contour of $\Delta \rho$ as the function of $(u,w)$, as depicted in Figs. \ref{rho3311r}, \ref{rho3311s}, and \ref{rho3311m} for $\beta=-1/\sqrt{3}$, $\beta=1/\sqrt{3}$, and $\beta=-\sqrt{3}$, respectively. The effect of $\delta$ is quite similar to the 3-4-1-1 model and obviously different from \cite{dong2016}.

\begin{figure}[!h]
\begin{center}
\includegraphics[scale=0.35]{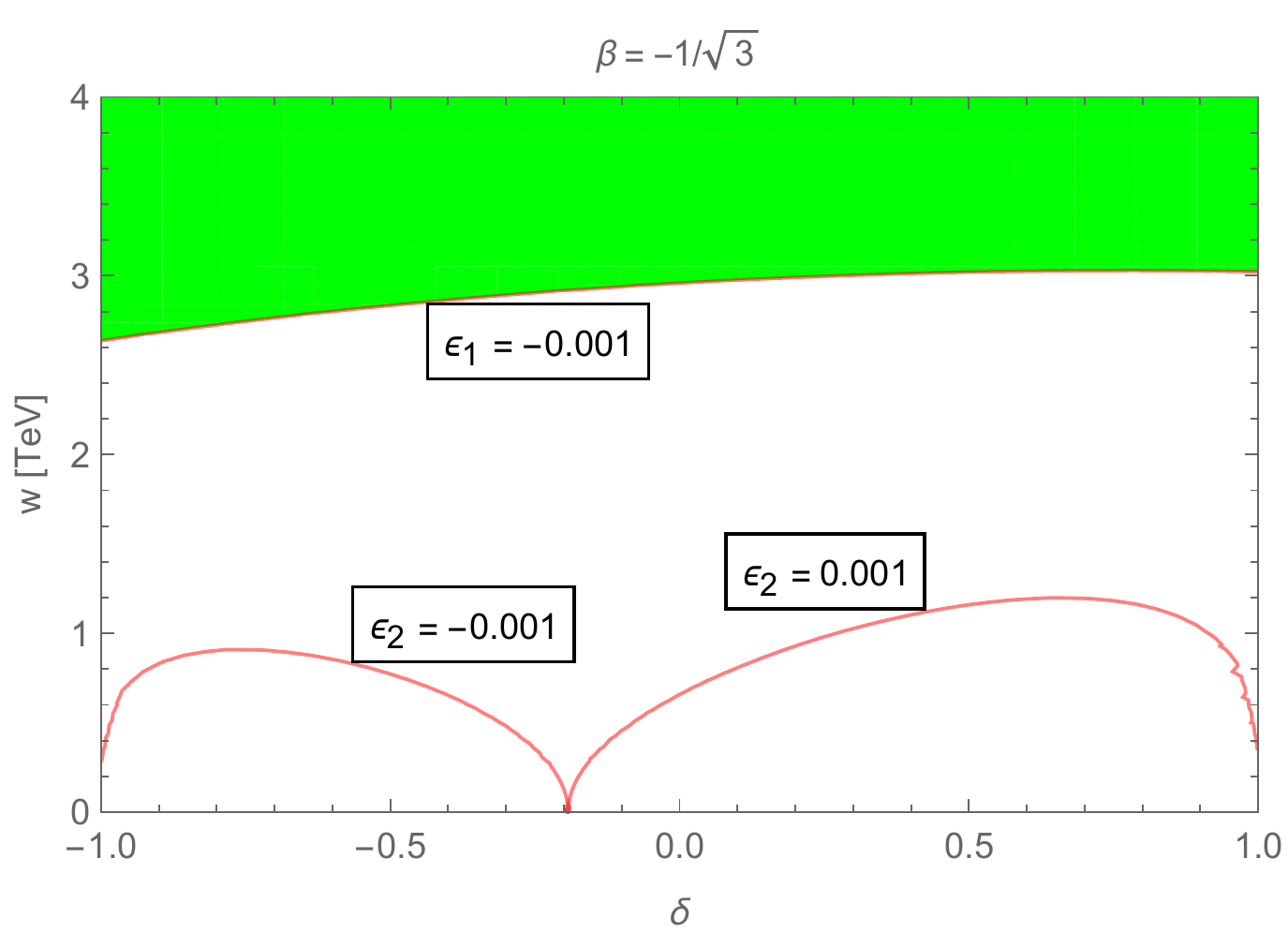}
\includegraphics[scale=0.35]{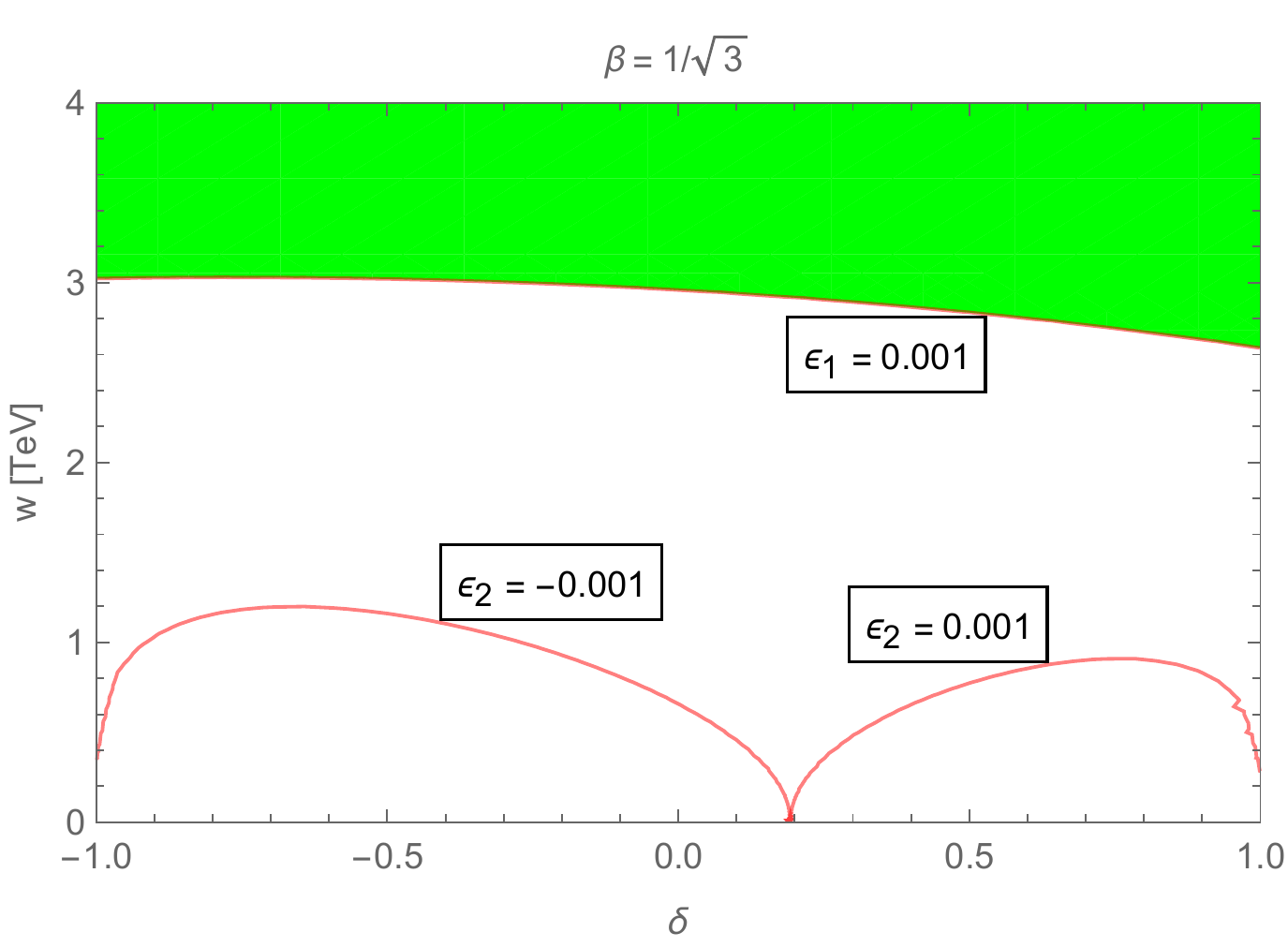}
\includegraphics[scale=0.35]{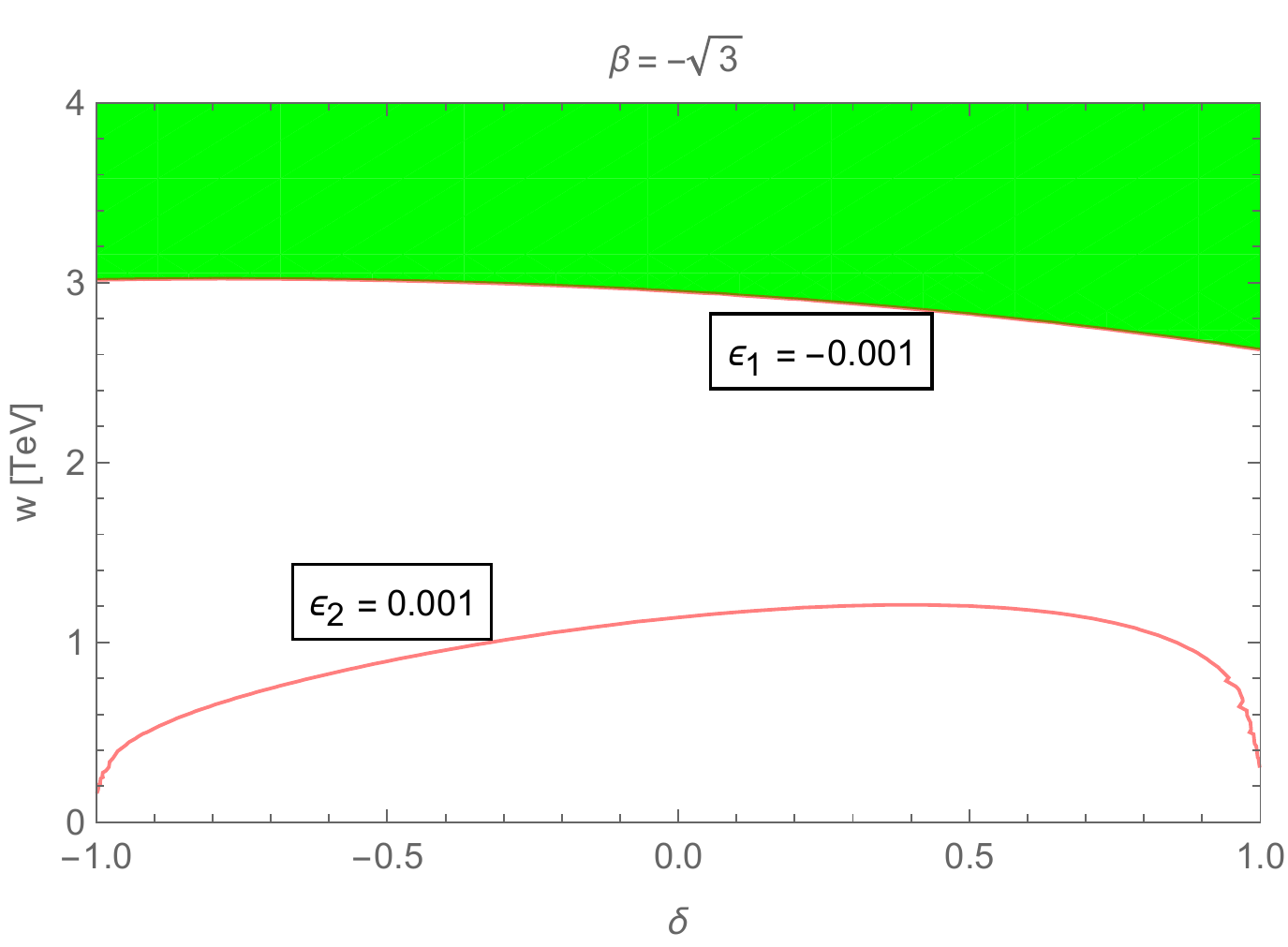}
\caption[]{\label{mixing} The bounds on the new physics scales as functions of $\delta$, contour by $|\epsilon_{1,2}|=10^{-3}$, for the three kinds of the models $\beta=-1/\sqrt{3}$, $\beta=1/\sqrt{3}$, and $\beta=-\sqrt{3}$, respectively.}
\end{center}
\end{figure}

The new physics contribution is safe, given that $|\epsilon_{1,2}|=10^{-3}$. Without loss of generality, we impose $u=v=246/\sqrt{2}$ as well as the given values of $\Lambda = 2w, t_N, \beta, b$ are used. In Fig. \ref{mixing}, $\epsilon_{1,2}$ are contoured as the functions of ($w,\delta$) for $\beta = -1/\sqrt3$, $\beta = 1/\sqrt3$, and $\beta = -\sqrt3$. It is clear that the new physics regime significantly changes when $\delta$ varies, in contradiction to \cite{dong2016}.

The meson mixing is described via the effective interaction \cite{dong2016}
\bea
 \mathcal{L}^{\mathrm{eff}}_{\mathrm{FCNC}}=(\bar{q}'_{iL}\ga^\mu q'_{jL})^2 [(V^*_{qL})_{3i}(V_{qL})_{3j}]^2\left(\fr{g^2_2}{m^2_{Z_2}}+\fr{g^2_3}{m^2_{Z_3}}\right),
\eea
where
 \bea
g_2&=&\fr{g}{\sqrt3}\left(\fr{1}{\sqrt{1-\beta^2t_W^2}}c_\xi+\fr{bt_N-\delta\beta t_X}{\sqrt{1-\delta^2}}s_\xi\right),\\
g_3&=&g_2(c_\xi\rightarrow s_\xi, s_\xi\rightarrow -c_\xi).
\eea
The $B^0_s-\bar{B}^0_s$ bound leads to \cite{dong2016}
\bea
\sqrt{\fr{g_2^2}{m_{Z_2}^2}+\fr{g_3^2}{m_{Z_3}^2}}<\fr{1}{3.9\text{ TeV}}.\label{FCNC3311}
\eea
When $w\ll \Lambda$, the above bound translates to $w>3.9$ TeV, independent of $\beta$, $b$, $g$, $g_X$, $g_N$, and $\delta$. When $w\sim \La$, using the existing values of parameters, the bound for both scales is similar to the previous case, which is quite in agreement with the conclusion in \cite{dong2016}.

\section{\label{con} Conclusion}

We have proved that the 3-4-1-1 model provides dark matter candidates naturally, besides supplying small neutrino masses via the seesaw mechanism induced by the gauge symmetry breaking.

The kinetic mixing effects are evaluated, yielding the new physics scales at TeV scale, in agreement with the collision bound. The kinetic mixing and symmetry breaking effects are canceled out only in the new gauge sector and differs between the 3-4-1-1 and 3-3-1-1 models.    
  
Similar to the 3-3-1-1 model \cite{a3311}, the 3-4-1-1 model can address the question of cosmic inflation as well as asymmetric dark and normal matter, which attracts much attention.

\section*{Acknowledgments}

\appendix

\section{\label{appa}Anomaly checking}

The anomalies that cause troublesome include $[SU(3)_C]^2U(1)_X$, $[SU(3)_C]^2U(1)_N$, $[SU(4)_L]^2U(1)_X$, $[SU(4)_L]^2U(1)_N$, $[\text{Gravity}]^2U(1)_X$, $[\text{Gravity}]^2U(1)_N$, $[U(1)_X]^2U(1)_N$, $U(1)_X[U(1)_N]^2$, $[U(1)_X]^3$, and $[U(1)_N]^3$. Let us verify each of them. 

\bea [SU(3)_C]^2U(1)_X&\sim& \sum_{\mathrm{quarks}} (X_{q_L}-X_{q_R}) \crn
&=& 4X_{Q_3}+2\times 4 X_{Q_\al}-3X_{u_a}-3X_{d_a}-X_{J_3}-X_{K_3}-2X_{J_\al}-2X_{K_\al}\crn
&=&4\left(\frac{p+q+5/3}{4}\right)+8\left(-\fr{p+q+1/3}{4}\right)-3\left(\frac 2 3\right)-3\left(\frac {-1}{3}\right)-\left(q+\frac 2 3\right)\crn
&&-\left(p+\frac 2 3\right)-2\left(-q-\frac 1 3\right)-2\left(-p-\frac 1 3\right)=0. \eea

\bea [SU(3)_C]^2U(1)_N&\sim& \sum_{\mathrm{quarks}} (N_{q_L}-N_{q_R}) \crn
&=& 4N_{Q_3}+2\times 4 N_{Q_\al}-3N_{u_a}-3N_{d_a}-N_{J_3}-N_{K_3}-2N_{J_\al}-2N_{K_\al}\crn
&=&4\left(\fr{m+n+10/3}{4}\right)+8\left(-\fr{m+n+2/3}{4}\right)-3\left(\frac 1 3\right)-3\left(\frac {1}{3}\right)-\left(n+\frac 4 3\right)\crn
&&-\left(m+\frac 4 3\right)-2\left(-n-\frac 2 3\right)-2\left(-m-\frac 2 3\right)=0. \eea

\bea [SU(4)_L]^2 U(1)_X &\sim& \sum_{\mathrm{(anti)quadruplets}} X_{F_L}= 3X_{\psi_a}+3X_{Q_3}+2\times 3 X_{Q_\al} \crn
&=& 3\left(\frac{p+q-1}{4}\right)+3\left(\frac{p+q+5/3}{4}\right)+6\left(-\frac{p+q+1/3}{4}\right)=0.  \eea 

\bea [SU(4)_L]^2 U(1)_N &\sim& \sum_{\mathrm{(anti)quadruplets}} N_{F_L}= 3N_{\psi_a}+3N_{Q_3}+2\times 3 N_{Q_\al} \crn
&=& 3\left(\frac{m+n-2}{4}\right)+3\left(\frac{m+n+10/3}{4}\right)+6\left(-\frac{m+n+2/3}{4}\right)=0.  \eea 

\bea [\mathrm{Gravity}]^2U(1)_X&\sim&\sum_{\mathrm{fermions}}(X_{f_L}-X_{f_R})\crn
&=&3\times 4 X_{\psi_a}+3\times 4 X_{Q_3}+2\times 3 \times 4 X_{Q_\al}-3\times 3 X_{u_a}-3\times 3 X_{d_a}\crn
&&-3X_{J_3}-3X_{K_3}-2\times 3 X_{J_\al}-2\times 3 X_{K_\al}-3X_{E_a}-3X_{F_a}-3X_{e_a}-3X_{\nu_a}\crn
&=&12 \left(\frac{p+q-1}{4}\right)+12\left(\frac{p+q+5/3}{4}\right)+24\left(-\frac{p+q+1/3}{4}\right)-9 \left(\fr 2 3\right)\crn
&&-9 \left(\fr {-1}{3}\right) -3\left(q+\frac 2 3\right)-3\left(p+\frac 2 3\right)-6 \left(-q-\frac 1 3\right)-6 \left(-p-\frac 1 3\right)\crn
&&-3q-3p-3(-1)-3(0)=0.\eea

\bea [\mathrm{Gravity}]^2U(1)_N&\sim&\sum_{\mathrm{fermions}}(N_{f_L}-N_{f_R})\crn
&=&3\times 4 N_{\psi_a}+3\times 4 N_{Q_3}+2\times 3 \times 4 N_{Q_\al}-3\times 3 N_{u_a}-3\times 3 N_{d_a}\crn
&&-3N_{J_3}-3N_{K_3}-2\times 3 N_{J_\al}-2\times 3 N_{K_\al}-3N_{E_a}-3N_{F_a}-3N_{e_a}-3N_{\nu_a}\crn
&=&12 \left(\frac{m+n-2}{4}\right)+12\left(\frac{m+n+10/3}{4}\right)+24\left(-\frac{m+n+2/3}{4}\right)\crn
&&-9\left(\fr 1 3\right)-9 \left(\fr 1 3\right) -3\left(n+\frac 4 3\right)-3\left(m+\frac 4 3\right)-6 \left(-n-\frac 2 3\right)-6 \left(-m-\frac 2 3\right)\crn
&&-3n-3m-3(-1)-3(-1)=0.\eea

\bea [U(1)_X]^2U(1)_N&=&\sum_{\mathrm{fermions}}(X^2_{f_L}N_{f_L}-X^2_{f_R}N_{f_R})=3\times 4 X^2_{\psi_a}N_{\psi_a}+3\times 4 X^2_{Q_3} N_{Q_3}\crn 
&&+2\times 3\times 4 X^2_{Q_{\al}}N_{Q_{\al}}-3\times 3 X^2_{u_a}N_{u_a}-3\times 3 X^2_{d_a}N_{d_a}-3X^2_{J_3} N_{J_3}-3X^2_{K_3} N_{K_3}\crn
&&-2\times 3 X^2_{J_\al}N_{J_\al}-2\times 3 X^2_{K_\al}N_{K_\al}-3X^2_{E_a} N_{E_a}-3X^2_{F_a} N_{F_a}-3X^2_{e_a} N_{e_a}-3X^2_{\nu_a}N_{\nu_a}\crn
&=&12 \left(\frac{p+q-1}{4}\right)^2\left(\frac{m+n-2}{4}\right)+12 \left(\frac{p+q+5/3}{4}\right)^2\left(\frac{m+n+10/3}{4}\right)\crn
&&+24 \left(-\frac{p+q+1/3}{4}\right)^2\left(-\frac{m+n+2/3}{4}\right)-9\left(\fr 2 3\right)^2\left(\fr 1 3\right)-9 \left(\fr {-1}{3}\right)^2\left(\fr 1 3\right)\crn
&&-3\left(q+\frac 2 3\right)^2\left(n+\frac 4 3\right)-3\left(p+\frac 2 3\right)^2\left(m+\frac 4 3\right)-6\left(-q-\frac 1 3\right)^2\left(-n-\frac 2 3\right)\crn
&&-6\left(-p-\frac 1 3\right)^2\left(-m-\frac 2 3\right)-3q^2n-3p^2m-3(-1)^2(-1)-3(0)^2(-1)=0. \eea

\bea [U(1)_X]U(1)_N^2&=&\sum_{\mathrm{fermions}}(X_{f_L}N^2_{f_L}-X_{f_R}N^2_{f_R})=3\times 4 X_{\psi_a}N^2_{\psi_a}+3\times 4 X_{Q_3} N^2_{Q_3}\crn 
&&+2\times 3\times 4 X_{Q_{\al}}N^2_{Q_{\al}}-3\times 3 X_{u_a}N^2_{u_a}-3\times 3 X_{d_a}N^2_{d_a}-3X_{J_3} N^2_{J_3}-3X_{K_3} N^2_{K_3}\crn
&&-2\times 3 X_{J_\al}N^2_{J_\al}-2\times 3 X_{K_\al}N^2_{K_\al}-3X_{E_a} N^2_{E_a}-3X_{F_a} N^2_{F_a}-3X_{e_a} N^2_{e_a}-3X_{\nu_a}N^2_{\nu_a}\crn
&=&12 \left(\frac{p+q-1}{4}\right)\left(\frac{m+n-2}{4}\right)^2+12 \left(\frac{p+q+5/3}{4}\right)\left(\frac{m+n+10/3}{4}\right)^2\crn
&&+24 \left(-\frac{p+q+1/3}{4}\right)\left(-\frac{m+n+2/3}{4}\right)^2-9\left(\fr 2 3\right)\left(\fr 1 3\right)^2-9 \left(\fr {-1}{3}\right)\left(\fr 1 3\right)^2\crn
&&-3\left(q+\frac 2 3\right)\left(n+\frac 4 3\right)^2-3\left(p+\frac 2 3\right)\left(m+\frac 4 3\right)^2-6\left(-q-\frac 1 3\right)\left(-n-\frac 2 3\right)^2\crn
&&-6\left(-p-\frac 1 3\right)\left(-m-\frac 2 3\right)^2-3qn^2-3pm^2-3(-1)(-1)^2-3(0)(-1)^2=0. \eea 

\bea [U(1)_X]^3&=&\sum_{\mathrm{fermions}}(X^3_{f_L}-X^3_{f_R})=3\times 4 X^3_{\psi_a}+3\times 4 X^3_{Q_3}+2\times 3\times 4 X^3_{Q_\al}-3\times 3 X^3_{u_a}\crn
&&-3\times 3 X^3_{d_a}-3X^3_{J_3}-3X^3_{K_3}-2\times 3 X^3_{J_\al}-2\times 3 X^3_{K_\al}-3X^3_{E_a}-3X^3_{F_a}-3X^3_{e_a}-3X^3_{\nu_a}\crn
&=&12 \left(\frac{p+q-1}{4}\right)^3+12 \left(\frac{p+q+5/3}{4}\right)^3+24 \left(-\frac{p+q+1/3}{4}\right)^3-9\left(\frac 2 3\right)^3\crn
&&-9\left(\frac {-1} {3}\right)^3-3\left(q+\frac 2 3\right)^3-3\left(p+\frac 2 3\right)^3-6 \left(-q-\frac 1 3\right)^3-6 \left(-p-\frac 1 3\right)^3\crn
&&-3q^3-3p^3-3(-1)^3-3(-0)^3=0.\eea  

\bea [U(1)_N]^3&=&\sum_{\mathrm{fermions}}(N^3_{f_L}-N^3_{f_R})=3\times 4 N^3_{\psi_a}+3\times 4 N^3_{Q_3}+2\times 3\times 4 N^3_{Q_\al}-3\times 3 N^3_{u_a}\crn
&&-3\times 3 N^3_{d_a}-3N^3_{J_3}-3N^3_{K_3}-2\times 3 N^3_{J_\al}-2\times 3 N^3_{K_\al}-3N^3_{E_a}-3N^3_{F_a}-3N^3_{e_a}-3N^3_{\nu_a}\crn
&=&12 \left(\frac{m+n-2}{4}\right)^3+12 \left(\frac{m+n+10/3}{4}\right)^3+24 \left(-\frac{m+n+2/3}{4}\right)^3-9\left(\frac 1 3\right)^3\crn
&&-9\left(\frac 1 3\right)^3-3\left(n+\frac 4 3\right)^3-3\left(m+\frac 4 3\right)^3-6\left(-n-\frac 2 3\right)^3-6\left(-m-\frac 2 3\right)^3\crn
&&-3n^3-3m^3-3(-1)^3-3(-1)^3=0.\eea  

This again confirms that the embedding coefficients $(\beta,\gamma,b,c)$ are independent of the anomalies.

\end{document}